\documentclass[10pt]{amsart}
\usepackage{amsmath}
\usepackage{amssymb}
\usepackage{amsfonts}
\usepackage{amsthm,amsxtra}
\usepackage{amsfonts}
\usepackage{nicefrac}

\theoremstyle{plain}
\newtheorem{theorem}{Theorem}[section]
\newtheorem{corollary}{Corollary}[section]
\newtheorem{lemma}{Lemma}[section]
\newtheorem{proposition}{Proposition}[section]

\theoremstyle{definition}
\newtheorem{definition}{Definition}[section]

\theoremstyle{remark}
\newtheorem{remark}{Remark}[section]

\newcommand{\PV}{${\rm P}_{\rm V}\;$}

\newcommand{\PVI}{${\rm P}_{\rm VI}\;$}

\newcommand{\ZZ}{\mathbb Z}
\newcommand{\RR}{\mathbb R}
\newcommand{\CC}{\mathbb C}
\newcommand{\TT}{\mathbb T}

\newcommand{\shalf}{{\scriptstyle \frac{1}{2}}}
\newcommand{\half}{\nicefrac{1}{2}}
\newcommand{\thalf}{\nicefrac{3}{2}}

\begin{document}

\vspace{4cm}
\noindent
{\bf Discrete Painlev\'e equations, Orthogonal Polynomials on the Unit Circle and 
$N$-recurrences for averages over $U(N)$ -- \PVI $\tau$-functions}

\vspace{5mm}
\noindent
P.J.~Forrester and N.S.~Witte

\noindent
Department of Mathematics and Statistics\\
University of Melbourne, Victoria 3010, Australia \\
email: p.forrester@ms.unimelb.edu.au; n.witte@ms.unimelb.edu.au

\small
\begin{quote}
The theory of orthogonal polynomials on the unit circle is developed for a
general class of weights leading to systems of recurrence relations and derivatives 
of the polynomials and their associated functions, and to functional-difference
equations of certain coefficient functions appearing in the theory.
A natural formulation of the Riemann-Hilbert problem is presented which has
as its solution the above system of orthogonal polynomials and associated functions. 
In particular for the case of regular semi-classical weights on the unit circle
$ w(z) = \prod^m_{j=1}(z-z_j(t))^{\rho_j} $, consisting of $ m \in \mathbb{Z}_{> 0} $
singularities, difference equations with respect to the orthogonal polynomial
degree $ n $ (Laguerre-Freud equations) and differential equations with respect to the 
deformation variables $ z_j(t) $ (Schlesinger equations) are derived completely 
characterising the system. It is shown in the 
simplest non-trivial case of $ m=3 $ that quite generally and simply the difference
equations are equivalent to the discrete Painlev\'e equation associated with the 
degeneration of the rational surface $ D^{(1)}_4 \to D^{(1)}_5 $ and no other. In
a three way comparison with other methods employed on this problem - the Toeplitz
lattice and Virasoro constraints, the isomonodromic deformation of $ 2\times 2 $
linear Fuchsian differential equations, and the algebraic approach based upon the 
affine Weyl group symmetry - it is shown all are entirely equivalent, when reduced
in order by exact summation, to the above discrete Painlev\'e equation through
explicit transformation formulae. The fundamental matrix integrals over the unitary group
$ U(N) $ arising in the theory are given by the generalised hypergeometric function 
$ {{}^{\vphantom{(1)}}_2}F^{(1)}_1 $. From the general results flow
a number of applications to physical models and we give the simplest, lowest order
recurrence relations for the gap probabilities and moments of characteristic 
polynomials of the circular unitary ensemble ($ {\rm CUE}_N $) of random matrices
and the diagonal spin-spin correlation function of the square lattice Ising model.  
\end{quote}


\tableofcontents

\section{Introduction}
\setcounter{equation}{0}

With $-\pi < \theta \le \pi$, $z_l := e^{i \theta_l}$ the unitary group
$U(N)$ with Haar (uniform) measure   
has eigenvalue probability density function (see e.g.~\cite[Chapter 2]{rmt_Fo})
\begin{equation}\label{1.3}
   {1 \over (2 \pi )^N N!} \prod_{1 \le j < k \le N} | z_k - z_j |^2.
\end{equation} 
Our interest is in averages over $U(N)$ of class functions $ w(U) $ which have 
the factorization property $ \prod_{l=1}^N w(z_l) $ for 
$ \{z_1,\dots, z_N \} \in {\rm Spec}(U(N)) $. 
Introducing the Fourier components $\{w_l\}_{l=0,\pm1,\dots}$ of the weight
$ w(z) $ by $ w(z) = \sum_{l=-\infty}^\infty w_l z^l $, due to
the well known identity
\cite{ops_Sz}
\begin{equation}\label{1.3a}
  \Big \langle \prod_{l=1}^N w(z_l) \Big \rangle_{U(N)} =
  \det[ w_{i-j} ]_{i,j=1,\dots,N}, 
\end{equation}
we are equivalently studying Toeplitz determinants. As an explicit example
consider the unitary average
\begin{equation}\label{VI_Toep}
  T_N(t;\omega_1,\omega_2,\mu;\xi) :=
  \Big \langle \prod_{l=1}^N(1 - \xi \chi_{(\pi - \phi, \pi)}^{(l)})
  e^{ \omega_2 \theta_l} |1 + z_l |^{2 \omega_1} \Big ( {1 \over t z_l} \Big )^\mu
  ( 1 + t z_l)^{2 \mu}
  \Big \rangle_{U(N)} \Big |_{t = e^{i \phi}} ,
\end{equation}
with $\chi_{(\pi - \phi, \pi)}^{(l)} = 1$ for  $\theta_l \in (\pi - \phi, \pi)$
and $\chi_{(\pi - \phi, \pi)}^{(l)} = 0$ otherwise.
For special choices of the parameters (\ref{VI_Toep}) occurs in a variety of
problems from mathematical physics. Thus the case $(\omega_1,\omega_2,\mu) =
(0,0,0)$ is the generating function for the probability that the interval
$(\pi - \phi, \pi)$ contains exactly $k$ eigenvalues in Dyson's circular
unitary ensemble (which is equivalent to the unitary group
with Haar measure), while the
case $(\omega_1,\omega_2,\mu) = (1,0,1)$ is (apart from a simple factor) the 
generating function for the probability density function of the event
that two eigenvalues in the circular unitary ensemble of $ (N+2) \times (N+2) $ 
matrices are an angle $\phi$ apart with exactly $k$ eigenvalues in
between. The case $\xi = 2$, $\omega_2 = 0$, $\mu = \omega_1 = 1/2$ of (\ref{VI_Toep})
corresponds to the density matrix for the impenetrable Bose gas
\cite{Le_1964,FFGW_2002a}. Furthermore
in the case $\xi = 0$ one sees that (\ref{VI_Toep}) includes as special cases
\begin{align}
  & \Big \langle \prod_{l=1}^N z_l^{1/4} |1 + z_l |^{-1/2}
  (1 + k^{-2}z_l)^{1/2} \Big \rangle_{U(N)}, \qquad
  1/k^2 \le 1, \label{1.4a}
  \\
  & \Big \langle \prod_{l=1}^N (1 + 1/z_l)^{v'}(1 + q^2 z_l)^v  
  \Big \rangle_{U(N)}, \qquad q^2 < 1. \label{1.4b} 
\end{align}
The average (\ref{1.4a}) is equivalent to the Toeplitz determinant given
by Onsager for the diagonal spin-spin correlation in the two-dimensional
Ising model \cite{McCW_1973}, while (\ref{1.4b}) occurs as a cumulative
probability density in the study of processes relating to increasing
subsequences \cite{BR_2001,BO_2000}.

In \cite{FW_2002b} the average (\ref{VI_Toep}) was characterized as a $\tau$-function 
in Okamoto's Hamiltonian theory of the sixth Painlev\'e (\PVI\!) equation, 
up to a change of variables and/or multiplication by an elementary function.
The $\tau$-function $ \tau(t) $ is defined to be 
\begin{equation}
    H = {d \over dt}\log \tau ,
\end{equation}
where the Hamiltonian $ H(q,p;\alpha,t) $ is a rational function of 
the co-ordinates and momenta $ q,p $ and of parameters 
$ \alpha = (\alpha_0,\alpha_1,\alpha_2,\alpha_3,\alpha_4) $,
($ \alpha_0+\alpha_1+2\alpha_2+\alpha_3+\alpha_4 = 1 $),
and independent variable (deformation variable) $ t $. 
The dynamics $ T_t : q(0),p(0) \mapsto q(t),p(t) $ is governed by the Hamilton 
equations
\begin{equation}\label{VI_Hdyn}
  {dq \over dt} =  {\partial H\over \partial p} , \qquad
  {dp \over dt} = -{\partial H\over \partial q}
\end{equation}
where eliminating $ p(t) $ gives the sixth Painlev\'e equation in $ q(t) $.
The above identification implies that the logarithmic derivative of $T_N$ 
with respect to $ t $ is an auxiliary Hamiltonian for the \PVI system and
it was shown in \cite{FW_2002b} that it satisfies a difference equation related to 
a particular discrete Painlev\'e equation with respect to increments in unit 
amounts of one of the parameters $ \alpha $ (or in particular $ \mu $). One of the
objectives of this paper is to derive discrete Painlev\'e type recurrences
for $T_N$ directly with respect to increments in $N$ only.
In the algebraic approach \cite{FW_2003a} the strategy is to choose
a particular shift operator or Schlesinger transformation $ L $ constructed from
compositions of fundamental reflection operators and Dynkin diagram automorphisms
of the \PVI symmetry group 
$ W_a(D^{(1)}_4) = \langle s_0,s_1,s_2,s_3,s_4,r_1,r_3,r_4 \rangle $ and use it 
to generate sequences of parameters and dynamical variables 
$ L^n : \{\alpha, q, p, H, \tau \} \mapsto \{\alpha_n, q_n, p_n, H_n, \tau_n \} $
for $ n=0,1,\ldots $. This operator is chosen to increment the parameters such
that only $ N \mapsto N+1 $ and thus $ \tau_N $ is essentially the average
(\ref{VI_Toep}). In all our applications the initial point $ \alpha(N=0) $ is located on
a reflection hyperplane in the space $ \alpha \in \CC^4 $ and this characterises
the entire sequence as classical solutions to the \PVI system.
The initial member $ \tau_0=1 $ and the first nontrivial member 
$ \tau_1 $ is a solution of the Gauss hypergeometric differential equation.

The discrete Painlev\'e equation that is fundamental in the \PVI system is 
that associated with the degeneration of the rational surface 
$ D^{(1)}_4 \to D^{(1)}_5 $ 
\begin{align}
  g_{n+1}g_n 
  & = t{(f_n+1-\alpha_2)(f_n+1-\alpha_0-\alpha_2) \over f_n(f_n+\alpha_3)} 
  \label{dPV:a} \\
  f_n+f_{n-1} & = -\alpha_3 + {\alpha_1 \over g_n-1} +
                  {\alpha_4 t \over g_n-t} ,
  \label{dPV:b}
\end{align}
where 
$ \alpha_1 \mapsto \alpha_1+1,\alpha_2 \mapsto \alpha_2-1,\alpha_4 \mapsto \alpha_4+1 $
as $ n \mapsto n+1 $.
We label the particular discrete equations that arise according to the 
unambiguous algebraic-geometric classification of Sakai \cite{Sa_2001} by
association with a degeneration of a particular rational surface into 
another, rather than the historical names employed (discrete fifth Painlev\'e
equation, {${\rm dP}_{\rm V}$}). In its full generality the average (\ref{VI_Toep})
was first characterised in terms of (\ref{dPV:a}) and (\ref{dPV:b}) in \cite{FW_2003a}.
In fact this discrete Painlev\'e equation (but different in detail than the ones 
we will present below including that in \cite{FW_2003a}) for the average 
(\ref{1.4b}) is already known from the work of Borodin \cite{Bo_2001}. In 
contrast recurrences for the average (\ref{VI_Toep}) with $ \xi=0 $ have also 
been obtained recently by Adler and van Moerbeke from their theory of the 
Toeplitz lattice and its Virasoro algebra \cite{AvM_2002} that do not appear 
to relate to (\ref{dPV:a}) and (\ref{dPV:b}) \cite{BB_2002}. Another of our 
objectives is to show that indeed the recurrences of \cite{AvM_2002} are of 
the discrete Painlev\'e type, by deriving explicit transformation formulae 
between them and (\ref{dPV:a}),(\ref{dPV:b}).

Because we are dealing with Toeplitz determinants with symbols $ w(z) $ satisfying
certain analytic conditions it is immediate from the theory of Szeg\"o,
Geronimus and others that orthogonal polynomial systems on the unit circle 
$\{ \phi_n(z) \}^{\infty}_{n=0} $ with respect to such weights are relevant.
For example if $ \phi \in [0,2\pi) $, $ \omega_1,\omega_2,\mu,\xi \in \RR $, $ \xi < 1 $,
and $ 2\omega_1 > -1, 2\mu > -1 $ then $ w(e^{i\theta}) $ where
\begin{equation*}
   w(z) = z^{-\mu-\omega}(1+z)^{2\omega_1}(1+tz)^{2\mu}
   \begin{cases}
     1 & \theta \notin (\pi-\phi,\pi) \\
     1-\xi & \theta \in (\pi-\phi,\pi)
   \end{cases} ,
\end{equation*}
is a real, positive weight defining a measure with an infinite number of points of 
increase and thus an orthogonal polynomial system on $ \TT $ exists with respect 
to this weight by Favard's theorem. 
To characterise the averages (\ref{VI_Toep}) using orthogonal polynomial 
theory we have found it necessary to substantially develop the general theory of
such systems. To a large extent this task has been completed for orthogonal
polynomial systems defined on the line in the works of Bauldry \cite{Ba_1990}, 
Bonan and Clark \cite{BC_1990}, Belmehdi and Ronveaux \cite{BR_1994},
Magnus \cite{Ma_1999,Ma_1995a,Ma_1994} 
but had remained incomplete for those systems on the unit circle \cite{IW_2001}.
To this end we have derived closed systems of differential relations for the 
polynomials, their reciprocal polynomials 
$\{ \phi^*_n(z) \}^{\infty}_{n=0} $, and associated functions 
$\{ \epsilon_n(z) \}^{\infty}_{n=0} $, $\{ \epsilon^*_n(z) \}^{\infty}_{n=0} $
in Proposition \ref{ops_spectralD}.
In the notation of (\ref{ops_Ydefn}) and Corollary 2.3 let
\begin{equation*}
   Y_n(z;t) := 
   \begin{pmatrix}
          \phi_n(z)   &  \epsilon_n(z)/w(z) \cr
          \phi^*_n(z) & -\epsilon^*_n(z)/w(z) \cr
   \end{pmatrix}  ,
\end{equation*}
and set
\begin{equation*}
   {d \over dz}Y_{n} := A_n Y_{n} .
\end{equation*} 
Entries in the matrix $ A_n $ are fixed by four coefficient functions
$ \Omega_n(z),\Omega^*_n(z), \Theta_n(z), \Theta^*_n(z) $ in (\ref{ops_YzDer})
and complete sets of difference and functional relations for these coefficient 
functions are given in Proposition \ref{ops_Linear1} and Corollary \ref{ops_Linear2}. 
We also formulate a $ 2 \times 2$ matrix Riemann-Hilbert problem in Proposition
\ref{ops_RHP} for general classes of weights which parallels the case for orthogonal 
polynomials on the line \cite{IKF_1991,FIK_1991,FIK_1992,De_1999} and whose 
solution is simply related to $ Y_n $. 
For our particular applications the weight (\ref{VI_Toep}) is a member of the regular 
semi-classical class 
\begin{equation*}
  w(z) = \prod^m_{j=1}(z-z_j(t))^{\rho_j}, \quad \rho_j \in \CC,
\end{equation*}
with an arbitrary number $ m $ of isolated singularities located at $ z_j(t) $. 
A key feature of such weights is that
\begin{equation*}
  {1\over w(z)}{d \over dz}w(z) = {2V(z) \over W(z)} ,
\end{equation*}
where the polynomials $ {\rm deg}V(z) < m, {\rm deg}W(z)=m $.
The coefficient functions for regular semi-classical weights are polynomials of 
$ z $ with bounded degree 
$ {\rm deg}\Omega_n(z)={\rm deg}\Omega^*_n(z)=m-1, 
  {\rm deg}\Theta_n(z)={\rm deg}\Theta^*_n(z)=m-2 $
(see Proposition \ref{ops_SCpoly}).
In addition evaluations of these functions at the singular points satisfy bilinear 
relations (see Proposition \ref{ops_Bilinear}) which lead directly to one of the 
pair of coupled discrete Painlev\'e equations. 
Deformation derivatives of the linear system of differential equations above 
with respect to arbitrary trajectories of the singularities are given in Proposition
\ref{ops_deformD} which can summarised as
\begin{equation*}
   {d \over dt}Y_{n} := B_n Y_{n}
   = \left\{ B_{\infty} - \sum^{m}_{j=1}{A_{nj} \over z-z_j}{d \over dt}z_j
        \right\} Y_{n} ,\quad \text{where} \quad
   A_n = \sum^{m}_{j=1}{A_{nj} \over z-z_j} ,
\end{equation*} 
and consequently systems of Schlesinger equations for the elements of $ A_{nj} $
(or the coefficient functions evaluated at $ z_j $) are given in 
(\ref{ops_Schl:a}-\ref{ops_Schl:c}). 
It is quite natural that systems governed by regular
semi-classical weights preserve the monodromy data of the solutions $ Y_n $ about 
each singularity $ z_j $ with respect to arbitrary deformations.

Another theme we wish to develop is the evaluation of the above Toeplitz
determinants in terms of generalised hypergeometric functions. Given a partition
$ \kappa = (\kappa_1,\kappa_2, \ldots, \kappa_N) $ such that 
$ \kappa_1 \geq \kappa_2 \geq \cdots \geq \kappa_N \geq 0 $ one defines the
generalised, multi-variable hypergeometric function through a series representation
\cite{Ya_1992,Ka_1993}
\begin{equation}
  {}^{\vphantom{(1)}}_{p}F^{(1)}_{q}(a_1,\ldots,a_p;b_1,\ldots,b_q;t_1,\ldots,t_N)
  = \sum^{\infty}_{\kappa \geq 0} 
    {[a_1]^{(1)}_{\kappa}\cdots [a_p]^{(1)}_{\kappa} \over 
     [b_1]^{(1)}_{\kappa}\cdots [b_q]^{(1)}_{\kappa}}
    {s_{\kappa}(t_1,\ldots,t_N) \over h_{\kappa}}
\end{equation}
for $ p,q \in \ZZ_{\geq 0} $. Here the generalised Pochhammer symbols are
\begin{equation}
   [a]^{(1)}_{\kappa} := \prod^{N}_{j=1}(a-j+1)_{\kappa_j} ,
\end{equation}
the hook length is
\begin{equation}
   h_{\kappa} = \prod_{(i,j) \in \kappa} [a(i,j)+l(i,j)+1] ,
\end{equation}
where $ a(i,j), l(i,j) $ are the arm and leg lengths of the $ (i,j) $th
box in the Young diagram of the partition $ \kappa $, and
$ s_{\kappa}(t_1,\ldots,t_N) $ is the Schur symmetric polynomial of 
$ N $ variables. The superscript $ (1) $ distinguishes these functions from the 
single variable $ N=1 $ functions and also indicates that they are a special 
case of a more general function parameterised by an arbitrary complex number 
$ d \neq 1 $.

In Section 2 we derive systems of differential-difference
and functional relations for orthogonal polynomials and associated functions
on the unit circle for a general class of weights and formulate the Riemann-Hilbert 
problem. 
In Section 3 we specialise to regular semi-classical weights and derive bilinear
difference equations. In addition we calculate the deformation derivatives of the 
orthogonal polynomial system, derive a system of Schlesinger equations
and show the deformations are of the isomonodromic type. The foregoing theory is 
utilised in the simplest case of three singularities and the $ N $-recurrences 
derived in Section 4. In Section 5
the connection of the orthogonal polynomial theory with the Okamoto $\tau$-function 
theory is established. Application of our recurrences to the physical 
models described previously are considered in Section 6.

\section{Orthogonal Polynomials on the Unit Circle and Riemann-Hilbert Problem}
\label{OPSsection}
\setcounter{equation}{0}

We consider a complex function for our formal weight $ w(z) $, analytic in the cut
complex $ z $-plane and which possesses a Fourier expansion
\begin{equation}
  w(z) = \sum_{k=-\infty}^{\infty} w_{k}z^k, \quad
  w_{k} = \int_{\TT} {d\zeta \over 2\pi i\zeta} w(\zeta)\zeta^{-k},
\end{equation}
where $ \TT $ denotes the unit circle $ |\zeta|=1 $, with 
$ z=e^{i\theta}, \theta \in (-\pi,\pi] $. Hereafter we will assume that
$ z^jw(z), z^jw'(z) \in L(\TT) $ for all $ j \in \ZZ $. The doubly infinite 
sequence $ \{ w_k \}^{\infty}_{k=-\infty} $ are the trigonometric moments of the 
distribution $ w(e^{i\theta})d\theta/2\pi $ and define the trigonometric moment 
problem. Define the Toeplitz determinants
\begin{align}
   I^{\epsilon}_{n}[w] 
  & := \det \left[ \int_{\TT} {d\zeta \over 2\pi i\zeta} w(\zeta)\zeta^{\epsilon+j-k}
           \right]_{0 \leq j,k \leq n-1},
  \nonumber \\
  & = \det \left[ w_{-\epsilon+j-k} \right]_{0 \leq j,k \leq n-1},
  \nonumber \\
  & = {1 \over n!} \int_{\TT^n}
      \prod^{n}_{l=1} {d\zeta_l \over 2\pi i\zeta_l} w(\zeta_l)\zeta_l^{\epsilon}
      \prod_{1 \leq j<k \leq n} |\zeta_{j}-\zeta_{k}|^2,
\end{align}
where $ \epsilon = 0,\pm 1 $. The last equality shows the determinants are equivalent to 
the CUE averages defined earlier and their generalisations. In certain circumstances the 
weight is real and positive, $ \overline{w(z)} = w(z) $ where the bar denotes complex
conjugate, for example when all the monodromy parameters are real and certain
independent deformation variables $ t \in \TT $,  and so the Toeplitz matrix $ I^0_n[w] $
is Hermitian (see the example of the weight in section \ref{PVIsection}) but in general 
this will not be true.

Furthermore consider the system of orthogonal polynomials 
$ \{ \phi_n(z) \}_{n \in \ZZ_{\geq 0}} $ defined with respect to the weight $ w(z) $
on the unit circle, assuming that none of the $ I^{0}_{n}[w] $ vanish. This
system is taken to be orthonormal 
\begin{equation}
  \int_{\TT} {d\zeta \over 2\pi i\zeta} w(\zeta)\phi_m(\zeta)\overline{\phi_n(\zeta)}  
   = \delta_{m,n}
\label{ops_onorm}
\end{equation}
and the leading and trailing coefficients are defined by
\begin{equation}
   \phi_n(z) = \kappa_n z^n + l_n z^{n-1}+ m_n z^{n-2} + \ldots + \phi_n(0)
             = \sum^{n}_{j=0} c_{n,j}z^j,
\end{equation}
where $ \kappa_n $ is chosen to be real and positive without loss of generality.
We also define the reciprocal polynomial by
\begin{equation}
   \phi^{*}_n(z) := z^n\bar{\phi}_n(1/z) = \sum^{n}_{j=0} \bar{c}_{n,j}z^{n-j} .
\end{equation}
The orthogonal polynomials are defined up to an overall factor by the orthogonality
with respect to the monomials
\begin{equation}
  \int_{\TT} {d\zeta \over 2\pi i\zeta} w(\zeta)\phi_n(\zeta)\overline{\zeta^j}  
   = 0 \qquad 0 \leq j \leq n-1 ,
\label{ops_orthog:a}
\end{equation}
whereas their reciprocal polynomials are similarly defined by
\begin{equation}
  \int_{\TT} {d\zeta \over 2\pi i\zeta} w(\zeta)\phi^*_n(\zeta)\overline{\zeta^j}  
   = 0 \qquad 1 \leq j \leq n .
\label{ops_orthog:b}
\end{equation}

The system is alternatively defined by the sequence of ratios $ r_n = \phi_n(0)/\kappa_n $,
known as reflection coefficients because of their role in the scattering theory
formulation of OPS on the unit circle, together with a companion quantity
$ \bar{r}_n $ (notwithstanding the notation, only when $ w(z) $ is real does 
$ \bar{r}_n $ equal the complex conjugate of $ r_n $). From the Szeg\"o theory 
\cite{ops_Sz} $ r_n $ and $ \bar{r}_n $ are related to the above Toeplitz 
determinants by
\begin{equation}
  r_n = (-1)^n{ I^{1}_{n}[w] \over I^{0}_{n}[w]}, \quad
  \bar{r}_n = (-1)^n{ I^{-1}_{n}[w] \over I^{0}_{n}[w]}.
\end{equation}
The Toeplitz determinants of central interest can then be recovered from
\begin{equation}
   {I^{0}_{n+1}[w] I^{0}_{n-1}[w] \over (I^{0}_{n}[w])^2}
   = 1 - r_{n}\bar{r}_n.
\label{ops_I0}
\end{equation}
Further additional identities from the Szeg\"o theory that relate the leading 
coefficients back to the reflection coefficients are
\begin{align}
   \kappa_n^2 & = \kappa_{n-1}^2 + |\phi_n(0)|^2,
   \label{ops_kappa}\\
   {l_{n} \over \kappa_{n}} & = \sum^{n-1}_{j=0} r_{j+1}\bar{r}_{j},
   \label{ops_l}\\
   {m_{n} \over \kappa_{n}} 
    & = \sum^{n-1}_{j=1} r_{j+1}\Big[ \bar{r}_{j-1} 
                       + \bar{r}_{j}{l_{j-1} \over \kappa_{j-1}} \Big].
   \label{ops_m}
\end{align}
Some useful relations for the leading coefficients of the product of a monomial
and an orthogonal polynomial or its derivative are 
\begin{align}
\begin{split}
  z\phi_n(z) & = {\kappa_{n} \over \kappa_{n+1}} \phi_{n+1}(z) 
       + \left({l_{n} \over \kappa_{n}}-{l_{n+1} \over \kappa_{n+1}}\right) \phi_{n}(z)
  \\
  & \quad
       + \left\{ {l_{n} \over \kappa_{n-1}}
                 \left( {l_{n+1} \over \kappa_{n+1}}-{l_{n} \over \kappa_{n}}\right)
                + {m_{n} \over \kappa_{n-1}}-{m_{n+1} \over \kappa_{n+1}}
                  {\kappa_{n} \over \kappa_{n-1}} \right\} \phi_{n-1}(z)
       + \pi_{n-2}
  \\
  z^2\phi_n(z) & ={\kappa_{n} \over \kappa_{n+2}} \phi_{n+2}(z) 
       + \left({l_{n} \over \kappa_{n+1}}-{l_{n+2} \over \kappa_{n+2}}
               {\kappa_{n} \over \kappa_{n+1}}\right) \phi_{n+1}(z)
  \\
  & \quad
       + \left\{ {l_{n+1} \over \kappa_{n+1}}
                 \left( {l_{n+2} \over \kappa_{n+2}}-{l_{n} \over \kappa_{n}}\right)
                + {m_{n} \over \kappa_{n}}-{m_{n+2} \over \kappa_{n+2}}
                  \right\} \phi_{n}(z) + \pi_{n-1} 
  \\
  \phi'_n(z) 
  & = n{\kappa_{n} \over \kappa_{n-1}} \phi_{n-1}(z) + \pi_{n-2} 
  \\
  z\phi'_n(z)
  & = n \phi_n(z) - {l_{n} \over \kappa_{n-1}} \phi_{n-1}(z)
                  + \pi_{n-2}
  \\
  z^2\phi'_n(z)
  & = n{\kappa_{n} \over \kappa_{n+1}} \phi_{n+1}(z)
         + \left\{(n-1){l_{n} \over \kappa_{n}}-n{l_{n+1} \over \kappa_{n+1}}
           \right\} \phi_n(z) + \pi_{n-1}
\end{split}
\label{ops_prod}
\end{align}
where $ ' $ denotes the derivative with respect to $ z $ and
where $ \pi_{n} $ denotes an arbitrary polynomial of the linear space of polynomials 
with degree at most $ n $.
 
Fundamental consequences of the orthogonality condition are the mixed linear 
recurrence relations
\begin{align}
  \kappa_n  \phi_{n+1}(z)
   & = \kappa_{n+1}z \phi_{n}(z)+\phi_{n+1}(0) \phi^*_n(z)
  \label{ops_rr:a} \\
  \kappa_n \phi^*_{n+1}(z)
   & = \kappa_{n+1} \phi^*_{n}(z)+\bar{\phi}_{n+1}(0) z\phi _n(z)
  \label{ops_rr:b}
\end{align}
as well as the three-term recurrences
\begin{align}
  \kappa_n\phi_n(0)\phi_{n+1}(z) + \kappa_{n-1}\phi_{n+1}(0)z\phi_{n-1}(z)
   & = [\kappa_{n}\phi_{n+1}(0)+\kappa_{n+1}\phi_{n}(0)z]\phi_n(z)
  \label{ops_ttr:a} \\
  \kappa_n\bar{\phi}_n(0)\phi^*_{n+1}(z) + \kappa_{n-1}\bar{\phi}_{n+1}(0)z\phi^*_{n-1}(z)
   & = [\kappa_{n}\bar{\phi}_{n+1}(0)z+\kappa_{n+1}\bar{\phi}_{n}(0)]\phi^*_n(z)
  \label{ops_ttr:b} 
\end{align}

The analogue of the Christoffel-Darboux summation formula is
\begin{align}
  \sum^{n}_{j=0} \phi_j(z)\overline{\phi_j(\zeta)} 
  & =
  { \phi^*_n(z)\overline{\phi^*_n(\zeta)}
   -z\bar{\zeta}\phi_n(z)\overline{\phi_n(\zeta)} \over 1-z\bar{\zeta} }
  \label{ops_CD:a} \\
  & = 
  { \phi^*_{n+1}(z)\overline{\phi^*_{n+1}(\zeta)}
   -\phi_{n+1}(z)\overline{\phi_{n+1}(\zeta)} \over 1-z\bar{\zeta} } ,
  \label{ops_CD:b}
\end{align}
for $ z\bar{\zeta} \not= 1 $.

Equations (\ref{ops_ttr:a},\ref{ops_ttr:b}) being second order linear difference 
equations admit other linearly independent solutions $ \psi_n(z), \psi^*_n(z) $, 
and we define two such polynomial solutions, the polynomials of the second kind 
or associated polynomials
\begin{equation}
  \psi_n(z) 
  := \int_{\TT}{d\zeta \over 2\pi i\zeta}{\zeta+z \over \zeta-z}w(\zeta)
               [\phi_n(\zeta)-\phi_n(z)],
       \quad n \geq 1, \quad \psi_0 := 1,
\label{ops_psi:a}
\end{equation}
and its reciprocal polynomial $ \psi^*_n(z) $. 
The integral formula for $ \psi^*_{n} $ is
\begin{equation}
  \psi^*_n(z) 
  := -\int_{\TT}{d\zeta \over 2\pi i\zeta}{\zeta+z \over \zeta-z}w(\zeta)
                [z^n\overline{\phi_n(\zeta)}-\phi^*_n(z)],
       \quad n \geq 1, \quad \psi^*_0 := 1.
\label{ops_psi:b}
\end{equation} 
A central object in the theory is the Carath\'eodory function, or generating 
function of the Toeplitz elements
\begin{equation}
   F(z) := \int_{\TT}{d\zeta \over 2\pi i\zeta}{\zeta+z \over \zeta-z}w(\zeta)
\label{ops_Cfun}
\end{equation}
which has the expansions inside and outside the unit circle
\begin{equation}
   F(z) = \begin{cases}
            1 + 2 \sum^{\infty}_{k=1}w_{k} z^k, 
           & \text{if $ |z| < 1 $}, \\
           -1 - 2 \sum^{\infty}_{k=1}w_{-k} z^{-k}, 
           & \text{if $ |z| > 1 $}.
          \end{cases}
\end{equation}
Having these definitions one requires two non-polynomial solutions 
$ \epsilon_n(z), \epsilon^*_n(z) $ to the recurrences and these are constructed 
as linear combinations of the polynomial solutions according to 
\begin{align}
   \epsilon_n(z) := \psi_n(z)+F(z)\phi_n(z)
   & = \int_{\TT}{d\zeta \over 2\pi i\zeta}{\zeta+z \over \zeta-z}w(\zeta)
                   \phi_n(\zeta)
   \label{ops_eps:a} \\
   \epsilon^*_n(z) := \psi^*_n(z)-F(z)\phi^*_n(z)
   & = -z^n\int_{\TT}{d\zeta \over 2\pi i\zeta}{\zeta+z \over \zeta-z}w(\zeta)
                   \overline{\phi_n(\zeta)}
   \nonumber \\
   & = {1 \over \kappa_n} 
          -\int_{\TT}{d\zeta \over 2\pi i\zeta}{\zeta+z \over \zeta-z}w(\zeta)
                   \phi^*_n(\zeta)
   \label{ops_eps:b}
\end{align}

\begin{theorem}[\cite{Ge_1961},\cite{Ge_1962},\cite{Ge_1977},\cite{JNT_1989}]
$ \psi_n(z), \psi^*_{n}(z)$ satisfy the three-term recurrence relations 
(\ref{ops_ttr:a}, \ref{ops_ttr:b}) and along with 
$ \epsilon_{n}(z), \epsilon^*_{n}(z) $ satisfy a variant of 
(\ref{ops_rr:a},\ref{ops_rr:b}) namely
\begin{align}
  \kappa_n  \epsilon_{n+1}(z)
   & = \kappa_{n+1}z \epsilon_{n}(z)-\phi_{n+1}(0) \epsilon^*_n(z)
  \label{ops_rre:a} \\
  \kappa_n \epsilon^*_{n+1}(z)
   & = \kappa_{n+1} \epsilon^*_{n}(z)-\bar{\phi}_{n+1}(0) z\epsilon _n(z)
  \label{ops_rre:b}
\end{align}
\end{theorem}

\begin{theorem}[\cite{Ge_1961}]
The Casoratians of the polynomial solutions 
$ \phi_n, \phi^*_n, \psi_n, \psi^*_{n} $ are
\begin{align}
   \phi_{n+1}(z)\psi_n(z) - \psi_{n+1}(z)\phi_n(z) 
  & = \phi_{n+1}(z)\epsilon_n(z) - \epsilon_{n+1}(z)\phi_n(z)
    = 2{\phi_{n+1}(0) \over \kappa_n}z^n
  \label{ops_Cas:a} \\
   \phi^*_{n+1}(z)\psi^*_n(z) - \psi^*_{n+1}(z)\phi^*_n(z) 
  & = \phi^*_{n+1}(z)\epsilon^*_n(z) - \epsilon^*_{n+1}(z)\phi^*_n(z) 
    = 2{\bar{\phi}_{n+1}(0) \over \kappa_n}z^{n+1} 
  \label{ops_Cas:b} \\
   \phi_{n}(z)\psi^*_n(z) + \psi_{n}(z)\phi^*_n(z) 
  & = \phi_{n}(z)\epsilon^*_n(z) + \epsilon_{n}(z)\phi^*_n(z)
    = 2z^n   
  \label{ops_Cas:c}
\end{align} 
\end{theorem}

We will require the leading order terms in expansions of 
$ \phi_n(z), \phi^*_n(z), \epsilon_n(z), \epsilon^*_{n}(z) $ both inside and 
outside the unit circle.
\begin{corollary}
The orthogonal polynomials $ \phi_n(z), \phi^*_n(z) $ have the following 
expansions
\begin{align}
   \phi_n(z) & = 
   \begin{cases}
      \phi_n(0)
       + \dfrac{1}{\kappa_{n-1}}
         (\kappa_n\phi_{n-1}(0)+\phi_n(0)\bar{l}_{n-1})z
       + {\rm O}(z^2) & |z| < 1 \\
      \kappa_n z^n + l_n z^{n-1} + {\rm O}(z^{n-2}) & |z| > 1 
   \end{cases} \label{ops_phiexp:a} \\
   \phi^*_n(z) & = 
   \begin{cases}
      \kappa_n + \bar{l}_n z + {\rm O}(z^{2}) & |z| < 1 \\
      \bar{\phi}_n(0) z^n
       + \dfrac{1}{\kappa_{n-1}}
         (\kappa_n\bar{\phi}_{n-1}(0)+\bar{\phi}_n(0)l_{n-1})z^{n-1}
       + {\rm O}(z^{n-2}) & |z| > 1 \\
   \end{cases} \label{ops_phiexp:b}
\end{align}
whilst the associated functions have the expansions
\begin{align}
   \dfrac{\kappa_n}{2}\epsilon_n(z) & = 
   \begin{cases}
      z^n
       - \dfrac{\bar{l}_{n+1}}{\kappa_{n+1}}z^{n+1}
       + {\rm O}(z^{n+2}) & |z| < 1 \\
         \dfrac{\phi_{n+1}(0)}{\kappa_{n+1}}z^{-1}
       + \left(\dfrac{\kappa^2_n}{\kappa^2_{n+1}}
               \dfrac{\phi_{n+2}(0)}{\kappa_{n+2}}
               -\dfrac{\phi_{n+1}(0)}{\kappa_{n+1}}
                \dfrac{l_{n+1}}{\kappa_{n+1}}
         \right)z^{-2} + {\rm O}(z^{-3}) & |z| > 1 
   \end{cases} \label{ops_epsexp:a} \\
   \dfrac{\kappa_n}{2}\epsilon^*_n(z) & = 
   \begin{cases}
         \dfrac{\bar{\phi}_{n+1}(0)}{\kappa_{n+1}}z^{n+1}
       + \left(\dfrac{\kappa^2_n}{\kappa^2_{n+1}}
                \dfrac{\bar{\phi}_{n+2}(0)}{\kappa_{n+2}}
               -\dfrac{\bar{\phi}_{n+1}(0)}{\kappa_{n+1}}
                \dfrac{\bar{l}_{n+1}}{\kappa_{n+1}}
         \right)z^{n+2} + {\rm O}(z^{n+3}) & |z| < 1 \\
       1 - \dfrac{l_{n+1}}{\kappa_{n+1}}z^{-1} + \left(
         \dfrac{l_{n+2}l_{n+1}}{\kappa_{n+2}\kappa_{n+1}}
        -\dfrac{m_{n+2}}{\kappa_{n+2}}
         \right)z^{-2} + {\rm O}(z^{-3}) & |z| > 1
   \end{cases} \label{ops_epsexp:b}
\end{align}
\end{corollary}

The $z$-derivatives or spectral derivatives of the orthogonal polynomials in 
general are related to two consecutive polynomials \cite{IW_2001} and we generalise
this with the following parameterisation.
\begin{proposition}\label{ops_spectralD}
The derivatives of the orthogonal polynomials and associated functions are
expressible as linear combinations in a related way ($ ' := d/dz $), 
\begin{align}
   W(z)\phi'_n(z) & = 
   \Theta_n(z) \phi_{n+1}(z) - (\Omega_n(z)+V(z)) \phi_n(z)
   \label{ops_zD:a} \\
   W(z)\phi^*_n{\!'}(z) & = 
   -\Theta^*_n(z) \phi^*_{n+1}(z) + (\Omega^*_n(z)-V(z)) \phi^*_n(z)
   \label{ops_zD:c} \\
   W(z)\epsilon'_n(z) & = 
   \Theta_n(z) \epsilon_{n+1}(z) - (\Omega_n(z)-V(z)) \epsilon_n(z)
   \label{ops_zD:b} \\
   W(z)\epsilon^*_n{\!'}(z) & = 
   -\Theta^*_n(z) \epsilon^*_{n+1}(z) + (\Omega^*_n(z)+V(z)) \epsilon^*_n(z)
   \label{ops_zD:d} 
\end{align}
with coefficient functions $ W(z), V(z) $ independent of $ n $.
\end{proposition}
\begin{proof}
The first, (\ref{ops_zD:a}), was found in \cite{IW_2001} where the coefficients
were taken to be (their notation $ A_n, B_n $ should not be confused with our 
use of it subsequently)
\begin{align}
   A_n & = -{\kappa_{n-1}\phi_{n+1}(0) \over \kappa_{n}\phi_{n}(0)}
            {z\Theta_n(z) \over W(z)}
  \\
   B_n & = {1 \over W(z)}\left( \Omega_n(z)+V(z)
            - \left[{\phi_{n+1}(0) \over \phi_{n}(0)}+{\kappa_{n+1} \over \kappa_{n}}z 
              \right]\Theta_n(z) \right) .
\end{align}
The other differential relations can be found in an analogous manner.
\end{proof}

The coefficient functions 
$ \Theta_n(z), \Omega_n(z), \Theta^*_n(z), \Omega^*_n(z) $
satisfy coupled linear recurrence relations themselves, one of
which was reported in \cite{IW_2001}. The full set are given in the following 
proposition.
\begin{proposition}\label{ops_Linear1}
The coefficient functions satisfy the coupled linear recurrence relations 
\begin{equation}
  \Omega_n(z) + \Omega_{n-1}(z) 
  - \left( {\phi_{n+1}(0) \over \phi_{n}(0)}+{\kappa_{n+1} \over \kappa_{n}}z
    \right)\Theta_n(z) + (n-1){W(z) \over z} = 0
\label{ops_rrCf:a}
\end{equation}
\begin{multline}
   \left( {\phi_{n+1}(0) \over \phi_{n}(0)}+{\kappa_{n+1} \over \kappa_{n}}z
   \right) (\Omega_{n-1}(z) - \Omega_{n}(z)) 
   \\
   + {\kappa_{n}\phi_{n+2}(0) \over \kappa_{n+1}\phi_{n+1}(0)}z\Theta_{n+1}(z)
   - {\kappa_{n-1}\phi_{n+1}(0) \over \kappa_{n}\phi_{n}(0)}z\Theta_{n-1}(z)
   - {\phi_{n+1}(0) \over \phi_{n}(0)}{W(z) \over z} = 0
\label{ops_rrCf:b}
\end{multline}
\begin{equation}
  \Omega^*_n(z) + \Omega^*_{n-1}(z) 
  - \left( {\kappa_{n+1} \over \kappa_{n}}+{\bar{\phi}_{n+1}(0) \over \bar{\phi}_{n}(0)}z
    \right)\Theta^*_n(z) - n{W(z) \over z} = 0
\label{ops_rrCf:c}
\end{equation}
\begin{multline}
   \left( {\kappa_{n+1} \over \kappa_{n}}+{\bar{\phi}_{n+1}(0) \over \bar{\phi}_{n}(0)}z
   \right) (\Omega^*_{n-1}(z) - \Omega^*_{n}(z)) 
   \\
   + {\kappa_{n}\bar{\phi}_{n+2}(0) \over \kappa_{n+1}\bar{\phi}_{n+1}(0)}
      z\Theta^*_{n+1}(z)
   - {\kappa_{n-1}\bar{\phi}_{n+1}(0) \over \kappa_{n}\bar{\phi}_{n}(0)}
      z\Theta^*_{n-1}(z)
   + {\kappa_{n+1} \over \kappa_{n}}{W(z) \over z} = 0
\label{ops_rrCf:d}
\end{multline}
\begin{equation}
  \Omega_{n+1}(z) + \Omega^*_{n}(z) 
  - \left( {\phi_{n+2}(0) \over \phi_{n+1}(0)}+{\kappa_{n+2} \over \kappa_{n+1}}z
    \right)\Theta_{n+1}(z)
  + {\kappa_{n+1} \over \kappa_{n}}(z\Theta_n(z)-\Theta^*_n(z)) = 0
\label{ops_rrCf:e}
\end{equation}
\begin{multline}
   \Omega_{n}(z) - \Omega_{n+1}(z) 
   + {\kappa_{n+2} \over \kappa_{n+1}}
    \left( z+{\bar{\phi}_{n+1}(0) \over \kappa_{n+1}}{\phi_{n+2}(0) \over \kappa_{n+2}}
    \right)\Theta_{n+1}(z) \\
   + {|\phi_{n+1}(0)|^2 \over \kappa_{n+1}\kappa_{n}}\Theta^*_n(z)
   - {\kappa_{n+1} \over \kappa_{n}}z\Theta_{n}(z)
   - {W(z) \over z} = 0
\label{ops_rrCf:f}
\end{multline}
\begin{multline}
  \Omega^*_{n+1}(z) + \Omega_{n}(z) 
  - \left( {\kappa_{n+2} \over \kappa_{n+1}}
           +{\bar{\phi}_{n+2}(0) \over \bar{\phi}_{n+1}(0)}z
    \right)\Theta^*_{n+1}(z) \\
  - {\kappa_{n+1} \over \kappa_{n}}(z\Theta_n(z)-\Theta^*_n(z)) - {W(z) \over z}= 0
\label{ops_rrCf:g}
\end{multline}
\begin{multline}
   \Omega^*_{n}(z) - \Omega^*_{n+1}(z)
   + {\kappa_{n+2} \over \kappa_{n+1}}
    \left( 1+{\phi_{n+1}(0) \over \kappa_{n+1}}{\bar{\phi}_{n+2}(0) \over \kappa_{n+2}}z
    \right)\Theta^*_{n+1}(z) \\
   + {|\phi_{n+1}(0)|^2 \over \kappa_{n+1}\kappa_{n}}z\Theta_n(z)
   - {\kappa_{n+1} \over \kappa_{n}}\Theta^*_{n}(z) = 0
\label{ops_rrCf:h}
\end{multline}
\end{proposition} 
\begin{proof}
The first (\ref{ops_rrCf:a}) was found in \cite{IW_2001} by a direct evaluation
of the left-hand side using integral definitions of the coefficient functions,
however all of the relations follow from the compatibility of the differential
relations and the recurrence relations. 
Thus (\ref{ops_rrCf:a},\ref{ops_rrCf:b}) follow from the compatibility of 
(\ref{ops_zD:a}) and (\ref{ops_ttr:a}), 
(\ref{ops_rrCf:c}),(\ref{ops_rrCf:d}) follow from 
(\ref{ops_zD:c}) and (\ref{ops_ttr:b}),
(\ref{ops_rrCf:e},\ref{ops_rrCf:f}) follow from the combination of
(\ref{ops_zD:a},\ref{ops_zD:c}) and (\ref{ops_rr:a}), and 
(\ref{ops_rrCf:g},\ref{ops_rrCf:h}) follow from the combination of
(\ref{ops_zD:a},\ref{ops_zD:c}) and (\ref{ops_rr:b}).
\end{proof}

\begin{remark}
The relations given above are obviously not all independent, as for example we
note that (\ref{ops_rrCf:a}) can derived from (\ref{ops_rrCf:e}) with the use
of (\ref{ops_rrCf:j}) below.
\end{remark}

\begin{corollary}\label{ops_Linear2}
Some additional identities satisfied by the coefficient functions are the following
\begin{gather}
  {\phi_{n+1}(0) \over \phi_{n}(0)}\Theta_{n}(z)
 - {\kappa_{n} \over \kappa_{n-1}}z\Theta_{n-1}(z) 
 = {\bar{\phi}_{n+1}(0) \over \bar{\phi}_{n}(0)}z\Theta^*_n(z)
  - {\kappa_{n} \over \kappa_{n-1}} \Theta^*_{n-1}(z)
 \label{ops_rrCf:i} \\  
  \Omega^*_{n}(z)-\Omega_{n}(z)
  = - {\kappa_{n+1} \over \kappa_{n}}(z\Theta_{n}(z)-\Theta^*_n(z))+n{W(z) \over z}
 \label{ops_rrCf:j} \\
  \Omega^*_{n}(z)+\Omega_{n}(z)  
  = \left(1-{\phi_{n+1}(0)\bar{\phi}_{n+1}(0) \over \kappa^2_{n+1}}\right)
    \left[{\phi_{n+2}(0) \over \phi_{n+1}(0)}\Theta_{n+1}(z)
          + {\kappa_{n+1} \over \kappa_{n}} \Theta^*_{n}(z)\right] 
    + {W(z) \over z} 
 \label{ops_rrCf:k}
\end{gather}
\end{corollary}

For a general system of orthogonal polynomials on the unit circle the coupled 
recurrence relations and spectral differential relations 
can be reformulated in terms of first order $2\times 2$ matrix equations (or
alternatively as second order scalar equations). Here we define our
matrix variables and derive such matrix relations, and this serves as an 
introduction to a characterisation of the general orthogonal polynomial system on
the unit circle as the solution to a $2\times 2$ matrix Riemann-Hilbert problem.

Firstly we note that the recurrence relations for the associated functions 
$ \epsilon_n(z), \epsilon^*_n(z) $ given in (\ref{ops_rre:a},\ref{ops_rre:b}) differ
from those of the polynomial systems (\ref{ops_rr:a},\ref{ops_rr:b}) by a reversal of
the signs of $ \phi_n(0), \bar{\phi}_n(0) $. We can compensate for this by constructing
the $ 2\times 2 $ matrix
\begin{equation}
   Y_n(z) := 
   \begin{pmatrix}
          \phi_n(z)   & \dfrac{\epsilon_n(z)}{w(z)} \cr
          \phi^*_n(z) & -\dfrac{\epsilon^{\vphantom{I}*}_n(z)}{w(z)} \cr
   \end{pmatrix} ,
\label{ops_Ydefn}
\end{equation}
and note from (\ref{ops_Cas:b}) that $ \det Y_n = -2z^n/w(z) $.

\begin{corollary}
The recurrence relations for a general system of orthogonal polynomials 
(\ref{ops_rr:a},\ref{ops_rr:b}) and their associated functions 
(\ref{ops_rre:a},\ref{ops_rre:b}) are equivalent to the matrix recurrence
\begin{equation}
   Y_{n+1} := M_n Y_{n}
   = {1\over \kappa_n}
       \begin{pmatrix}
              \kappa_{n+1} z   & \phi_{n+1}(0) \cr
              \bar{\phi}_{n+1}(0) z & \kappa_{n+1} \cr
       \end{pmatrix} Y_{n} , 
\label{ops_Yrecur}
\end{equation} 
with according to (\ref{ops_kappa}) $, \det M_n = z $.
\end{corollary}

\begin{corollary}
The system of spectral derivatives for a general system of orthogonal polynomials
and associated functions (\ref{ops_zD:a}-\ref{ops_zD:d}) are equivalent to 
the matrix differential equation
\begin{align}
   Y'_{n} := & A_n Y_{n}
   \nonumber \\
   = & {1\over W(z)}
       \begin{pmatrix}
              -\left[ \Omega_n(z)+V(z)
                     -\dfrac{\kappa_{n+1}}{\kappa_n}z\Theta_n(z)
               \right]
            & \dfrac{\phi_{n+1}(0)}{\kappa_n}\Theta_n(z)
            \cr
              -\dfrac{\bar{\phi}_{n+1}(0)}{\kappa_n}z\Theta^*_n(z)
            &  \Omega^*_n(z)-V(z)
                     -\dfrac{\kappa_{n+1}}{\kappa_n}\Theta^*_n(z)
            \cr
       \end{pmatrix} Y_{n} . 
\label{ops_YzDer}
\end{align} 
\end{corollary}
\begin{proof}
This follows from (\ref{ops_zD:a}-\ref{ops_zD:d}) and employing 
(\ref{ops_rr:a},\ref{ops_rr:b},\ref{ops_rre:a},\ref{ops_rre:b}).
\end{proof}

\begin{remark}
Compatibility of the relations (\ref{ops_Yrecur}) and (\ref{ops_YzDer}) leads to 
\begin{equation}
   M'_n = A_{n+1}M_n-M_nA_n ,
\end{equation}
and upon examining the $ 11 $-component of this we recover the linear recurrence 
(\ref{ops_rrCf:f}), the $ 12 $-component yields (\ref{ops_rrCf:e}), whilst the
$ 21 $-component gives (\ref{ops_rrCf:g}) and the $ 22 $-component implies 
(\ref{ops_rrCf:h}).
\end{remark}

\begin{remark}
There are, in a second-order difference equation such as (\ref{ops_rre:a}) or 
(\ref{ops_rre:b}), other forms of the matrix variables and equations and these
alternative forms will appear in our subsequent work. Defining
\begin{equation}
   X_n(z;t) := 
   \begin{pmatrix}
          \phi_{n+1}(z)   & \dfrac{\epsilon_{n+1}(z)}{w(z)} \cr
          \phi_n(z) & \dfrac{\epsilon_n(z)}{w(z)} \cr
   \end{pmatrix} ,
   \quad
   X^*_n(z;t) := 
   \begin{pmatrix}
          \phi^*_{n+1}(z)   & \dfrac{\epsilon^*_{n+1}(z)}{w(z)} \cr
          \phi^*_n(z) & \dfrac{\epsilon^*_n(z)}{w(z)} \cr
   \end{pmatrix} ,
\label{ops_Xdefn}
\end{equation}
we find the spectral derivatives to be
\begin{equation}
   W(z)X'_{n} = \begin{pmatrix}
              \Omega_n(z)-V(z)
                     +n\dfrac{W(z)}{z}
            & -\dfrac{\kappa_n\phi_{n+2}(0)}{\kappa_{n+1}\phi_{n+1}(0)} z\Theta_{n+1}(z)
            \cr
              \Theta_n(z)
            & -\Omega_n(z)-V(z)
            \cr
                \end{pmatrix} X_{n} , 
\label{ops_XzDer:a}
\end{equation} 
\begin{equation}
   W(z)X^*_{n}{\!'} = \begin{pmatrix}
              -\Omega^*_n(z)-V(z)
                     +(n+1)\dfrac{W(z)}{z}
            & \dfrac{\kappa_n\bar{\phi}_{n+2}(0)}{\kappa_{n+1}\bar{\phi}_{n+1}(0)}
              z\Theta^*_{n+1}(z)
            \cr
              -\Theta^*_n(z)
            & \Omega^*_n(z)-V(z)
            \cr
                      \end{pmatrix} X^*_{n} . 
\label{ops_XzDer:b}
\end{equation} 
Another system is based upon the definition
\begin{equation}
   Z_n(z;t) := 
   \begin{pmatrix}
          \phi_{n+1}(z)   & \dfrac{\epsilon_{n+1}(z)}{w(z)} \cr
          \phi^*_n(z) & -\dfrac{\epsilon^*_n(z)}{w(z)} \cr
   \end{pmatrix} ,
   \quad
   Z^*_n(z;t) := 
   \begin{pmatrix}
          \phi^*_{n+1}(z)   & -\dfrac{\epsilon^*_{n+1}(z)}{w(z)} \cr
          \phi_n(z) & \dfrac{\epsilon_n(z)}{w(z)} \cr
   \end{pmatrix} ,
\label{ops_Zdefn}
\end{equation}
and in this case the spectral derivatives are
\begin{multline}
   W(z)Z'_{n} \\
     = \begin{pmatrix}
              -\Omega^*_n(z)-V(z)
              +\dfrac{\kappa_n}{\kappa_{n+1}}\Theta^*_n(z)
                     +(n+1)\dfrac{W(z)}{z}
            & \dfrac{\kappa_n\phi_{n+2}(0)}{\kappa^2_{n+1}}\Theta_{n+1}(z)
            \cr
              -\dfrac{\bar{\phi}_{n+1}(0)}{\kappa_{n+1}}\Theta^*_n(z)
            & \Omega^*_n(z)-V(z)
              -\dfrac{\kappa_n}{\kappa_{n+1}}\Theta^*_n(z)
            \cr
       \end{pmatrix} Z_{n} , 
\label{ops_ZzDer:a}
\end{multline} 
\begin{multline}
   W(z)Z^*_{n}{\!'} 
     = \begin{pmatrix}
              \Omega_n(z)-V(z)
              -\dfrac{\kappa_n}{\kappa_{n+1}}z\Theta_n(z)
            & -\dfrac{\kappa_n\bar{\phi}_{n+2}(0)}{\kappa^2_{n+1}}z^2\Theta^*_{n+1}(z)
            \cr
              \dfrac{\phi_{n+1}(0)}{\kappa_{n+1}}\Theta_n(z)
            & -\Omega_n(z)-V(z)
              +\dfrac{\kappa_n}{\kappa_{n+1}}z\Theta_n(z)
            \cr
       \end{pmatrix} Z^*_{n} . 
\label{ops_ZzDer:b}
\end{multline} 
\end{remark}

We end this section with a characterisation of a general system of orthogonal 
polynomials on the unit circle (and their associated functions) as a solution to 
a particular Riemann-Hilbert problem.

\begin{proposition}\label{ops_RHP}
Consider the following Riemann-Hilbert problem for a $ 2 \times 2 $ matrix function
$ Y : {\CC} \to SL(2,\CC) $ defined in the following statements
\begin{enumerate}
  \item
   $ Y(z) $ is analytic in $ \{z: |z| > 1\}\cup\{z: |z|< 1\} $,
  \item
   on $ z \in \Sigma $ where $ \Sigma $ is the oriented unit circle in a 
   counter-clockwise sense and $ +(-) $ denote the left(right)-hand side or 
   interior(exterior) 
   \begin{equation}
      Y_{+}(z) = Y_{-}(z)
       \begin{pmatrix}
              1 & w(z)/z \cr
              0 & 1 \cr
       \end{pmatrix} ,
   \label{ops_RHjump}
   \end{equation}
  \item
   as $ z \to \infty $
   \begin{equation}
     Y(z) = \left( \mathbb{I} + {\rm O}(z^{-1}) \right)
       \begin{pmatrix}
              z^n             & {\rm O}(z^{-2}) \cr
              {\rm O}(z^{n}) & -z^{-1}\cr
       \end{pmatrix} ,
   \label{ops_RHzLarge}
   \end{equation}
  \item
   as $ z \to 0 $
   \begin{equation}
     Y(z) = \left( \mathbb{I} + {\rm O}(z) \right)
       \begin{pmatrix}
              {\rm O}(1) & {\rm O}(z^{n-1}) \cr
              {\rm O}(1) & {\rm O}(z^{n})   \cr
       \end{pmatrix} .
   \label{ops_RHzSmall}
   \end{equation}
\end{enumerate}
It is assumed that the weight function $ w(z) $ satisfies the restrictions given
at the beginning of this section. Then the unique solution to this Riemann-Hilbert 
problem is 
\begin{equation}
   Y(z) = 
       \begin{pmatrix}
               \dfrac{\phi_n(z)}{\kappa_n}
            &  \dfrac{\epsilon_n(z)}{2\kappa_n z} \\
               \kappa_n\phi^*_n(z) 
            & -\dfrac{\kappa_n\epsilon^{\vphantom{I}*}_n(z)}{2z}
       \end{pmatrix} , \quad n \geq 1 .
\end{equation}
\end{proposition}
\begin{proof}
We firstly note from the jump condition (\ref{ops_RHjump}) that $ Y_{11}, Y_{21} $ 
are entire $ z \in \CC $. From the $ 11 $-entry of the asymptotic condition 
(\ref{ops_RHzLarge}) it is clear that $ Y_{11} = \pi_n(z) $ a polynomial of degree
at most $ n $. Similarly $ Y_{21} = \sigma_n(z) $ from an observation of the
$ 21 $-component. From the $ 12 $- and $ 22 $-components of the jump condition
we deduce 
\begin{equation}
   Y_{+12}-Y_{-12} = {w(z) \over z}Y_{11} ,
  \qquad
   Y_{+22}-Y_{-22} = {w(z) \over z}Y_{21} ,
\end{equation}
and therefore
\begin{equation}
   Y_{12} 
  = \int_{\TT}{d\zeta \over 2\pi i\zeta}{w(\zeta)\pi_n(\zeta) \over \zeta-z} ,
  \qquad
   Y_{22} 
  = \int_{\TT}{d\zeta \over 2\pi i\zeta}{w(\zeta)\sigma_n(\zeta) \over \zeta-z} .
\end{equation}
Consider the large $ z $ expansion of $ Y_{12} $ implied by the first of these 
formulae
\begin{equation}
   Y_{12} = -z^{-1}\int_{\TT}{d\zeta \over 2\pi i\zeta}w(\zeta)\pi_n(\zeta)
   + {\rm O}(z^{-2}) .
\end{equation}
According to (\ref{ops_RHzLarge}) the integral vanishes and so $ \pi_n(\zeta) $ 
is orthogonal to the monomial $ \zeta^0 $. Now take the small $ z $ expansion
\begin{equation}
   Y_{12} 
   = \sum^{n-2}_{l=0}z^{l}\int_{\TT}{d\zeta \over 2\pi i\zeta}
     w(\zeta)\pi_n(\zeta)\overline{\zeta^{l+1}}
   + {\rm O}(z^{n-1}) .
\end{equation}
From the $ 12 $-component of the condition (\ref{ops_RHzSmall}) we observe 
that all terms in the sum vanish and we conclude the $ \pi_n(\zeta) $ is 
orthogonal to the monomials $ \zeta, \ldots, \zeta^{n-1} $ and the first 
term which survives has the monomial $ \zeta^{n} $. Thus 
$ \pi_n(z) \propto \phi_n(z) $, and from the explicit coefficient in the 
$ 11 $-entry of (\ref{ops_RHzLarge}) $ \pi_n(z) $ is the monic orthogonal polynomial
$ \phi_n(z)/\kappa_n $. We turn our attention to $ Y_{22} $ and examine
the small $ z $ expansion
\begin{equation}
   Y_{22} 
   = \sum^{n-1}_{l=0}z^{l}\int_{\TT}{d\zeta \over 2\pi i\zeta}
     w(\zeta)\sigma_n(\zeta)\overline{\zeta^{l+1}}
   + {\rm O}(z^{n}) .
\end{equation}
The $ 22 $-component of (\ref{ops_RHzSmall}) tells us that all terms in the
sum vanish and consequently $ \sigma_n(\zeta) $ is orthogonal to all monomials
$ \zeta, \ldots, \zeta^{n} $. Therefore $ \sigma_n(\zeta) \propto \phi^*_n(z) $
and we can determine the proportionality constant from the $ 22 $-component of 
the asymptotic formula (\ref{ops_RHzLarge}) and comparing it with
\begin{equation}
   Y_{22} = -z^{-1}\int_{\TT}{d\zeta \over 2\pi i\zeta}w(\zeta)\sigma_n(\zeta)
   + {\rm O}(z^{-2}),
\end{equation}
to conclude $ \sigma_n(\zeta) = \kappa_n\phi^*_n(z) $. Finally we note that
\begin{equation}
  \int_{\TT}{d\zeta \over 2\pi i\zeta}{w(\zeta)\phi_n(\zeta) \over \zeta-z}
  = {1\over 2z}\epsilon_n(z),
  \qquad
  \int_{\TT}{d\zeta \over 2\pi i\zeta}{w(\zeta)\phi^*_n(\zeta) \over \zeta-z}
  = -{1\over 2z}\epsilon^*_n(z),
\end{equation}
when $ n > 0 $. We also point out $ \det Y = -z^{n-1} $.
\end{proof}

\begin{remark}
Our original matrix solution $ Y_n $ specified by (\ref{ops_Ydefn}) is related 
to the solution of the above Riemann-Hilbert problem by
\begin{equation}
    Y_n = 
       \begin{pmatrix}
              \kappa_n & 0 \cr
              0 & \dfrac{1}{\kappa_n} \cr
       \end{pmatrix}
        Y 
       \begin{pmatrix}
              1 & 0 \cr
              0 & \dfrac{2z}{w(z)} \cr
       \end{pmatrix} , \qquad
    Y = 
       \begin{pmatrix}
              \dfrac{1}{\kappa_n} & 0 \cr
              0 & \kappa_n \cr
       \end{pmatrix}
        Y_n 
       \begin{pmatrix}
              1 & 0 \cr
              0 & \dfrac{w(z)}{2z} \cr
       \end{pmatrix} .
\end{equation}
Our formulation of the Riemann-Hilbert problem differs from those given in 
studies concerning orthogonal polynomial systems on the unit circle with
more specialised weights \cite{Ba_2001},\cite{BDMcLMZ_2001},\cite{BDJ_2000a},
\cite{BDJ_1999}. We have chosen this 
formulation as it is closest to that occurring for orthogonal polynomial
systems of the line \cite{De_1999}, the jump matrix is independent of the index $ n $
which only appears in the asymptotic condition and it is simply related to
our matrix formulation (\ref{ops_Ydefn}).
\end{remark}

\section{Regular Semi-classical Weights and Isomonodromic Deformations}
\label{SC+IDsection}
\setcounter{equation}{0}

All of the above results apply for a general class of weights on the unit circle
but now we want to consider an additional restriction, namely the special 
structure of regular or generic semi-classical weights.
\begin{definition}[\cite{Ma_1995a}]
The log-derivative of a regular or generic semi-classical weight function is 
rational in $ z $ with
\begin{equation}
   W(z)w'(z) = 2V(z)w(z)
\label{ops_scwgt}
\end{equation}
where $ V(z), W(z) $ are polynomials with the following properties
\begin{enumerate}
 \item
  $ {\rm deg}(W) \geq 2 $,
 \item
  $ {\rm deg}(V) < {\rm deg}(W) $,
 \item
  the $ m $ zeros of $ W(z) $, $ \{z_1, z_2, \ldots ,z_m\} $ are distinct,
 \item
  the residues $ \rho_k = 2V(z_k)/W'(z_k) \notin \ZZ $. 
\end{enumerate}
\end{definition}
The terminology regular refers to the connection of this definition with systems
of linear second order differential equations in the complex plane which possess 
only isolated regular singularities, and we will see the appearance of these
later. An explicit example of such a weight is that of the form
$ w(z) = \prod^m_{j=1}(z-z_j)^{\rho_j} $ with $ z_j \neq z_k $ for $ j \neq k $, 
which are also known as generalised Jacobi weights. In addition we will assume 
the polynomials defined above take the following forms
\begin{equation}
   W(z) = \prod^m_{j=1}(z-z_j), \qquad
   {2V(z) \over W(z)} = \sum^m_{j=1}{\rho_j \over z-z_j} .
\label{ops_scwgt2}
\end{equation}
The above definition is restrictive and has been generalised by relaxing some of 
the conditions in a series of works \cite{Ma_1987}, \cite{MR_1992}, \cite{MR_1998}.
In these works the orthogonal polynomial systems were characterised by integral 
representations of semi-classical linear functionals with respect to certain paths 
in the complex plane.

It follows from these definitions that the Carath\'eodory function satisfies
an inhomogeneous form of (\ref{ops_scwgt}).
\begin{lemma}[\cite{La_1972b},\cite{Ma_1995a}]
The Carath\'eodory function (\ref{ops_Cfun}) satisfies the first order linear
ordinary differential equation 
\begin{equation}
   W(z)F'(z) = 2V(z)F(z)+U(z) ,
\label{ops_FD}
\end{equation}
where $ U(z) $ is a polynomial in $ z $. 
\end{lemma}

This lemma leads to the following important result.
\begin{proposition}\label{ops_SCpoly}
The coefficient functions 
$ \Theta_n(z), \Theta^*_n(z), \Omega_n(z),\Omega^*_n(z) $ are polynomials in
$ z $ of degree $ m-2, m-2, m-1, m-1 $ respectively. Specifically these have 
leading and trailing expansions of the form
\begin{multline}
  \Theta_n(z) =
 (n+1+\sum^m_{j=1}\rho_j){\kappa_n \over \kappa_{n+1}}z^{m-2}
 \\
 + \bigg\{ -[(n+1+\sum^m_{j=1}\rho_j)\sum^m_{j=1}z_j - \sum^m_{j=1}\rho_j z_j]
             {\kappa_n \over \kappa_{n+1}}
            +(n+2+\sum^m_{j=1}\rho_j){\kappa^3_n \over \kappa^2_{n+1}\kappa_{n+2}}
             {\phi_{n+2}(0) \over \phi_{n+1}(0)}
 \\
 -(n+\sum^m_{j=1}\rho_j){\phi_{n+1}(0)\bar{\phi}_{n}(0) \over 
                                   \kappa_{n+1}\kappa_{n}}
           -2{\kappa_nl_{n+1} \over \kappa^2_{n+1}}   
   \bigg\}z^{m-3} + {\rm O}(z^{m-4}) 
\label{ops_Thexp:a}
\end{multline}
\begin{multline}
 \Theta_n(z) =
 [2V(0)-nW'(0)]{\phi_{n}(0) \over \phi_{n+1}(0)} 
 \\
 + \bigg\{ [2V'(0)-\shalf nW''(0)]{\phi_{n}(0) \over \phi_{n+1}(0)}
          + [2V(0)-(n-1)W'(0)]{\kappa_n\phi_{n-1}(0) \over \kappa_{n-1}\phi_{n+1}(0)}
 \\
 + \left( [(n+1)W'(0)-2V(0)]{\bar{l}_{n+1} \over \kappa_{n+1}}
                   -[(n-1)W'(0)-2V(0)]{\bar{l}_{n-1} \over \kappa_{n+1}} \right)
           {\phi_{n}(0) \over \phi_{n+1}(0)} 
   \bigg\}z + {\rm O}(z^{2})
\label{ops_Thexp:b}
\end{multline}
\begin{multline}
  \Theta^*_n(z)=
 -(n+\sum^m_{j=1}\rho_j){\bar{\phi}_{n}(0) \over \bar{\phi}_{n+1}(0)}z^{m-2} 
 \\
 + \bigg\{  [(n+\sum^m_{j=1}\rho_j)\sum^m_{j=1}z_j - \sum^m_{j=1}\rho_j z_j]
             {\bar{\phi}_{n}(0) \over \bar{\phi}_{n+1}(0)}
            +(n+1+\sum^m_{j=1}\rho_j){\bar{\phi}_{n}(0) \over \bar{\phi}_{n+1}(0)}
             {l_{n+1} \over \kappa_{n+1}}
 \\ 
 -(n-1+\sum^m_{j=1}\rho_j){\kappa_n\bar{\phi}_{n-1}(0)+\bar{\phi}_{n}(0)l_{n-1}
                               \over \bar{\phi}_{n+1}(0)}
      \bigg\}z^{m-3} + {\rm O}(z^{m-4})
\label{ops_ThSexp:a}
\end{multline}
\begin{multline}
  \Theta^*_n(z) =
 -[2V(0)-(n+1)W'(0)]{\kappa_n \over \kappa_{n+1}}
 \\
 + \bigg\{-[2V'(0)-\shalf (n+1)W''(0)]{\kappa_{n} \over \kappa_{n+1}}
           -[2V(0)-nW'(0)]{\bar{l}_n \over \kappa_{n+1}}
 \\
 + [(n+2)W'(0)-2V(0)]\left( 
     {\kappa^3_{n} \over \kappa_{n+2}\kappa^2_{n+1}}
     {\bar{\phi}_{n+2}(0) \over \bar{\phi}_{n+1}(0)}
                   -{\kappa_n \over \kappa_{n+1}}{\bar{l}_{n+1} \over \kappa_{n+1}}
                       \right)
   \bigg\}z + {\rm O}(z^{2})
\label{ops_ThSexp:b}
\end{multline}
\begin{multline}
  \Omega_n(z) =
 (1+\half\sum^m_{j=1}\rho_j)z^{m-1}
 \\
 + \bigg\{ -\half(\sum^m_{j=1}\rho_j)(\sum^m_{j=1}z_j)
            +\half \sum^m_{j=1}\rho_j z_j 
            -\sum^m_{j=1}z_j
 \\
 + (n+2+\sum^m_{j=1}\rho_j)
           {\kappa^2_n \over \kappa_{n+2}\kappa_{n+1}}
           {\phi_{n+2}(0) \over \phi_{n+1}(0)}
     - {l_{n+1} \over \kappa_{n+1}} \bigg\}z^{m-2} + {\rm O}(z^{m-3})
\label{ops_Omexp:a}
\end{multline}
\begin{multline}
  \Omega_n(z) =
 V(0)-nW'(0)
 \\
 + \bigg\{ V'(0)-\shalf nW''(0)
           +\left( V(0){\kappa_n \over \kappa_{n+1}}
                    +[V(0)-nW'(0)]{\kappa_{n+1} \over \kappa_{n}} \right)
            {\phi_{n}(0) \over \phi_{n+1}(0)}
 \\
 + [V(0)-nW'(0)]{\bar{l}_n \over \kappa_n}
   - [V(0)-(n+1)W'(0)]{\bar{l}_{n+1} \over \kappa_{n+1}} \bigg\}z + {\rm O}(z^2) 
\label{ops_Omexp:b}
\end{multline}
\begin{multline}
  \Omega^*_n(z) = 
 -\half\sum^m_{j=1}\rho_j z^{m-1}
 \\
 + \bigg\{  \half(\sum^m_{j=1}\rho_j)(\sum^m_{j=1}z_j)
            -\half \sum^m_{j=1}\rho_j z_j
            -(n+\sum^m_{j=1}\rho_j){\kappa_n \over \kappa_{n+1}}
             {\bar{\phi}_{n}(0) \over \bar{\phi}_{n+1}(0)}
            +{l_{n+1} \over \kappa_{n+1}} \bigg\}z^{m-2}
 \\
 + {\rm O}(z^{m-3})
\label{ops_OmSexp:a}
\end{multline}
\begin{multline}
  \Omega^*_n(z) =
 (n+1)W'(0)-V(0)
 \\
 + \bigg\{ \shalf (n+1)W''(0)-V'(0) 
           +[(n+2)W'(0)-2V(0)]
            {\kappa^2_{n} \over \kappa_{n+2}\kappa_{n+1}}
            {\bar{\phi}_{n+2}(0) \over \bar{\phi}_{n+1}(0)}
           -W'(0){\bar{l}_{n+1} \over \kappa_{n+1}} \bigg\}z
 \\
 + {\rm O}(z^2)
\label{ops_OmSexp:b}
\end{multline}
\end{proposition}

\begin{proof}
Following the approach of Laguerre \cite{La_1972b} we write $ F(z) $ in terms of
$ \phi_n(z), \psi_n(z), \epsilon_n(z) $ and use (\ref{ops_FD}) to deduce
\begin{align}
  0 & = WF'-2VF-U
  \\
  & = W\left({\epsilon_n-\psi_n \over \phi_n}\right)' 
        -2V{\epsilon_n-\psi_n \over \phi_n}-U
  \\
  & = {W(\psi_n\phi'_n-\phi_n\psi'_n)+2V\phi_n\psi_n-U\phi^2_n \over \phi^2_n}
       + W\left({\epsilon_n \over \phi_n}\right)'-2V{\epsilon_n \over \phi_n} .
\end{align}
The numerator of the first term is independent of $ \epsilon_n $, and so
is a polynomial in $ z $, and we denote this by
\begin{equation}
  2{\phi_{n+1}(0) \over \kappa_n} z^n\Theta_n(z) =
  W(-\phi_n\epsilon'_n+\epsilon_n\phi'_n)+2V\phi_n\epsilon_n . 
\label{ops_Thdfn}
\end{equation}
Given that this is a polynomial we can determine its degree and minimum power 
of $ z $ by utilising the expansions of $ \phi_n, \epsilon_n $ both inside
and outside the unit circle, namely (\ref{ops_phiexp:a},\ref{ops_epsexp:a}).
We find the degree of the right-hand side is $ n+m-2 $ so that $ \Theta_n(z) $ 
is a polynomial of degree $ m-2 $. Developing the expansions further we arrive 
at (\ref{ops_Thexp:a}). An identical argument applies to the other combination
\begin{equation}
   2{\bar{\phi}_{n+1}(0) \over \kappa_n} z^{n+1}\Theta^*_n(z) =
  W(\phi^*_n\epsilon^*_n{\!'}-\epsilon^*_n\phi^*_n{\!'})-2V\phi^*_n\epsilon^*_n 
\label{ops_ThSdfn}
\end{equation}
and $ \Theta^*_n(z) $ is also a polynomial of degree $ m-2 $ with the expansion
(\ref{ops_Thexp:b}). To establish (\ref{ops_Omexp:a}) we utilise the other form 
of $ \Theta_n(z) $ and (\ref{ops_Cas:a}) to deduce
\begin{align}
  W(\psi_n\phi'_n-\phi_n\psi'_n)+2V\phi_n\psi_n-U\phi^2_n
  & = 2{\phi_{n+1}(0) \over \kappa_n}z^n\Theta_n(z)
  \\
  & = [\phi_{n+1}\psi_n-\psi_{n+1}\phi_n]\Theta_n(z) .
\end{align}
Separating those terms with $ \phi_n $ and $ \psi_n $ as factors we have
\begin{equation}
   \left\{ \Theta_n(z)\phi_{n+1}-W\phi'_n-V\phi_n \right\} \psi_n
 = \left\{ \Theta_n(z)\psi_{n+1}-W\psi'_n+V\psi_n-U\phi_n \right\} \phi_n ,
\end{equation}
so that this polynomial contains both $ \phi_n $ and $ \psi_n $ as factors and 
can be written as $ \Omega_n\phi_n\psi_n $ with $ \Omega_n(z) $ a polynomial 
of bounded degree. This latter polynomial can be defined as
\begin{align}
  2{\phi_{n+1}(0) \over \kappa_n} z^n\Omega_n(z) 
  & = W(\psi_{n+1}\phi'_n-\phi_{n+1}\psi'_n)
      +V(\phi_n\psi_{n+1}+\psi_n\phi_{n+1})-U\phi_n\phi_{n+1}
  \\
  & = W(\epsilon_{n+1}\phi'_n-\phi_{n+1}\epsilon'_n)
      +V(\phi_n\epsilon_{n+1}+\epsilon_n\phi_{n+1}) .
\label{ops_Omdfn}
\end{align}
Again employing the expansions (\ref{ops_phiexp:a},\ref{ops_epsexp:a}) we
determine the degree of $ \Omega_n(z) $ to be $ m-1 $ and the expansion
(\ref{ops_Omexp:a}) follows. Starting with the alternative definition of
$ \Theta^*_n(z) $ and (\ref{ops_Cas:b})
\begin{align}
  W(\phi^*_n\psi^*_n{\!'}-\psi^*_n\phi^*_n{\!'})-2V\phi^*_n\psi^*_n-U\phi^{*2}_n
  & = 2{\bar{\phi}_{n+1}(0) \over \kappa_n} z^{n+1}\Theta^*_n(z)
  \\
  & = [\phi^*_{n+1}\psi^*_n-\psi^*_{n+1}\phi^*_n]\Theta^*_n(z) .
\end{align}
and using the above argument we identify for the polynomial $ \Omega^*_n(z) $
\begin{align}
  2{\bar{\phi}_{n+1}(0) \over \kappa_n} z^{n+1}\Omega^*_n(z) 
  & = W(-\psi^*_{n+1}\phi^*_n{\!'}+\phi^*_{n+1}\psi^*_n{\!'})
      -V(\phi^*_n\psi^*_{n+1}+\psi^*_n\phi^*_{n+1})-U\phi^*_n\phi^*_{n+1}
  \\
  & = W(-\epsilon^*_{n+1}\phi^*_n{\!'}+\phi^*_{n+1}\epsilon^*_n{\!'})
      -V(\phi^*_n\epsilon^*_{n+1}+\epsilon^*_n\phi^*_{n+1}) .
\label{ops_OmSdfn}
\end{align}
The degree of $ \Omega^*_n(z) $ to be $ m-1 $ and has the expansion
(\ref{ops_OmSexp:a}). 
\end{proof}

\begin{remark}
Solving for $ \phi'_n $ and $ \epsilon'_n $ between (\ref{ops_Thdfn}) and 
(\ref{ops_Omdfn}) leads to (\ref{ops_zD:a}) and (\ref{ops_zD:b}),
whilst solving for $ \phi^*_n{\!'} $ and $ \epsilon^*_n{\!'} $ using
(\ref{ops_ThSdfn},\ref{ops_OmSdfn}) yields (\ref{ops_zD:c}) and 
(\ref{ops_zD:d}).
\end{remark}

Furthermore, in the case of a regular semi-classical weight function, the matrix 
$ A_n(z;t) $ has the partial fraction decomposition
\begin{equation}
   A_n(z;t) := \sum^{m}_{j=1}{ A_{nj}(t) \over z-z_j} ,
\end{equation}
under the assumptions following (\ref{ops_scwgt}) and the residue matrices are
given by
\begin{equation}
   A_{nj} = {\rho_j \over 2V(z_j)}
       \begin{pmatrix}
              -\Omega_n(z_j)-V(z_j)
              +\dfrac{\kappa_{n+1}}{\kappa_n}z_j\Theta_n(z_j)
            & \dfrac{\phi_{n+1}(0)}{\kappa_n}\Theta_n(z_j)
            \cr
              -\dfrac{\bar{\phi}_{n+1}(0)}{\kappa_n}z_j\Theta^*_n(z_j)
            &  \Omega^*_n(z_j)-V(z_j)
                     -\dfrac{\kappa_{n+1}}{\kappa_n}\Theta^*_n(z_j)
            \cr
       \end{pmatrix} .
\end{equation}
Using the identity (\ref{ops_rrCf:j}) we note that $ {\rm Tr} A_{nj} = -\rho_j $ and 
$ {\rm Tr} A_n(z;t) = -w'(z)/w(z) $.

Bilinear residue formulae relating products of a polynomial and an associated 
function evaluated at a singular point will arise in the theory of the 
deformation derivatives later and we give a complete list of results for 
these.
  
\begin{corollary}
Bilinear residues are related to the coefficient function residues in the 
following equations
\begin{align}
   \phi_n(z_j)\epsilon_n(z_j) 
  & =  2{\phi_{n+1}(0) \over \kappa_n}z^n_j{\Theta_n(z_j) \over 2V(z_j)}
  \label{ops_BilRes:a} \\
   \phi^*_n(z_j)\epsilon^*_n(z_j) 
  & = -2{\bar{\phi}_{n+1}(0) \over \kappa_n}z^{n+1}_j{\Theta^*_n(z_j) \over 2V(z_j)}
  \label{ops_BilRes:b} \\
   \phi_{n+1}(z_j)\epsilon_n(z_j) 
  & =  2{\phi_{n+1}(0) \over \kappa_n}z^n_j{\Omega_n(z_j)+V(z_j) \over 2V(z_j)}
  \label{ops_BilRes:c} \\
   \phi_{n}(z_j)\epsilon_{n+1}(z_j) 
  & =  2{\phi_{n+1}(0) \over \kappa_n}z^n_j{\Omega_n(z_j)-V(z_j) \over 2V(z_j)}
  \label{ops_BilRes:d} \\
   \phi^*_{n}(z_j)\epsilon^*_{n+1}(z_j) 
  & = -2{\bar{\phi}_{n+1}(0) \over \kappa_n}z^{n+1}_j
        {\Omega^*_n(z_j)+V(z_j) \over 2V(z_j)}
  \label{ops_BilRes:e} \\
   \phi^*_{n+1}(z_j)\epsilon^*_{n}(z_j) 
  & = -2{\bar{\phi}_{n+1}(0) \over \kappa_n}z^{n+1}_j
        {\Omega^*_n(z_j)-V(z_j) \over 2V(z_j)}
  \label{ops_BilRes:f} \\
   \phi_n(z_j)\epsilon^*_n(z_j) 
  & = -{z^n_j \over V(z_j)}\left[ \Omega_n(z_j)-V(z_j)
        -{\kappa_{n+1} \over \kappa_n}z_j\Theta_n(z_j) \right]
  \label{ops_BilRes:g} \\
  & = -{z^n_j \over V(z_j)}\left[ \Omega^*_n(z_j)-V(z_j)
        -{\kappa_{n+1} \over \kappa_n}\Theta^*_n(z_j) \right]
  \label{ops_BilRes:h} \\
   \phi^*_n(z_j)\epsilon_n(z_j) 
  & =  {z^n_j \over V(z_j)}\left[ \Omega_n(z_j)+V(z_j)
        -{\kappa_{n+1} \over \kappa_n}z_j\Theta_n(z_j) \right]
  \label{ops_BilRes:i} \\
  & =  {z^n_j \over V(z_j)}\left[ \Omega^*_n(z_j)+V(z_j)
        -{\kappa_{n+1} \over \kappa_n}\Theta^*_n(z_j) \right] .
  \label{ops_BilRes:j}
\end{align}
\end{corollary}
\begin{proof}
These are all found by evaluating one of (\ref{ops_Thdfn}), (\ref{ops_ThSdfn}),
(\ref{ops_Omdfn}), or (\ref{ops_OmSdfn}) at $ z=z_j $ and using 
(\ref{ops_Cas:a}-\ref{ops_Cas:c}).
\end{proof}

\begin{remark}
The initial members of the sequences of coefficient functions 
$ \{\Theta_n\}^{\infty}_{n=0} $, $ \{\Theta^*_n\}^{\infty}_{n=0} $,
$ \{\Omega_n\}^{\infty}_{n=0} $, $ \{\Omega^*_n\}^{\infty}_{n=0} $ are given by
\begin{align}
  \Theta_0(z) = & 2V(z)-U(z)
  \label{ops_Theta0} \\
  \Theta_1(z) = & \kappa^2_1 z^2(2V(z)-U(z)) - 2\kappa_1\phi_1(0)zU(z)
              -2\kappa_1\phi_1(0)W(z) - \phi^2_1(0)(2V(z)+U(z))
  \label{ops_Theta1} \\
  \Theta^*_0(z) = & -2V(z)-U(z)
  \label{ops_ThetaS0} \\
  \Theta^*_1(z) = & \bar{\phi}^2_1(0) z^2(2V(z)-U(z)) - 2\kappa_1\bar{\phi}_1(0)zU(z) 
              -2\kappa_1\bar{\phi}_1(0)W(z) - \kappa^2_1 (2V(z)+U(z))
  \label{ops_ThetaS1} \\
  \Omega_0(z) = & {\kappa_1 \over 2\phi_1(0)}z(2V(z)-U(z)) - {U(z) \over 2}
  \label{ops_Omega0} \\
  \Omega^*_0(z) = & -{\kappa_1 \over 2\bar{\phi}_1(0)z}(2V(z)+U(z)) - {U(z) \over 2} .
  \label{ops_OmegaS0}
\end{align}
\end{remark}
 
One can take combinations of the above functional-difference equations and 
construct exact differences when $ z $ is evaluated at the singular points 
of the weight, i.e. $ W(z)=0 $. The integration of the system is given in the
following proposition.

\begin{proposition}\label{ops_Bilinear}
At all the singular points $ z_j, j=1,\ldots m $, with the exception of $ z_j=0 $,
the coefficient functions satisfy the bilinear identities
\begin{gather}
  \Omega^2_n(z_j)
  = {\kappa_n \phi_{n+2}(0) \over \kappa_{n+1} \phi_{n+1}(0)}z_j
      \Theta_n(z_j)\Theta_{n+1}(z_j)+V^2(z_j)
  \label{ops_OTeq:a} \\
  \Omega^{*2}_n(z_j)
  = {\kappa_n \bar{\phi}_{n+2}(0) \over \kappa_{n+1} \bar{\phi}_{n+1}(0)}z_j
      \Theta^*_n(z_j)\Theta^*_{n+1}(z_j)+V^2(z_j)
  \label{ops_OTeq:b} \\
  \left[\Omega_{n-1}(z_j)
    -{\kappa^2_{n-1} \over \kappa^2_{n}}{\phi_{n+1}(0) \over \phi_{n}(0)}\Theta_n(z_j)
  \right]^2
  = {\phi_{n+1}(0)\bar{\phi}_{n}(0) \over \kappa^2_{n}}
      \Theta_n(z_j)\Theta^*_{n-1}(z_j)+V^2(z_j)
  \label{ops_OTeq:c} \\
  \left[\Omega^*_{n-1}(z_j)
    -{\kappa^2_{n-1} \over \kappa^2_{n}}{\bar{\phi}_{n+1}(0) \over \bar{\phi}_{n}(0)}
      z_j\Theta^*_n(z_j)
  \right]^2
  = {\kappa_{n-1}\bar{\phi}_{n+1}(0)\phi_{n}(0) \over \kappa^3_{n}}
      z^2_j\Theta^*_n(z_j)\Theta_{n-1}(z_j)+V^2(z_j)
  \label{ops_OTeq:d} \\
    {\phi_{n+1}(0)\bar{\phi}_{n+1}(0) \over \kappa^2_{n}}
      \Theta_n(z_j)\Theta^*_{n}(z_j)+V^2(z_j)
  = \left[\Omega_{n}(z_j)-{\kappa_{n+1} \over \kappa_{n}}z_j\Theta_n(z_j)\right]^2
  \label{ops_OTeq:e} \\
  \phantom{
    {\phi_{n+1}(0)\bar{\phi}_{n+1}(0) \over \kappa^2_{n}}
      \Theta_n(z_j)\Theta^*_{n}(z_j)+V^2(z_j) }
  = \left[\Omega^*_{n}(z_j)-{\kappa_{n+1} \over \kappa_{n}}\Theta^*_n(z_j)\right]^2
  \label{ops_OTeq:f}
\end{gather}
\end{proposition}
\begin{proof}[First Proof]
We take the first pair of identities (\ref{ops_OTeq:a}) and (\ref{ops_OTeq:b})
as an example for our first proof.
Multiplying the $ \Omega_n, \Omega_{n-1} $ terms of (\ref{ops_rrCf:a}) by the
corresponding terms of (\ref{ops_rrCf:b}), evaluated at a singular point
$ z=z_j $, one has an exact difference
\begin{equation}
  \Omega^2_n(z_j)-\Omega^2_{n-1}(z_j) =
  {\kappa_n \phi_{n+2}(0) \over \kappa_{n+1} \phi_{n+1}(0)}z_j
      \Theta_n(z_j)\Theta_{n+1}(z_j)
 -{\kappa_{n-1} \phi_{n+1}(0) \over \kappa_n \phi_{n}(0)}z_j
      \Theta_{n-1}(z_j)\Theta_{n}(z_j) ,
\end{equation}
assuming none of the $ z_j $ coincide with $ -r_{n+1}/r_n $ for any $ n $.
Upon summing this relation the summation constant is calculated to be
\begin{equation}
   \Omega^2_0(z_j)-{\kappa_0 \phi_{2}(0) \over \kappa_{1} \phi_{1}(0)}z_j
      \Theta_0(z_j)\Theta_{1}(z_j)
   = V^2(z_j) ,
\end{equation}
by using the initial members of the coefficient function sequences in 
(\ref{ops_Omega0},\ref{ops_Theta0},\ref{ops_Theta1}). The result is 
(\ref{ops_OTeq:a}), whilst the second relation follows from an identical argument 
applied to (\ref{ops_rrCf:c},\ref{ops_rrCf:d}).
\end{proof}

\begin{proof}[Second Proof]
The three pairs of formulae (\ref{ops_OTeq:a},\ref{ops_OTeq:b}), 
(\ref{ops_OTeq:c},\ref{ops_OTeq:d}) and (\ref{ops_OTeq:e},\ref{ops_OTeq:f})
arise from the fact that at a singular point $ z_j $ the determinant of the 
matrix spectral derivative must vanish. Thus (\ref{ops_OTeq:a}) and 
(\ref{ops_OTeq:b}) express the condition that the determinant of the matrix on
the right-hand sides of (\ref{ops_XzDer:a}) and (\ref{ops_XzDer:b}) vanish
respectively. It can be shown that the same condition applied to the right-hand 
sides of (\ref{ops_ZzDer:b}) and (\ref{ops_ZzDer:a}) implies (\ref{ops_OTeq:c}) 
and (\ref{ops_OTeq:d}) respectively when one takes into account the identities 
(\ref{ops_rrCf:f}), (\ref{ops_rrCf:i}), (\ref{ops_rrCf:j}) and (\ref{ops_rrCf:c}).
The last pair are a consequence of $ \det(WA_n(z_j;t)) = 0 $ along with the 
identity (\ref{ops_rrCf:j}). 
\end{proof}

\begin{proof}[Third Proof]
All the bilinear identities in Proposition \ref{ops_Bilinear} can be easily
derived from the residue formulae (\ref{ops_BilRes:a}-\ref{ops_BilRes:j}) by 
multiplying any two of the above formulae and then factoring the resulting 
product in a different way. Thus (\ref{ops_OTeq:a}) arises from multiplying
(\ref{ops_BilRes:c}) and (\ref{ops_BilRes:d}) and then factoring the product
in order to employ (\ref{ops_BilRes:a}). Equation (\ref{ops_OTeq:c}) comes from
multiplying (\ref{ops_BilRes:a}) and (\ref{ops_BilRes:b}) with $ n \mapsto n-1 $,
using the recurrences (\ref{ops_rr:b}), (\ref{ops_rre:b}) with $ n \mapsto n-1 $
to solve for $ \phi^*_{n-1}(z_j), \epsilon^*_{n-1}(z_j) $ and employing 
(\ref{ops_BilRes:a}) along with (\ref{ops_BilRes:c}) and (\ref{ops_BilRes:d}) setting 
$ n \mapsto n-1 $. Equation (\ref{ops_OTeq:e}) is derived by multiplying 
(\ref{ops_BilRes:a}) and (\ref{ops_BilRes:b}) and then factoring using
(\ref{ops_BilRes:g}) and (\ref{ops_BilRes:i}). The reciprocal versions follow
from similar reasoning.
\end{proof}

\begin{remark}
It is clear from the first proof that the bilinear identities given in Proposition 
\ref{ops_Bilinear} can be straightforwardly generalised to ones that are functions 
of $ z $ rather than evaluated at special $ z$ values. 
They can be derived directly from Proposition \ref{ops_Linear1}, so apply in situations where
the weights are not semi-classical, and contain additional
terms with a factor of $ W(z) $ and sums of products of other coefficients ranging from 
$ j = 1,\ldots n $. 
However because we will have no use for such relations we refrain from writing these down.
\end{remark}

\begin{remark}
If $ z=0 $ is a singular point then the limit as $ z \to 0 $ may be taken in the
product of (\ref{ops_rrCf:a},\ref{ops_rrCf:b}), however this does not lead to
any new independent relation but simply recovers
\begin{equation*}
   \Omega_n(0) = V(0)-nW'(0) .
\end{equation*}
\end{remark}

We now consider the dynamics of deforming the semi-classical weight 
(\ref{ops_scwgt2}) through a $t$-dependence of the singular points $ z_j(t) $,
\begin{equation}
   {\dot{w} \over w} = -\sum^m_{j=1}\rho_j{\dot{z}_j \over z-z_j} ,
\label{ops_tDwgt}
\end{equation}
where $ \dot{} := d/dt $.
Given this motion of the singularities we consider the $t$-derivatives of
the orthogonal polynomial system.
  
\begin{proposition}\label{ops_deformD}
The deformation derivative of a semi-classical orthogonal polynomial is
\begin{multline}
  \dot{\phi}_n(z) =
  \Big\{-{\dot{\kappa}_n \over \kappa_n}
        -\sum^m_{j=1}\rho_j{\dot{z}_j \over z_j}
        +\half\sum^m_{j=1}\rho_j{\dot{z}_j \over z_j}
              z^{-n}_j \epsilon_n(z_j)\phi^*_n(z_j){z \over z-z_j}
  \Big\}\phi_n(z) 
  \\
  -\Big\{\half\sum^m_{j=1}\rho_j{\dot{z}_j \over z_j}
              z^{1-n}_j \epsilon_n(z_j)\phi_n(z_j){1 \over z-z_j}
  \Big\}\phi^*_n(z) ,
\label{ops_tD:a}
\end{multline}
whilst that of a reciprocal polynomial is
\begin{multline}
  \dot{\phi}^*_n(z) =
  \Big\{-{\dot{\kappa}_n \over \kappa_n}
        +\half\sum^m_{j=1}\rho_j{\dot{z}_j \over z_j}
              z^{1-n}_j \epsilon^*_n(z_j)\phi_n(z_j){1 \over z-z_j}
  \Big\}\phi^*_n(z) 
  \\
  -\Big\{\half\sum^m_{j=1}\rho_j{\dot{z}_j \over z_j}
              z^{-n}_j \epsilon^*_n(z_j)\phi^*_n(z_j){z \over z-z_j}
  \Big\}\phi_n(z) .
\label{ops_tD:b}
\end{multline}
The deformation derivative of an associated function is
\begin{multline}
  \dot{\epsilon}_n(z) =
  \Big\{-{\dot{\kappa}_n \over \kappa_n}
        -\half\sum^m_{j=1}\rho_j{\dot{z}_j \over z_j}
              z^{-n}_j \epsilon^*_n(z_j)\phi_n(z_j){z \over z-z_j}
  \Big\}\epsilon_n(z) 
  \\
  +\Big\{\half\sum^m_{j=1}\rho_j{\dot{z}_j \over z_j}
              z^{1-n}_j \epsilon_n(z_j)\phi_n(z_j){1 \over z-z_j}
  \Big\}\epsilon^*_n(z) ,
\label{ops_tD:c}
\end{multline}
and that of a reciprocal associated function is
\begin{multline}
  \dot{\epsilon}^*_n(z) =
  \Big\{-{\dot{\kappa}_n \over \kappa_n}
        -\half\sum^m_{j=1}\rho_j{\dot{z}_j \over z_j}
              z^{1-n}_j \epsilon_n(z_j)\phi^*_n(z_j){1 \over z-z_j}
  \Big\}\epsilon^*_n(z) 
  \\
  +\Big\{\half\sum^m_{j=1}\rho_j{\dot{z}_j \over z_j}
              z^{-n}_j \epsilon^*_n(z_j)\phi^*_n(z_j){z \over z-z_j}
  \Big\}\epsilon_n(z) .
\label{ops_tD:d}
\end{multline}
\end{proposition}

\begin{proof}
Differentiating the orthonormality condition 
\begin{equation*}
   \int{d\zeta \over 2\pi i\zeta}w(\zeta)
               \phi_{n}(\zeta)\overline{\phi_{n-i}(\zeta)} = \delta_{i,0} .
\end{equation*}
and using (\ref{ops_tDwgt}) we find
\begin{equation*}
  0 = {\dot{\kappa}_n \over \kappa_n}\delta_{i,0}
      + \int{d\zeta \over 2\pi i\zeta}w(\zeta)\dot{\phi}_n\overline{\phi_{n-i}}
      - \sum_j\rho_j \dot{z}_j \int{d\zeta \over 2\pi i\zeta}w(\zeta)
        {1 \over \zeta-z_j}\phi_n\overline{\phi_{n-i}}, \quad i=0,\ldots,n
\end{equation*}
Now 
\begin{align*}
  \int{d\zeta \over 2\pi i\zeta}w(\zeta){1 \over \zeta-z}
  \phi_n(\zeta)\overline{\phi_{n-i}(\zeta)}
  & = \int{d\zeta \over 2\pi i\zeta}w(\zeta)\phi_n(\zeta)
          {\bar{\phi}_{n-i}(\zeta^{-1})-\bar{\phi}_{n-i}(z^{-1}) \over \zeta-z}
  \\
  & \qquad
      + \bar{\phi}_{n-i}(z^{-1})
      \int{d\zeta \over 2\pi i\zeta}w(\zeta){\phi_{n}(\zeta) \over \zeta-z}
  \\
  & = -{1 \over z}\delta_{i,0}
      + {\bar{\phi}_{n-i}(z^{-1}) \over 2z}\epsilon_n(z), \quad n > 0
\end{align*}
so that 
\begin{equation*}
  0 = \left({\dot{\kappa}_n \over \kappa_n}+\sum_j \rho_j{\dot{z}_j \over z_j}
      \right)\delta_{i,0}
  - \half\sum^m_{j=1}\rho_j{\dot{z}_j \over z_j}
    z^{i-n}_j\phi^*_{n-i}(z_j)\epsilon_n(z_j)
  + \int{d\zeta \over 2\pi i\zeta}w(\zeta)\dot{\phi_n}\overline{\phi_{n-i}} .
\end{equation*}
In addition we can represent $ \bar{\phi}_{n-i}(z) $ as
\begin{align*}
  \bar{\phi}_{n-i}(z) & = \sum^n_{j=0} \delta_{i,j}\bar{\phi}_{n-j}(z)
  \\
  & = \sum^n_{j=0} \int{d\zeta \over 2\pi i\zeta}w(\zeta)
      \overline{\phi_{n-i}(\zeta)}\phi_{n-j}(\zeta)\bar{\phi}_{n-j}(z) 
  \\
  &  = \int{d\zeta \over 2\pi i\zeta}w(\zeta)\overline{\phi_{n-i}(\zeta)}
       \sum^n_{j=0} \phi_{n-j}(\zeta)\overline{\phi_{n-j}(\bar{z})}
  \\
  & = \int{d\zeta \over 2\pi i\zeta}w(\zeta)\overline{\phi_{n-i}(\zeta)}
      {\phi^*_n(\zeta)\overline{\phi^*_n(\bar{z})}
       -\zeta z\phi_n(\zeta)\overline{\phi_n(\bar{z})} \over 1-\zeta z} .
\end{align*}
Writing the Kronecker delta in a similar way the whole expression becomes
\begin{multline*}
  0 = \int{d\zeta \over 2\pi i\zeta}w(\zeta)\overline{\phi_{n-i}(\zeta)}
      \bigg\{ \dot{\phi}_n(\zeta) 
            +\left({\dot{\kappa}_n \over \kappa_n}
                   +\sum^m_{j=1}\rho_j{\dot{z}_j \over z_j}\right)\phi_n(\zeta)
  \\
  -\half\sum^m_{j=1}\rho_j{\dot{z}_j \over z_j}\epsilon_n(z_j)
   {\phi^*_n(\zeta)z^{-n}_j\phi_n(z_j)
       -\zeta z^{-1}_j\phi_n(\zeta)z^{-n}_j\phi^*_n(z_j) \over 1-\zeta z^{-1}_j}
      \bigg\}
\end{multline*}
for all $ 0 \leq i \leq n $ and (\ref{ops_tD:a}) then follows. The second 
relation follows by an identical argument applied to 
\begin{equation*}
   \int{d\zeta \over 2\pi i\zeta}w(\zeta)
               \phi_{n-i}(\zeta)\overline{\phi_{n}(\zeta)} = \delta_{i,0} .
\end{equation*}
The derivatives of the associated functions (\ref{ops_tD:c}), (\ref{ops_tD:d})
follow from differentiating the definitions (\ref{ops_eps:a}), (\ref{ops_eps:b})
and employing the first two results of the proposition along with the relation
(\ref{ops_Cas:c}).
\end{proof}

\begin{corollary}\label{cor_rdot}
The $t$-derivatives of the reflection coefficients are 
\begin{align}
   {\dot{r}_n \over r_n} & =
   \half\sum^m_{j=1}\rho_j{\dot{z}_j \over z_j}
   {\Omega_{n-1}(z_j)-V(z_j) \over V(z_j)} 
   \label{ops_rdot} \\
   {\dot{\bar{r}}_n \over \bar{r}_n} & =
   \half\sum^m_{j=1}\rho_j{\dot{z}_j \over z_j}
   {\Omega^*_{n-1}(z_j)+V(z_j) \over V(z_j)} 
   \label{ops_rCdot}
\end{align}
\end{corollary}

\begin{proof}
An alternative form to (\ref{ops_tD:a}) is
\begin{equation}
  \dot{\phi}_n(\zeta) =
  -\left({\dot{\kappa}_n \over \kappa_n}
         +\sum^m_{j=1}\rho_j{\dot{z}_j \over z_j}\right)\phi_n(\zeta)
  +\half\sum^m_{j=1}\rho_j{\dot{z}_j \over z_j}\epsilon_n(z_j)
   \sum^n_{l=0} \bar{\phi}_{n-l}(z^{-1}_j)\phi_{n-l}(\zeta) ,
\end{equation}
and by examining the coefficients of $ \zeta^n, \zeta^0 $ we deduce that
\begin{equation*}
   {\dot{r}_n \over r_n} = \half{\kappa_{n-1} \over \phi_n(0)}
    \sum^m_{j=1}\rho_j{\dot{z}_j \over z_j}
    z^{1-n}_j\epsilon_n(z_j)\phi_{n-1}(z_j) .
\end{equation*}
Noting that the derivative term of (\ref{ops_Omdfn}) vanishes when
$ z=z_j $ and employing (\ref{ops_Cas:a}) we arrive at (\ref{ops_rdot}).
The second equation, (\ref{ops_rCdot}), follows by identical reasoning.
\end{proof}

Sums of the bilinear residues over the singular points are related to deformation 
derivatives in the following way,
\begin{align}
  2{\dot{\kappa}_n \over \kappa_n} 
  & = -\sum^m_{j=1}\rho_j{\dot{z}_j \over z_j}
    + \half\sum^m_{j=1}\rho_j{\dot{z}_j \over z_j}
                       z^{-n}_j\epsilon_n(z_j)\phi^*_{n}(z_j) ,
  \nonumber \\
  & = - \half\sum^m_{j=1}\rho_j{\dot{z}_j \over z_j}
                       z^{-n}_j\epsilon^*_n(z_j)\phi_{n}(z_j) ,
  \label{ops_tDkappa} \\
  {\dot{\phi}_n(0) \over \phi_n(0)}+{\dot{\kappa}_n \over \kappa_n}
  + \sum^m_{j=1}\rho_j{\dot{z}_j \over z_j}
  & = \half{\kappa_n \over \phi_n(0)}\sum^m_{j=1}\rho_j{\dot{z}_j \over z_j}
                       z^{-n}_j\epsilon_n(z_j)\phi_{n}(z_j),
  \nonumber \\
  & = {\phi_{n+1}(0) \over \phi_n(0)}
      \sum^m_{j=1}{\rho_j \over 2V(z_j)}\dot{z}_jz^{-1}_j\Theta_n(z_j)
  \label{ops_tDphi} \\
  {\dot{\bar{\phi}}_n(0) \over \bar{\phi}_n(0)}+{\dot{\kappa}_n \over \kappa_n}
  & = - \half{\kappa_n \over \bar{\phi}_n(0)} \sum^m_{j=1}\rho_j{\dot{z}_j \over z_j}
                       z^{-n}_j\epsilon^*_n(z_j)\phi^*_{n}(z_j)
  \nonumber \\
  & = {\bar{\phi}_{n+1}(0) \over \bar{\phi}_n(0)}
      \sum^m_{j=1}{\rho_j \over 2V(z_j)}\dot{z}_j\Theta^*_n(z_j) .
  \label{ops_tDCphi}
\end{align}

For the regular semi-classical weights we can also formulate the system of
deformation derivatives as a $2\times 2$ matrix differential equation and
demonstrate that the system preserves the monodromy data with respect 
to the motion of the singularities $ z_j(t) $. 

\begin{corollary}
The deformation derivatives for a system of regular semi-classical orthogonal 
polynomials and associated functions (\ref{ops_tD:a}-\ref{ops_tD:d}) are 
equivalent to the matrix differential equation
\begin{equation}
   \dot{Y}_{n} := B_n Y_{n}
   = \left\{ B_{\infty} - \sum^{m}_{j=1}{\dot{z}_j \over z-z_j}A_{nj}
        \right\} Y_{n} . 
\label{ops_YtDer}
\end{equation} 
where
\begin{equation}
  B_{\infty} = 
       \begin{pmatrix}
               \dfrac{\dot{\kappa}_{n}}{\kappa_n}
            & 0
            \cr
               \dfrac{\kappa_n\dot{\bar{\phi}}_{n}(0)+\dot{\kappa}_{n}\bar{\phi}(0)}{
               \kappa^2_n}
            & -\dfrac{\dot{\kappa}_{n}}{\kappa_n}
            \cr
       \end{pmatrix} .
\end{equation}
\end{corollary}
\begin{proof}
This follows from a partial fraction decomposition of the system 
(\ref{ops_tD:a}-\ref{ops_tD:d}) and using (\ref{ops_tDkappa},\ref{ops_tDCphi}). 
\end{proof}

In the case of the pair (\ref{ops_Yrecur}), (\ref{ops_YtDer}) compatibility 
implies the relation
\begin{equation}
   \dot{M}_n = B_{n+1}M_n-M_nB_n ,
\end{equation}
however there are no new identities arising from this condition. Taking the 
$ 11 $-component of both sides of this equation we see that it is identically satisfied 
through the use of (\ref{ops_rrCf:f}) and (\ref{ops_tDCphi}). Or if we take the
$ 12 $-components then they are equal when use of made of (\ref{ops_rrCf:a}) and
(\ref{ops_tDkappa},\ref{ops_rdot}). In a similar way we find both sides of the 
$ 21 $-components are identical when we employ (\ref{ops_rrCf:c}) and 
(\ref{ops_tDCphi}). Finally the $ 22 $-components on both sides are the same 
after taking into account (\ref{ops_rrCf:h}) and (\ref{ops_tDkappa},\ref{ops_tDCphi}).

For the pair of linear differential relations (\ref{ops_YzDer}), (\ref{ops_YtDer}) 
compatibility leads us to the Schlesinger equations 
\begin{equation}
   \dot{A}_{nj} = \left[ B_{\infty},A_{nj} \right] 
   + \sum_{k \neq j} {\dot{z}_j-\dot{z}_k \over z_j-z_k} \left[ A_{nk}, A_{nj} \right] .
\end{equation}
Again there is not anything essentially new here, that couldn't be derived from the
system of deformation derivatives (\ref{ops_tD:a}-\ref{ops_tD:d}), but it is an
efficient way to compute the deformation derivatives of bilinear products.
Employing the explicit representations of our matrices $ A_k $ we find the 
following independent derivatives in component form
\begin{multline}
  {d \over dt}{\rho_j \over 2V(z_j)}
  \left[ \Omega_n(z_j)+V(z_j)-{\kappa_{n+1} \over \kappa_{n}}z_j\Theta_n(z_j) \right]
  = -{\rho_j \over 2V(z_j)}{\phi_{n+1}(0) \over \kappa^3_n}
     {d \over dt}(\kappa_n\bar{\phi}_n(0))\Theta_n(z_j)
  \\
    - {\rho_j \over 2V(z_j)}{|\phi_{n+1}(0)|^2 \over \kappa^2_{n}}
      \sum_{k \neq j}{\dot{z}_j-\dot{z}_k \over z_j-z_k} {\rho_k \over 2V(z_k)}
      \left[ z_k\Theta^*_n(z_k)\Theta_n(z_j)-z_j\Theta_n(z_k)\Theta^*_n(z_j) \right] ,
  \label{ops_Schl:a}
\end{multline}
\begin{multline}
  {d \over dt}{\rho_j \over 2V(z_j)}
  {\phi_{n+1}(0) \over \kappa_{n}}\Theta_n(z_j)
  = {\rho_j \over V(z_j)}{\phi_{n+1}(0) \over \kappa_n}\bigg\{
     {\dot{\kappa}_n \over \kappa_n}\Theta_n(z_j)
    + \sum_{k \neq j}{\dot{z}_j-\dot{z}_k \over z_j-z_k} {\rho_k \over 2V(z_k)}
  \\ \times
      \left[ \Theta_n(z_k)
             \big[\Omega_n(z_j)-{\kappa_{n+1} \over \kappa_{n}}z_j\Theta_n(z_j)\big]
            -\Theta_n(z_j)
             \big[\Omega_n(z_k)-{\kappa_{n+1} \over \kappa_{n}}z_k\Theta_n(z_k)\big] \right]
    \bigg\} ,
  \label{ops_Schl:b}
\end{multline}
\begin{multline}
  {d \over dt}{\rho_j \over 2V(z_j)}
  {\bar{\phi}_{n+1}(0) \over \kappa_{n}}z_j\Theta^*_n(z_j)
  = {\rho_j \over V(z_j)}{\bar{\phi}_{n+1}(0) \over \kappa_n}\bigg\{
     - {\dot{\kappa}_n \over \kappa_n}z_j\Theta^*_n(z_j)
  \\
     + {1\over \kappa_n\bar{\phi}_{n+1}(0)}{d \over dt}(\kappa_n\bar{\phi}_{n}(0))
       \big[\Omega^*_n(z_j)-{\kappa_{n+1} \over \kappa_{n}}\Theta^*_n(z_j)\big]
    - \sum_{k \neq j}{\dot{z}_j-\dot{z}_k \over z_j-z_k} {\rho_k \over 2V(z_k)}
  \\ \times
      \left[ z_k\Theta^*_n(z_k)
                \big[\Omega^*_n(z_j)-{\kappa_{n+1} \over \kappa_{n}}\Theta^*_n(z_j)\big]
            -z_j\Theta^*_n(z_j)
                \big[\Omega^*_n(z_k)-{\kappa_{n+1} \over \kappa_{n}}\Theta^*_n(z_k)\big] \right]
    \bigg\} .
  \label{ops_Schl:c}
\end{multline}

\begin{remark}
The fact that the deformation equations satisfy the Schlesinger system of 
partial differential equations should be of no great surprise as the isomonodromic
properties of the regular semi-classical weights are quite transparent. In the
neighbourhood of any isolated singularity $ |z-z_j| < \delta $ the 
Carath\'eodory function can be decomposed 
\begin{equation}
   F(z) = f_j(z) + C_jw(z) ,
\end{equation}
where $ f_j(z) $ is the unique, holomorphic function in this neighbourhood and
$ C_j $ a coefficient. The analytic continuation of $ Y_n $ around a closed loop 
enclosing the singularity is easily found and defines the monodromy matrix $ M_j $,
\begin{equation}
   \left .Y_n \right|_{z_j + \delta e^{2\pi i}} =
   \left .Y_n \right|_{z_j + \delta}M_j, \qquad
   M_j = 
       \begin{pmatrix}
              1 & C_j(1-e^{-2\pi i\rho_j}) \cr
              0 & e^{-2\pi i\rho_j} \cr
       \end{pmatrix} .
\end{equation}
From the definition (\ref{ops_Cfun}) one can confirm that
$ C_j = 1/i\sin(\pi\rho_j) $, so that the $ M_j $ is quite naturally independent of
the deformation variables $ z_j $ or $ t $ ($ \rho_j $ being constant).
\end{remark}

\section{The simplest Semi-classical Class: \PVI System}
\label{PVIsection}
\setcounter{equation}{0}

Here we consider the application of the general theory above to the simplest 
instance of the semi-classical weight, namely $ m=3 $ singular points with two
fixed at $ z = 0,-1 $ and the third a variable at $ z=-1/t $. 
 
Explicitly we consider the unitary group average (\ref{VI_Toep}) where 
$ t = e^{i\phi} $, $ \omega_1, \omega_2, \mu \in \CC $ 
($ \omega,\bar{\omega}=\omega_1\pm i\omega_2 $) and $ \xi \in \CC $. 
In the initial formulation $ t \in \TT $ but will be analytically continued off 
the unit circle. The weight function
\begin{equation}
   w(z) = t^{-\mu} z^{-\mu-\omega}(1+z)^{2\omega_1}(1+tz)^{2\mu}
   \begin{cases}
     1 & \theta \notin (\pi-\phi,\pi) \\
     1-\xi & \theta \in (\pi-\phi,\pi)
   \end{cases} ,
\label{VI_wgt}
\end{equation}
is known as a generalised Jacobi weight, with branch points at 
$ z = 0,-1, -1/t, \infty $. When $ t \in \TT $, $ \mu, \omega_1, \omega_2, \xi \in \mathbb{R} $
and $ \xi < 1 $ this weight is real and positive. The Toeplitz matrix is 
then hermitian and as a consequence $ \bar{r}_n $ is the complex conjugate of 
$ r_n $, but generally this is not the case.
The Toeplitz matrix elements can be evaluated in terms of the Gauss hypergeometric
function, and there are several forms this can take which exhibit manifest 
analyticity at either of the special points $ t = 0,1, \infty $.

\begin{lemma}
The Toeplitz matrix element $ w_n $ for the weight (\ref{VI_wgt}), under the restriction
$ \Re(\mu), \Re(\omega_1) > -\half $, is given in terms of hypergeometric functions 
analytic at $ t = 0,1 $
\begin{multline}
   t^{\mu}w_{n} = {\Gamma(2\omega_1+1) \over  
                    \Gamma(1+n+\mu+\omega)\Gamma(1-n-\mu+\bar{\omega})}
   {}_2F_1(-2\mu,-n-\mu-\omega;1-n-\mu+\bar{\omega};t) \\
 + {\xi \over 2\pi i} e^{\pm \pi i(n+\mu-\bar{\omega})}
    {\Gamma(2\mu+1)\Gamma(2\omega_1+1) \over \Gamma(2\mu+2\omega_1+2)}
    t^{n+\mu-\bar{\omega}}(1-t)^{2\mu+2\omega_1+1}
 \\
   \times {}_2F_1(2\mu+1,1+n+\mu+\omega;2\mu+2\omega_1+2;1-t)
\label{VI_toepM}
\end{multline}
where the $ \pm $ sign is taken accordingly as $ {\rm Im}(t) \gtrless 0 $.
This can also be written as
\begin{multline}
   t^{\mu}w_{n} = 
   \left\{ 1+\xi{e^{\pm\pi i(n+\mu-\bar{\omega})} \over 
           2i\sin\pi(n+\mu-\bar{\omega})}
   \right\}
   {\Gamma(2\omega_1+1) \over
                    \Gamma(1+n+\mu+\omega)\Gamma(1-n-\mu+\bar{\omega})} \\
   \times{}_2F_1(-2\mu,-n-\mu-\omega;1-n-\mu+\bar{\omega};t) \\
   -\xi{e^{\pm\pi i(n+\mu-\bar{\omega})} \over 2i\sin\pi(n+\mu-\bar{\omega})}
       {\Gamma(2\mu+1) \over
                    \Gamma(1+n+\mu-\bar{\omega})\Gamma(1-n+\mu+\bar{\omega})} \\
           \times t^{n+\mu-\bar{\omega}}(1-t)^{2\mu+2\omega_1+1}                               
   {}_2F_1(2\mu+1,1+n+\mu+\omega;1+n+\mu-\bar{\omega};t).
\label{VI_toepM2}
\end{multline}
\end{lemma}
\begin{proof}
This follows from the generalisation of the Euler integral for the Gauss hypergeometric
function and consideration of the consistent phases for the branch cuts linking the
singular points, see \cite{Kl_1933} pp. 91, section 17 "Verallgemeinerung der Eulersche
Integrale".
\end{proof}

\begin{remark}
The first of these forms was given in \cite{FW_2003a}.
\end{remark}

For the description of (\ref{VI_wgt}) in terms of the semi-classical form 
(\ref{ops_scwgt2}) we have
\begin{gather}
  m=3 , \quad \{z_j\}^{3}_{j=1} = \{0,-1,-1/t\}, \quad 
  \{\rho_j\}^{3}_{j=1} = \{-\mu-\omega,2\omega_1,2\mu\}, \\
  2V(0) = -(\mu+\omega)t^{-1}, \; W'(0) = t^{-1}, \quad 
   V(-t^{-1}) = \mu{1-t \over t^2}
\end{gather}
and the coefficient functions are
\begin{gather}
\begin{split}
  \Theta_N(z)
 = & {\kappa_N \over \kappa_{N+1}}\left[ (N+1+\mu+\bar{\omega})z
       -{r_N \over r_{N+1}}(N+\mu+\omega)t^{-1} \right]
\end{split}
 \label{VI_theta:a} \\
\begin{split}
  \Theta^*_N(z)
 = & {\kappa_N \over \kappa_{N+1}}\left[
       -{\bar{r}_N \over \bar{r}_{N+1}}(N+\mu+\bar{\omega})z
       +(N+1+\mu+\omega)t^{-1} \right]
\end{split}
 \label{VI_theta:b} \\
\begin{split}
 & \Omega_N(z)
   = [1+\shalf(\mu+\bar{\omega})]z^2
 \\
 & + \left\{ (N+2+\mu+\bar{\omega})(1-r_{N+1}\bar{r}_{N+1}){r_{N+2} \over r_{N+1}}
             -{l_{N+1} \over \kappa_{N+1}}
             +[1+\shalf (\mu+\bar{\omega})]{1+t \over t}-\omega_1-{\mu \over t} \right\}z
 \\
 &  - [N+\shalf (\mu+\omega)]t^{-1}
\end{split}
 \label{VI_omega:a} \\
\begin{split}
  \Omega^*_N(z)
 = & -\shalf (\mu+\bar{\omega})z^2
 \\
 & + \left\{ {l_{N+1} \over \kappa_{N+1}}-(N+\mu+\bar{\omega})
             (1-r_{N+1}\bar{r}_{N+1}){\bar{r}_{N} \over \bar{r}_{N+1}}
             -\shalf (\mu+\bar{\omega}){1+t \over t}+\omega_1+{\mu \over t} \right\}z
 \\
 & +[N+1+\shalf (\mu+\omega)]t^{-1}
\end{split}
 \label{VI_omega:b}
\end{gather}

Our objective is to show that the average (\ref{VI_Toep}) can be evaluated via 
recurrence relations for the reflection coefficients. We begin with some 
preliminary lemmas. 

\begin{lemma}
The reflection coefficients for the weight (\ref{VI_wgt}) satisfy the homogeneous
second-order difference equation
\begin{multline}
   (N+1+\mu+\bar{\omega})tr_{N+1}\bar{r}_{N} - (N-1+\mu+\bar{\omega})tr_{N}\bar{r}_{N-1}
  \\ = (N+1+\mu+\omega)\bar{r}_{N+1}r_{N} - (N-1+\mu+\omega)\bar{r}_{N}r_{N-1} .
\label{VI_2ndRR}
\end{multline}
\end{lemma}

\begin{proof}
The result of the above lemma, (\ref{VI_2ndRR}) can be found immediately from the
general theory of Section 2, in many ways. By equating coefficients of $ z $
in the functional-difference equation (\ref{ops_rrCf:g}) using (\ref{VI_theta:a},
\ref{VI_theta:b},\ref{VI_omega:a},\ref{VI_omega:b}), all are trivially satisfied
except for the $ z $ coefficient, which is precisely (\ref{VI_2ndRR}). Similarly
starting with (\ref{ops_rrCf:h}) and employing (\ref{VI_theta:a},
\ref{VI_theta:b},\ref{VI_omega:b}), one finds (\ref{VI_2ndRR}). Alternatively 
one could start with either (\ref{ops_rrCf:i}) or (\ref{ops_rrCf:k}) and arrive at the
same result
\end{proof}

\begin{corollary}
The sub-leading coefficients $ l_{N}, \bar{l}_{N} $ satisfy the linear inhomogeneous 
equation
\begin{equation}
   (N+\mu+\bar{\omega})tl_{N} - (N+\mu+\omega)\bar{l}_{N}
   = N\left[ \mu(t-1)+\bar{\omega}-\omega t \right]\kappa_{N} .
\label{VI_lRecur}
\end{equation}
\end{corollary}

\begin{proof}
By substituting the general expression for the first difference of $ l_N, \bar{l}_{N} $ 
using (\ref{ops_l}) in (\ref{VI_2ndRR}) one finds that it can be summed exactly to yield
\begin{multline}
  (N+1+\mu+\bar{\omega})t{l_{N+1} \over \kappa_{N+1}}
  - (N+1+\mu+\omega){\bar{l}_{N+1} \over \kappa_{N+1}} \\
  - (N+\mu+\bar{\omega})t{l_{N} \over \kappa_{N}}
  + (N+\mu+\omega){\bar{l}_{N} \over \kappa_{N}}
  = \mu(t-1)+\bar{\omega}-\omega t .
\label{VI_lRecurD}
\end{multline}
This can be summed once more to yield the stated result.
\end{proof}

\begin{remark}
One could alternatively proceed via the Freud approach \cite{Fr_1976} (see also
\cite{FW_2003b}) and consider the integral
\begin{equation}
    \int_{\TT} {dz \over 2\pi iz} (1+z)(1+tz)
  \left[ -{\mu+\omega \over z} + {2\omega_1 \over 1+z} + {2\mu t\over 1+tz} \right]
   w(z) \phi_{N}(z)\overline{\phi_{N}(z)}.
\end{equation}
Here we recognise the logarithmic derivative of the weight function in the 
integrand
\begin{equation}
   {w' \over w} = -{\mu+\omega \over z} + {2\omega_1 \over 1+z} + {2\mu t\over 1+tz},
\end{equation}
and by evaluating the integral in the two ways we find a linear equation
for $ l_{N} $, namely (\ref{VI_lRecurD}).
\end{remark}

\begin{lemma}
The sub-leading coefficients are related to the reflection coefficients by
\begin{align}
  \bar{l}_{N}/\kappa_{N} + tl_{N}/\kappa_{N} -N(t+1)
  & = {1-r_{N}\bar{r}_{N} \over r_{N}}
    \left[ (N+1+\mu+\bar{\omega})tr_{N+1} + (N-1+\mu+\omega)r_{N-1} \right]
  \label{VI_magnus:a} \\
  & = {1-r_{N}\bar{r}_{N} \over \bar{r}_{N}}
    \left[ (N+1+\mu+\omega)\bar{r}_{N+1} + (N-1+\mu+\bar{\omega})t\bar{r}_{N-1} \right]
\label{VI_magnus:b}
\end{align}
\end{lemma}

\begin{proof}
The first relation follows from a comparison of the coefficients of $ z $ for 
$ \Omega_N(z) $ given the two distinct expansions, the first by (\ref{ops_Omexp:a}) 
which reduces to (\ref{VI_omega:a}) and the second by the specialisation of 
(\ref{ops_Omexp:b}). The second relation follows from identical arguments applied to
$ \Omega^*_N(z) $ or by employing (\ref{VI_2ndRR}) in the first relation.
\end{proof}

\begin{remark}
The first result appears in the Magnus derivation \cite{Ma_2000} for the generalised
Jacobi weight, with $ \theta_1 = \pi-\phi, \theta_2 = \pi $, 
$ \alpha = \mu, \beta = \omega_1, \gamma = -\omega_2 $. Then Eq. (14) of that work 
is precisely (\ref{VI_magnus:a}).
\end{remark}

\begin{remark}
The Magnus relation (\ref{VI_magnus:a}) can also be found by employing the Freud 
method. In this one uses integration by parts on the integral
\begin{equation}
   \int_{\TT} {dz \over 2\pi iz} z^{-1}(1+z)(1+tz)
   w'(z) \phi_{N+1}(z)\overline{\phi_{N}(z)},
\end{equation}
and in the term involving $ \phi'_{N+1}(z) $ one employs (\ref{ops_zD:a})
for the derivative and (\ref{VI_theta:a}), (\ref{VI_omega:a}) for the coefficient
functions. Equating this expression to a direct evaluation of the integral then 
yields (\ref{VI_magnus:a}). 
\end{remark}

\begin{lemma}
The sub-leading coefficient $ l_N $ can be expressed in terms of the reflection 
coefficients in the following ways
\begin{gather}\label{VI_lSoln:a}
   2t{l_N \over \kappa_{N}} =
   (N+1+\mu+\bar{\omega})t({r_{N+1} \over r_{N}}-r_{N+1}\bar{r}_{N})
  +(N-1+\mu+\omega){r_{N-1} \over r_{N}}
  \\
  -(N-1+\mu+\bar{\omega})r_{N}\bar{r}_{N-1} + (N+\mu-\omega)t + N-\mu+\bar{\omega}
  \nonumber \\
  = (N+1+\mu+\omega){\bar{r}_{N+1} \over \bar{r}_{N}}
   +(N-1+\mu+\bar{\omega})t({\bar{r}_{N-1} \over \bar{r}_{N}}-r_{N}\bar{r}_{N-1})
  \label{VI_lSoln:b} \\
  -(N+1+\mu+\bar{\omega})tr_{N+1}\bar{r}_{N} + (N+\mu-\omega)t + N-\mu+\bar{\omega}
  \nonumber
\end{gather} 
as well as analogous expressions for $ \bar{l}_N $.
\end{lemma}

\begin{proof}
The first expression follows from a comparison of the $ z^0 $ coefficients for 
$ \Theta_N(z) $ evaluated using both (\ref{ops_Thexp:a}) and (\ref{ops_Thexp:b}).
The second relation follows from an applying the same reasoning to $ \Theta^*_N(z) $.
\end{proof}

We will refer to the order of a system of coupled difference equations with two 
variables $ r_n, \bar{r}_n $ say as $ q/p $ where $ q\in \mathbb{Z}_{\geq 0} $ 
refers to the order of $ r_n $ and $ p\in \mathbb{Z}_{\geq 0} $ refers to the 
order of $ \bar{r}_n $.

\begin{corollary}
The reflection coefficients of the OPS for the weight (\ref{VI_wgt}) satisfy the
$ 2/2 $ order recurrence relations
\begin{align}
   tr_{N}\bar{r}_{N-1} + r_{N-1}\bar{r}_{N} -t-1
   & = {1-r_{N}\bar{r}_{N} \over r_{N}}
       \left[ (N+1+\mu+\bar{\omega})tr_{N+1} + (N-1+\mu+\omega)r_{N-1} \right]
   \nonumber\\
   & \quad
     - {1-r_{N-1}\bar{r}_{N-1} \over \bar{r}_{N-1}}
       \left[ (N+\mu+\omega)\bar{r}_{N} + (N-2+\mu+\bar{\omega})t\bar{r}_{N-2} 
       \right]
   \label{VI_rRecur:a}
   \\
   & = {1-r_{N}\bar{r}_{N} \over \bar{r}_{N}}
       \left[ (N+1+\mu+\omega)\bar{r}_{N+1} + (N-1+\mu+\bar{\omega})t\bar{r}_{N-1}
       \right]
   \nonumber\\
   & \quad
     - {1-r_{N-1}\bar{r}_{N-1} \over r_{N-1}}
       \left[ (N+\mu+\bar{\omega})tr_{N} + (N-2+\mu+\omega)r_{N-2} \right]
   \label{VI_rRecur:b}
\end{align}
and those specifying the solution for (\ref{VI_Toep}) have the initial values
\begin{equation} 
   r_{0} = \bar{r}_{0} = 1, \quad
   r_{1} = -w_{-1}/w_0, \quad \bar{r}_{1} = -w_{1}/w_{0},
   \label{VI_rInitial}
\end{equation} 
where the Toeplitz matrix elements are given in (\ref{VI_toepM2}).
\end{corollary}

\begin{proof}
Solving (\ref{VI_magnus:a}) for the combination of $ l_{N}, \bar{l}_{N} $ and 
differencing this, one arrives at (\ref{VI_rRecur:a}). This however is of order 
$ 3/1 $ but by employing (\ref{VI_2ndRR}) we can reduce the order in $ r_{N} $ of 
the recurrence to second order. The other member of the pair (\ref{VI_rRecur:b}) 
is found in an identical manner starting with 
(\ref{VI_magnus:b}).
\end{proof}

\begin{remark}
The second-order difference (\ref{VI_rRecur:a}) also follows immediately from 
equating
the polynomials in $ z $ arising in the functional-difference (\ref{ops_rrCf:b}), 
after employing (\ref{VI_theta:a},\ref{VI_omega:a}). The other member of the pair, 
(\ref{VI_rRecur:b}), follows from the functional-difference (\ref{ops_rrCf:d}),
after using (\ref{VI_theta:b},\ref{VI_omega:b}).
\end{remark}

\begin{remark}
In their most general example Adler and van Moerbeke also considered this
weight. In terms of their variables we should set 
$ P_1 = P_2 = 0, d_1 = t^{-1/2}, d_2 = t^{1/2} $, and without loss of generality
$ \gamma''_1 = \gamma'_2 = 0 $. For the other parameters 
$ \gamma = \mu-\omega, \gamma'_1 = 2\omega_1, \gamma''_2 = 2\mu $.
There is a slight difference in the dependent variables due to the additional
factor of $ t $, so that we have the identification $ x_{N} = (-1)^Nt^{N/2}r_{N} $,
$ y_{N} = (-1)^Nt^{-N/2}\bar{r}_{N} $ and $ v_{N} = 1-r_{N}\bar{r}_{N} $. 
Generalising their working one finds that their Eq. (0.0.14) implies
\begin{multline}
   -(N+1+\mu+\bar{\omega})x_{N+1}y_{N} + (N+1+\mu+\omega)x_{N}y_{N+1} \\
   +(N-1+\mu+\bar{\omega})x_{N}y_{N-1} - (N-1+\mu+\omega)x_{N-1}y_{N} = 0.
\end{multline}
Now by transforming to our $ r_{N},\bar{r}_{N} $ and employing (\ref{ops_l})
one finds this is precisely (\ref{VI_2ndRR}), which we showed is solved by
(\ref{VI_lRecur}). Their inhomogeneous Eq. (0.0.15) now takes the form
\begin{multline}
   -v_{N}\left[ (N+1+\mu+\bar{\omega})x_{N+1}y_{N-1}+N+\mu+\omega \right] \\
   +v_{N-1}\left[ (N-2+\mu+\bar{\omega})x_{N}y_{N-2}+N-1+\mu+\omega \right] \\
   +x_{N}y_{N-1}(x_{N}y_{N-1}+t^{1/2}+t^{-1/2}) \\
  = -v_{1}\left[ (2+\mu+\bar{\omega})x_{2}+1+\mu+\omega \right]
      +x_{1}(x_{1}+t^{1/2}+t^{-1/2}) .
\end{multline}
Upon recasting this into our variables and manipulating, it then becomes
\begin{multline}
   tr_{N}\bar{r}_{N-1} + \bar{r}_{N}r_{N-1}-t-1
   - {1-r_{N}\bar{r}_{N} \over r_{N}}
       \left[ (N+1+\mu+\bar{\omega})tr_{N+1} + (N-1+\mu+\omega)r_{N-1} \right] \\
   \quad
   + {1-r_{N-1}\bar{r}_{N-1} \over \bar{r}_{N-1}}
     \left[ (N+\mu+\omega)\bar{r}_{N} + (N-2+\mu+\bar{\omega})t\bar{r}_{N-2} \right] \\
   = {1-(1-r_{1}\bar{r}_{1})\left[(2+\mu+\bar{\omega})tr_{2}+1+\mu+\omega \right]
             +r_{1}(tr_{1}-t-1) \over r_{N}\bar{r}_{N-1}} .
\end{multline}
However using the identity
\begin{equation*}
  c{}_2F_1(a,b;c;x) 
  = [c+(1+b-a)x]{}_2F_1(a,b+1;c+1;x) -\frac{b+1}{c+1}(1+c-a)x{}_2F_1(a,b+2;c+2;x)
\end{equation*}
we note that the right-hand side is identically zero for the initial conditions 
(\ref{VI_rInitial}) and the recurrence is not genuinely inhomogeneous, thus 
yielding our first relation above, (\ref{VI_rRecur:a}).
\end{remark}

We seek recurrences for $ r_N, \bar{r}_N $ which are of the form of the discrete
Painlev\'e system (\ref{dPV:a}), (\ref{dPV:b}). For this purpose a number of
distinct forms of the former will be presented.

\begin{proposition}
The reflection coefficients satisfy a system of a $ 2/0 $ order recurrence relation
\begin{multline}\label{VI_2+0rRecur:a}
  \Big\{ (1-r_N\bar{r}_N) \left[
  (N+1+\mu+\bar{\omega})(N+\mu+\bar{\omega})tr_{N+1}
     - (N+\mu+\omega)(N-1+\mu+\omega)r_{N-1} \right]
  \\ + N(N+2\omega_1)(t-1)r_N \Big\}
  \\ \times
  \Big\{ (1-r_N\bar{r}_N) \left[
  (N+1+\mu+\bar{\omega})(N+\mu+\bar{\omega})tr_{N+1}
     - (N+\mu+\omega)(N-1+\mu+\omega)r_{N-1} \right] 
  \\ + (N+2\mu)(N+2\mu+2\omega_1)(t-1)r_N \Big\}
  \\ 
  = -(2N+2\mu+2\omega_1)^2t(1-r_N\bar{r}_N)
  \\ \times
  \left[ (N+1+\mu+\bar{\omega})r_{N+1}+(N+\mu+\omega)r_{N} \right]
  \left[ (N+\mu+\bar{\omega})r_{N}+(N-1+\mu+\omega)r_{N-1} \right]
\end{multline}
and a $ 0/2 $ order recurrence relation which is just (\ref{VI_2+0rRecur:a}) with the 
replacements $ \omega \leftrightarrow \bar{\omega} $ and 
$ t^{\pm 1/2}r_j \mapsto t^{\mp 1/2}\bar{r}_j $ 
\end{proposition}

\begin{proof}
Consider first the specialisation of (\ref{ops_OTeq:a}) to our weight at hand
at the singular point $ z=-1 $, and we have
\begin{multline} \label{VI_Bilinear:a}
   \Big\{ {l_N \over \kappa_N} - Nt^{-1} 
          - (N+1+\mu+\bar{\omega}){\kappa^2_{N-1} \over \kappa^2_N}{r_{N+1} \over r_N} 
          + \omega_1(1-t^{-1}) \Big\}^2
   \\
   + {\kappa^2_{N-1} \over \kappa^2_N}
     \left[ N+\mu+\bar{\omega} + {(N-1+\mu+\omega)\over t}{r_{N-1} \over r_N}
     \right]
     \left[ {(N+\mu+\omega)\over t} + (N+1+\mu+\bar{\omega}){r_{N+1} \over r_N}
     \right]
   \\
   = \omega^2_1 \left({t-1 \over t}\right)^2 ,
\end{multline}
by using (\ref{VI_theta:a},\ref{VI_omega:a}). Similarly (\ref{ops_OTeq:a}) 
evaluated at $ z=-1/t $ yields
\begin{multline} \label{VI_Bilinear:b}
   \Big\{ {l_N \over \kappa_N} - N 
          - (N+1+\mu+\bar{\omega}){\kappa^2_{N-1} \over \kappa^2_N}{r_{N+1} \over r_N} 
          + \mu(t^{-1}-1) \Big\}^2
   \\
   + {\kappa^2_{N-1} \over \kappa^2_N}t^{-1}
     \left[ N+\mu+\bar{\omega} + (N-1+\mu+\omega){r_{N-1} \over r_N}
     \right]
     \left[ N+\mu+\omega + (N+1+\mu+\bar{\omega}){r_{N+1} \over r_N}
     \right]
   \\
   = \mu^2 \left({t-1 \over t}\right)^2 .
\end{multline}
The first relation follow by eliminating $ l_N $ between (\ref{VI_Bilinear:a}) and
(\ref{VI_Bilinear:b}), whereas the second follows from an identical analysis to that
employed in the proof of Proposition \ref{propVI_dPa} but starting with the bilinear 
identity (\ref{ops_OTeq:b}).
\end{proof}

\begin{proposition}
The reflection coefficients also satisfy a system of $ 1/1 $ order recurrence relations
the first of which is
\begin{multline}\label{VI_1+1rRecur:a}
  \Big\{ -(N+\mu+\omega)(1-r_N\bar{r}_N)t\left[
  (N+1+\mu+\bar{\omega})r_{N+1}\bar{r}_N + (N-1+\mu+\bar{\omega})r_N\bar{r}_{N-1} \right]
  \\ + 2(N+\mu+\omega)^2r^2_N\bar{r}^2_N - (N+\mu+\omega)^2(t+1)r_N\bar{r}_N
     - 2(N+\mu+\omega)\bar{\omega}(t-1)r_N\bar{r}_N
  \\
     + (\mu-\bar{\omega})(\mu+\bar{\omega})(t-1) \Big\}
  \\ \times
  \Big\{ -(N+\mu+\omega)(1-r_N\bar{r}_N)t\left[
  (N+1+\mu+\bar{\omega})r_{N+1}\bar{r}_N + (N-1+\mu+\bar{\omega})r_N\bar{r}_{N-1} \right]
  \\ + 2(N+\mu+\omega)^2r^2_N\bar{r}^2_N - (N+\mu+\omega)^2(t+1)r_N\bar{r}_N
     + 2(N+\mu+\omega)\omega(t-1)r_N\bar{r}_N
  \\
     + (\mu-\omega)(\mu+\omega)(t-1) \Big\}
  = -\left[ 2(N+\mu+\omega)r_N\bar{r}_N+\bar{\omega}-\omega \right]^2(1-r_N\bar{r}_N)
  \\ \times
  \left[ (N+1+\mu+\bar{\omega})tr_{N+1}+(N+\mu+\omega)r_{N} \right]
  \left[ (N+\mu+\omega)\bar{r}_{N}+(N-1+\mu+\bar{\omega})t\bar{r}_{N-1} \right]
\end{multline}
and the second is obtained from (\ref{VI_1+1rRecur:a}) with the replacements
$ \omega \leftrightarrow \bar{\omega} $ and
$ t^{\pm 1/2}r_j \leftrightarrow t^{\mp 1/2}\bar{r}_j $.    
\end{proposition}

\begin{proof}
The specialisation of (\ref{ops_OTeq:c}) to the weight (\ref{VI_wgt}) evaluated
at the singular point $ z=-1 $ is
\begin{multline} \label{VI_Bilinear:c}
   \Big\{ {l_N \over \kappa_N} - Nt^{-1} 
          + (N+\mu+\omega)t^{-1}{\kappa^2_{N-1} \over \kappa^2_N}
          + \omega_1(1-t^{-1}) \Big\}^2
   \\
   + {\kappa^2_{N-1} \over \kappa^2_N}
     \left[ (N+1+\mu+\bar{\omega})r_{N+1} + (N+\mu+\omega)t^{-1} r_N
     \right]
   \\ \times
     \left[ (N-1+\mu+\bar{\omega})\bar{r}_{N-1} + (N+\mu+\omega)t^{-1}\bar{r}_N
     \right] = \omega^2_1 \left({t-1 \over t}\right)^2 ,
\end{multline}
by using (\ref{VI_theta:a},\ref{VI_omega:a}). Similarly (\ref{ops_OTeq:c}) 
evaluated at $ z=-1/t $ yields
\begin{multline} \label{VI_Bilinear:d}
   \Big\{ {l_N \over \kappa_N} - N 
          + (N+\mu+\omega){\kappa^2_{N-1} \over \kappa^2_N}
          + \mu(t^{-1}-1) \Big\}^2
   \\
   + {\kappa^2_{N-1} \over \kappa^2_N}
     \left[ (N+1+\mu+\bar{\omega})r_{N+1} + (N+\mu+\omega)r_N
     \right]
   \\ \times
     \left[ (N-1+\mu+\bar{\omega})\bar{r}_{N-1} + (N+\mu+\omega)\bar{r}_N
     \right] = \mu^2 \left({t-1 \over t}\right)^2 .
\end{multline}
Again eliminating $ l_N $ between these two equations yields the recurrence relation
(\ref{VI_1+1rRecur:a}). The second follows in the same way starting with (\ref{ops_OTeq:d}).
\end{proof}

\begin{proposition}
The reflection coefficients satisfy an alternative system of $ 1/1 $ order recurrence 
relations the first of which is
\begin{multline}\label{VI_1+1rRecur:b}
  \Big[ (N+1+\mu+\bar{\omega})(N+\mu+\bar{\omega})tr_{N+1}\bar{r}_N 
  \\  - (N+1+\mu+\omega)(N+\mu+\omega)\bar{r}_{N+1}r_N
      + (\bar{\omega}-\mu)(\bar{\omega}+\mu)(t-1) \Big]
  \\ \times
  \Big[ (N+1+\mu+\bar{\omega})(N+\mu+\bar{\omega})tr_{N+1}\bar{r}_N 
  \\  - (N+1+\mu+\omega)(N+\mu+\omega)\bar{r}_{N+1}r_N
      + (\omega-\mu)(\omega+\mu)(t-1) \Big]
  \\ 
  = (\bar{\omega}-\omega)^2
  \\ \times
  \left[ (N+1+\mu+\bar{\omega})tr_{N+1}+(N+\mu+\omega)r_{N} \right]
  \left[ (N+1+\mu+\omega)\bar{r}_{N+1}+(N+\mu+\bar{\omega})t\bar{r}_{N} \right]
\end{multline}
and the second is again obtained from (\ref{VI_1+1rRecur:b}) with the replacements
$ \omega \leftrightarrow \bar{\omega} $ and
$ t^{\pm 1/2}r_j \leftrightarrow t^{\mp 1/2}\bar{r}_j $.    
\end{proposition}

\begin{proof}
The specialisation of (\ref{ops_OTeq:e}) to the weight (\ref{VI_wgt}) evaluated
at the singular point $ z=-1 $ is
\begin{multline} \label{VI_Bilinear:e}
   \Big\{ {\bar{l}_{N+1} \over \kappa_{N+1}}
          + (N+\mu+\omega)\bar{r}_{N+1}r_N
          + \omega_1+(\mu-i\omega_2)t \Big\}^2
   \\
   = \left[ (N+1+\mu+\bar{\omega})tr_{N+1} + (N+\mu+\omega)r_N \right]
   \\ \times
     \left[ (N+1+\mu+\omega)\bar{r}_{N+1} + (N+\mu+\bar{\omega})t\bar{r}_N \right]
     + \omega^2_1 (t-1)^2 ,
\end{multline}
by using (\ref{VI_theta:a},\ref{VI_omega:a}). Similarly (\ref{ops_OTeq:e}) 
evaluated at $ z=-1/t $ yields
\begin{multline} \label{VI_Bilinear:f}
   \Big\{ {\bar{l}_{N+1} \over \kappa_{N+1}}
          + (N+\mu+\omega)\bar{r}_{N+1}r_N
          + \bar{\omega}+\mu t \Big\}^2
   \\
   = \left[ (N+1+\mu+\bar{\omega})r_{N+1} + (N+\mu+\omega)r_N \right]
   \\ \times
     \left[ (N+1+\mu+\omega)\bar{r}_{N+1} + (N+\mu+\bar{\omega})\bar{r}_N \right]
     + \mu^2 (t-1)^2 ,
\end{multline}
Again eliminating $ \bar{l}_{N+1} $ between these two equations yields the recurrence relation
(\ref{VI_1+1rRecur:b}). The second follows in the same way starting with (\ref{ops_OTeq:f}).
\end{proof}

\begin{remark}
Note that the recurrence system (\ref{VI_2+0rRecur:a}) and its partner is quadratic
in $ r_{N+1} $, $ r_{N-1} $ and $ \bar{r}_{N+1} $, $\bar{r}_{N-1} $, the system
(\ref{VI_1+1rRecur:a}) and its partner is also quadratic in $ r_{N+1}, \bar{r}_{N-1} $
and $ \bar{r}_{N+1} $, $ r_{N-1} $, and likewise (\ref{VI_1+1rRecur:b}) is quadratic in 
$ r_{N+1} $, $ \bar{r}_{N+1} $. This renders them less useful in practical iterations
than the higher order systems that are linear in the highest difference. By raising the 
order of one of the variables by one we can obtain a recurrence linear in the highest
difference.
\end{remark}

\begin{corollary}
The reflection coefficients satisfy a system of a $ 2/1 $ order recurrence relation
\begin{multline}\label{VI_2+1rRecur:a}
  (N+1+\mu+\bar{\omega})(\bar{\omega}-\omega)t(1-r_N\bar{r}_N)r_{N+1}
 \\
 + (N-1+\mu+\omega)[2(N+\mu+\omega)r_N\bar{r}_N+\bar{\omega}-\omega]r_{N-1}
 \\
 - (N-1+\mu+\bar{\omega})(2N+2\mu+2\omega_1)tr^2_N\bar{r}_{N-1}
 \\
 + \big[ (\bar{\omega}-\omega)N(t+1)-(2\mu+2\omega_1)[\mu(1-t)+\omega t-\bar{\omega}]
   \big]r_n = 0
\end{multline}
and a $ 1/2 $ order recurrence relation which is again obtained from 
(\ref{VI_2+1rRecur:a}) with the replacements $ \omega \leftrightarrow \bar{\omega} $ and
$ t^{\pm 1/2}r_j \leftrightarrow t^{\mp 1/2}\bar{r}_j $.
\end{corollary}

\begin{proof}
The solutions for the sub-leading coefficient $ l_N, \bar{l}_N $ that arise from the 
simultaneous solution of (\ref{VI_Bilinear:c},\ref{VI_Bilinear:d}) and 
(\ref{VI_Bilinear:e},\ref{VI_Bilinear:f}) respectively are given by
\begin{align}
  t{l_N \over \kappa_N} 
  = & \Big\{ (N+\mu+\omega)t(1-r_N\bar{r}_N)
      \left[ (N+1+\mu+\bar{\omega})r_{N+1}\bar{r}_N+(N-1+\mu+\bar{\omega})r_N\bar{r}_{N-1}
      \right]
  \label{VI_lSoln:c} \\
    & +(N+\mu+\omega)[N(t+1)-\mu(1-t)-\omega t+\bar{\omega}]r_N\bar{r}_N
      +(\omega+\mu)[\mu(1-t)+\omega t-\bar{\omega}] \Big\}
  \nonumber \\
    & \div [2(N+\mu+\omega)r_N\bar{r}_N+\bar{\omega}-\omega]
  \nonumber \\
  t{l_N \over \kappa_N}
  = & \Big\{ (N+\mu+\omega)
      \left[ (N-1+\mu+\bar{\omega})tr_{N}\bar{r}_{N-1}-(N-1+\mu+\omega)\bar{r}_{N}r_{N-1}
      \right]
  \label{VI_lSoln:d} \\
    & + (\omega+\mu)[\mu(1-t)+\omega t-\bar{\omega}] \Big\}\Big/(\bar{\omega}-\omega),
    \quad \text{if $ \bar{\omega} \neq \omega $} ,
  \nonumber 
\end{align}
and the corresponding expression for $ \bar{l}_N/\kappa_N $ under the above replacements.
Equating these two forms then leads to (\ref{VI_2+1rRecur:a}).
\end{proof}

The systems of recurrences that we have found are in fact equivalent to the 
discrete Painlev\'e equation associated with the 
degeneration of the rational surface $ D^{(1)}_4 \to D^{(1)}_5 $ and we give
our first demonstration of this fact here.
\begin{proposition}\label{propVI_dPa}
The $ N $-recurrence for the reflection coefficients of the orthogonal 
polynomial system with the weight (\ref{VI_wgt}) is governed by either of two 
systems of coupled first order discrete Painlev\'e equations (\ref{dPV:a}), (\ref{dPV:b}).
This first is
\begin{align}                                                                 
  g_{N+1}g_N                                                                
  & = t{(f_N+N)(f_N+N+2\mu) \over f_N(f_N-2\omega_1)}
  \label{VI_gRecur} \\       
  f_N+f_{N-1}      
  & = 2\omega_1+{N-1+\mu+\omega \over g_N-1}
      +{(N+\mu+\bar{\omega})t \over g_N-t},
  \label{VI_fRecur}
\end{align}
subject to the initial conditions
\begin{equation}
  g_1 = t{ \mu+\omega +(1+\mu+\bar{\omega})r_1 \over 
           \mu+\omega +(1+\mu+\bar{\omega})tr_1 } ,
  \quad f_0 = 0 .
\end{equation}
The transformations relating these variables to the reflection coefficients are
given by
\begin{align}
   g_N & =
  t{ N-1+\mu+\omega +(N+\mu+\bar{\omega})
                     \dfrac{r_N}{r_{N-1}}
     \over 
     N-1+\mu+\omega +(N+\mu+\bar{\omega})t
                     \dfrac{r_N}{r_{N-1}} } ,
  \label{VI_gXfm} \\
   f_N & = 
  {1 \over 1-t}\left[ 
   t{l_N \over \kappa_N} - N -(N+1+\mu+\bar{\omega})(1-r_N\bar{r}_N)t{r_{N+1} \over r_{N}}
               \right] .
  \label{VI_fXfm}
\end{align}
The second system is
\begin{align}                                                                 
  \bar{g}_{N+1}\bar{g}_N                                                                
  & = t^{-1}{(\bar{f}_N+N)(\bar{f}_N+N+2\omega_1) \over \bar{f}_N(\bar{f}_N-2\mu)}
  \label{VI_gRecurC} \\       
  \bar{f}_N+\bar{f}_{N-1}      
  & = 2\mu+{N+\mu+\omega \over \bar{g}_N-1}
      +{(N-1+\mu+\bar{\omega})t^{-1} \over \bar{g}_N-t^{-1}},
  \label{VI_fRecurC}
\end{align}
subject to the initial conditions
\begin{equation}
  \bar{g}_1 = {\mu+\bar{\omega} +(1+\mu+\omega)t^{-1}\bar{r}_1 \over 
           \mu+\bar{\omega} +(1+\mu+\omega)\bar{r}_1 } ,
  \quad \bar{f}_0 = 0 .
\end{equation}
The transformations relating these variables to the reflection coefficients are
given by
\begin{align}
   \bar{g}_N & =
   { N-1+\mu+\bar{\omega} +(N+\mu+\omega)t^{-1}
                     \dfrac{\bar{r}_N}{\bar{r}_{N-1}}
     \over 
     N-1+\mu+\bar{\omega} +(N+\mu+\omega)
                     \dfrac{\bar{r}_N}{\bar{r}_{N-1}} } ,
  \label{VI_gXfmC} \\
   \bar{f}_N & = 
  {1 \over 1-t}\left[ 
   -t{l_N \over \kappa_N}+Nt 
      +(N-1+\mu+\bar{\omega})(1-r_N\bar{r}_N)t{\bar{r}_{N-1} \over \bar{r}_{N}}
               \right] .
  \label{VI_fXfmC}
\end{align} 
\end{proposition}

\begin{proof}
Consolidating each of (\ref{VI_Bilinear:a}) and (\ref{VI_Bilinear:b}) into two 
terms and taking their ratio then leads to (\ref{VI_gRecur}) after
utilising the definitions (\ref{VI_gXfm},\ref{VI_fXfm}). The second member of the
recurrence system (\ref{VI_fRecur}) follows from the relation
\begin{multline}
  {l_{N+1} \over \kappa_{N+1}} + {l_N \over \kappa_N}
  \\
  = (N+2+\mu+\bar{\omega})(1-r_{N+1}\bar{r}_{N+1}){r_{N+2} \over r_{N+1}}
  + (N+\mu+\omega)t^{-1}{r_{N} \over r_{N+1}} 
  - (N+1+\mu+\bar{\omega})r_{N+1}\bar{r}_N
  \\
  - 2\omega_1 - 2\mu t^{-1} + (N+1+\mu+\bar{\omega})(1+t^{-1}) ,
\end{multline}
which results from a combination of (\ref{VI_lSoln:a}) and (\ref{ops_l}),
and the definition (\ref{VI_fXfm}). All the results for the second system follow
by applying identical reasoning starting with (\ref{ops_OTeq:b}).
\end{proof}

\begin{remark}
Generalised hypergeometric function evaluations were given in \cite{FW_2002b} 
in the special case $ \xi = 0 $. In terms of our unitary group average one such 
evaluation reads
\begin{multline}
   \Big\langle \prod^{N}_{l=1}
   z_l^{-\mu-\omega} (1+z_l)^{2\omega_1}(1+tz_l)^{2\mu} \Big\rangle_{U(N)} \\
  = \prod^{N-1}_{j=0}{j!\Gamma(2\omega_1+j+1) \over 
                      \Gamma(1+\mu+\omega+j)\Gamma(1-\mu+\bar{\omega}+j)} \\
  \times
    {}^{\vphantom{(1)}}_{2}F^{(1)}_{1}(-2\mu,-\mu-\omega;N-\mu+\bar{\omega};t_1,\ldots,t_N)
       |_{t_1=\ldots =t_N=t} ,
\label{VI_2F1}
\end{multline}
subject to $ \Re(\omega_1) > -\half $ and $ |t| < 1 $.
Similarly, for the reflection coefficients we have
\begin{align}
   r_{N} 
  & = (-1)^{N}{(\mu+\omega)_{N} \over (1-\mu+\bar{\omega})_{N}}
 {{}^{\vphantom{(1)}}_{2}F^{(1)}_{1}(-2\mu,1-\mu-\omega;N+1-\mu+\bar{\omega};t_1,\ldots,t_N)
      \over 
  {}^{\vphantom{(1)}}_{2}F^{(1)}_{1}(-2\mu,-\mu-\omega;N-\mu+\bar{\omega};t_1,\ldots,t_N) 
  }\Bigg|_{t_1=\ldots =t_N=t},
  \label{VI_genH:a}\\
   \bar{r}_{N} 
  & = (-1)^{N}{(-\mu+\bar{\omega})_{N} \over (1+\mu+\omega)_{N}}
 {{}^{\vphantom{(1)}}_{2}F^{(1)}_{1}(-2\mu,-1-\mu-\omega;N-1-\mu+\bar{\omega};t_1,\ldots,t_N)
      \over 
  {}^{\vphantom{(1)}}_{2}F^{(1)}_{1}(-2\mu,-\mu-\omega;N-\mu+\bar{\omega};t_1,\ldots,t_N)
  }\Bigg|_{t_1=\ldots =t_N=t}.
  \label{VI_genH:b}
\end{align}
The analog of the Euler identity for this function is
\begin{equation}
  {}^{\vphantom{(1)}}_{2}F^{(1)}_{1}(-2\mu,-\mu-\omega;N-\mu+\bar{\omega};t_1,\ldots,t_N)
       |_{t_1=\ldots =t_N=1} =
    \prod^{N}_{j=1}{\Gamma(j+2\mu+2\omega_1)\Gamma(j-\mu+\bar{\omega}) \over  
                      \Gamma(j+2\omega_1)\Gamma(j+\mu+\bar{\omega})} ,
\end{equation}
when $ \Re(\mu+\omega_1) > -\half $, $ \Re(-\mu+\bar{\omega}) > -1 $, thus implying 
Gamma function evaluations in the special case $ t=1 $.
\end{remark}
 
\begin{remark}
The degeneration of the \PVI system to the \PV system is facilitated by the replacements
$ \omega+\mu \mapsto \nu, \bar{\omega}-\mu \mapsto \mu, t \mapsto t/2\mu $ 
and then taking the limit $ \mu \to \infty $. The coefficients of the orthogonal 
polynomials $ r_N, l_N $ remain of $ {\rm O}(1) $ in this limit. 
Then we see the explicit degeneration of the following equations - 
(\ref{VI_lRecur}) $\to$ Eq. (4.23)\cite{FW_2003b},
the recurrence relations (\ref{VI_1+1rRecur:a}) $\to$ Eq. (4.60)\cite{FW_2003b},
(\ref{VI_2+1rRecur:a}) $\to$ Eq. (4.9)\cite{FW_2003b} and its conjugate to 
Eq. (4.10)\cite{FW_2003b} modulo the identity Eq. (4.5)\cite{FW_2003b},
and the hypergeometric functions (\ref{VI_2F1}) $\to$ Eq. (4.24)\cite{FW_2003b},
(\ref{VI_genH:a}) $\to$ Eq. (4.26)\cite{FW_2003b}, and
(\ref{VI_genH:b}) $\to$ Eq. (4.27)\cite{FW_2003b}.
\end{remark}

\begin{remark}
Two simple cases exist for the special values of the argument $ t=0,1 $. In the first case,
$ t=0 $, we have 
\begin{gather}
  r_{N} = (-1)^{N}{(\mu+\omega)_{N} \over (1-\mu+\bar{\omega})_{N}} ,\quad
  \bar{r}_{N} = (-1)^{N}{(-\mu+\bar{\omega})_{N} \over (1+\mu+\omega)_{N}} ,\\
  l_{N} = -{(\mu+\omega)N \over (N-\mu+\bar{\omega})} ,
  \label{VI_t=zero}
\end{gather}
whereas for $ t=1 $ (and $ \xi $ is irrelevant) we have
\begin{gather}
  r_{N} = (-1)^{N}{(\mu+\omega)_{N} \over (1+\mu+\bar{\omega})_{N}} ,\quad
  \bar{r}_{N} = (-1)^{N}{(\mu+\bar{\omega})_{N} \over (1+\mu+\omega)_{N}} ,\\
  l_{N} = -{(\mu+\omega)N \over (N+\mu+\bar{\omega})} .
\end{gather}
\end{remark}

There is a specialisation of the generalised Jacobi weights leading to a formulation 
of the orthogonal polynomial system in terms of real variables, and a simple phase 
factor appearing in the reflection coefficients. This occurs when 
$ \mu, \omega_1 \in \mathbb{R} $, $ \omega_2 = 0 $ 
and $ |t| = 1 $ and is a special case of hermitian Toeplitz matrix elements.
In such a situation $ \bar{r}_n $ is no longer independent 
of $ r_n $ (it is the complex conjugate of $ r_n $) and the coupled systems of 
difference equations reduce to a single equation.

\begin{corollary}\label{realRC}
When $ \omega_2 = 0 $, $ t \in \TT $, $ \mu, \omega_1 \in \mathbb{R} $ with
$ \mu+\omega_1 \notin \mathbb{Z}_{<0} $ and $ \bar{r}_1 = t r_ 1 $ then the 
reflection coefficients are products of a real coefficient $ x_n \in \mathbb{R} $ 
and a phase factor so that $ r_n = t^{-n/2}x_n, \bar{r}_n = t^{n/2}x_n $.
\end{corollary}
\begin{proof}
Setting $ \omega_2=0 $ in (\ref{VI_2ndRR}) we note this can be rearranged as
\begin{equation}
   (n+1+\mu+\omega)\left[ t{r_{n+1} \over r_n} - {\bar{r}_{n+1} \over \bar{r}_n}
                   \right]
 + (n-1+\mu+\omega)\left[ {r_{n-1} \over r_n} - t{\bar{r}_{n-1} \over \bar{r}_n}
                   \right] = 0 .
\end{equation}
Given that $ \bar{r}_n=t^nr_n, \bar{r}_{n-1}=t^{n-1}r_{n-1} $ we use the above 
equality to show
\begin{equation}
   \bar{r}_{n+1} = t^{n+1}r_{n+1} ,
\end{equation}
and by induction on $ n $ the statement $ \bar{r}_n = t^nr_n $ must be true $ n \geq 0 $
as it holds for $ n=0,1 $. The corollary then follows.
\end{proof}

\section{The $\tau$-function Theory for \PVI}
\label{tausection}
\setcounter{equation}{0}

In the Okamoto theory for \PVI the Hamiltonian function which governs the evolution
of $ \{ q,p;H,t \} $ through the system (\ref{VI_Hdyn}) is
\begin{multline}
  K := t(t-1)H \\
    =  q(q-1)(q-t)p^2 
     - \left[ \alpha_4(q-1)(q-t)+\alpha_3q(q-t)+(\alpha_0-1)q(q-1) \right]p
    \\
     + \alpha_2(\alpha_1+\alpha_2)(q-t) ,
\label{VI_Ham}
\end{multline}
with parameters 
$ \alpha_0,\alpha_1,\alpha_2,\alpha_3,\alpha_4 \in \CC $
subject to the constraint
$ \alpha_0+\alpha_1+2\alpha_2+\alpha_3+\alpha_4 = 1 $. 
In the introduction it was remarked that the unitary group average (\ref{VI_Toep})
was shown to be a $\tau$-function for the sixth Painlev\'e system 
\cite{FW_2002b} and this can be achieved via two distinct methods.
In the first method \cite{FW_2003a} the connection with the \PVI $\tau$-function 
was established
for an average with respect to the Cauchy unitary ensemble (see Eqs. (1.12,1.19,3.20,3.28,3.30) 
of \cite{FW_2002b}) and using the stereographic projection this average was related 
to (\ref{VI_Toep}) with the parameters
\begin{equation}
  (\alpha_0,\alpha_1,\alpha_2,\alpha_3,\alpha_4) 
   = \Big( N+1+2\omega_1,N+2\mu,-N,-\mu-\omega,-\mu-\bar{\omega} \Big).
\label{VI_CyUEparam}
\end{equation} 
The appropriate 
sequence of the Hamiltonian variables $ \{ q_n,p_n,H_n,\tau_n \}_{n=0,1,\ldots} $
in which $ N $ is only incremented is generated by a shift operator 
$ L^{-1}_{01} = r_1s_0s_1s_2s_3s_4s_2 $ in terms of the reflection operators
and Dynkin diagram automorphisms of the extended affine Weyl group
$ W_a(D^{(1)}_4) $ (see \cite{FW_2002b}). It has the action
$ L^{-1}_{01}: 
  \alpha_0 \mapsto \alpha_0+1, \alpha_1 \mapsto \alpha_1+1, \alpha_2 \mapsto \alpha_2-1 $.
For such a sequence we have the following result.
\begin{lemma}[\cite{FW_2003a},\cite{FW_2002b}]
The sequence of auxiliary variables $ \{g_n,f_n\}_{n=0,1,\ldots} $ defined by
\begin{gather}
  g_n := {q_n \over q_n - 1},
  \label{VI_L01gDefn} \\
  f_n := q_n(q_n-1)p_n + (1-\alpha_2-\alpha_4)(q_n-1)-\alpha_3q_n
                    - \alpha_0{q_n(q_n-1) \over q_n-t}
  \label{VI_L01fDefn}
\end{gather}
generated by the shift operator $ L^{-1}_{01} $ 
satisfies the discrete Painlev\'e equations associated
with the degeneration of the rational surface $ D^{(1)}_4 \to D^{(1)}_5 $
\begin{align}
  g_{n+1}g_n 
  & = {t\over t-1}{(f_n+1-\alpha_2)(f_n+1-\alpha_2-\alpha_4) \over f_n(f_n+\alpha_3)} 
  \label{VI_L01gRecur} \\
  f_n+f_{n-1} & = -\alpha_3 + {\alpha_1 \over g_n-1} +
                  {\alpha_0 t \over t(g_n-1)-g_n}
  \label{VI_L01fRecur}
\end{align}
\end{lemma}

Applying this result to the average (\ref{VI_Toep}) implies a recurrence scheme to 
compute the latter.
\begin{proposition}[\cite{FW_2003a}]\label{VI_L01dPV}
Let $\{g_N\}_{N=0,1,\dots}$, $\{f_N\}_{N=0,1,\dots}$ satisfy the discrete
Painlev\'e coupled difference equations associated
with the degeneration of the rational surface $ D^{(1)}_4 \to D^{(1)}_5 $
\begin{align}
  g_{N+1}g_N 
  & = {t\over t-1}{(f_N+N+1)(f_N+N+1+\mu+\bar{\omega}) \over f_N(f_N-\mu-\omega)}
  \label{VI_L01Recur:a} \\
  f_N+f_{N-1} 
  & = \mu+\omega+{N+2\mu \over g_N-1}
      +{(N+1+2\omega_1)t \over t(g_N-1)-g_N},
  \label{VI_L01Recur:b}
\end{align}
where $ t=1/(1-e^{i\phi}) $ subject to the initial conditions
\begin{equation*}
  g_0 = {q_0 \over q_0-1}, \quad
  f_0 = (1+\mu+\bar{\omega})(q_0-1)+(\mu+\omega)q_0-(2\omega_1+1){q_0(q_0-1) \over q_0-t}
\end{equation*}
with
\begin{equation}
  q_0 = \frac{1}{2}\left( 1+\frac{i}{\mu}{d \over d\phi}\log e^{i\mu\phi}T_1(e^{i\phi})
                   \right) .
\label{VI_L01Tinitial}
\end{equation}
Define $\{ q_N, p_N \}_{N=0,1,\dots}$ by
\begin{align}
  q_N 
  & = {g_N \over g_N-1}, \\
  p_N 
  & = {(g_N-1)^2 \over g_N}f_N \\
  & \quad
        -(N+1+\mu+\bar{\omega}){g_N-1 \over g_N}-(\mu+\omega)(g_N-1)
                 +(N+1+2\omega_1){g_N-1 \over t+(1-t)g_N} .
\nonumber 
\end{align}
Then with $T_0(e^{i\phi}) = 1$ and $T_1(e^{i\phi})=w_0(e^{i\phi}) $ as given by
(\ref{VI_toepM},\ref{VI_toepM2}), $\{T_N\}_{N=2,3,\dots}$ is specified by 
the recurrence
\begin{multline}
  -(N+\mu+\omega)(N+\mu+\bar{\omega})
  {T_{N+1}T_{N-1} \over T_N^2 } \\
  = q_N(q_N-1)p^2_N+(2\mu+2\omega_1)q_Np_N-(\mu+\bar{\omega})p_N-N(N+2\mu+2\omega_1) .
\label{VI_L01Trecur}
\end{multline}
\end{proposition}

In the second method the connection with the \PVI $\tau$-function was established
for an average with respect to the Jacobi unitary ensemble (see Eqs. (1.21,3.7,3.28,3.31,3.32) 
of \cite{FW_2002b}) and using the projection $ (-1,1) \to \TT $ under the condition 
$ \xi = 0 $ this average was related to (\ref{VI_Toep}) with the parameters
\begin{equation}
  (\alpha_0,\alpha_1,\alpha_2,\alpha_3,\alpha_4) 
   = \Big( 1-\mu-\omega,N+2\mu,-N, -\mu-\bar{\omega},N+2\omega_1 \Big).
\label{VI_JUEparam}
\end{equation} 
Sequences of the Hamiltonian variables $ \{ q_n,p_n,H_n,\tau_n \}_{n=0,1,\ldots} $
are now generated by the shift operator 
$ L^{-1}_{14} = r_3s_1s_4s_2s_0s_3s_2 $. It has the action
$ L^{-1}_{14}: 
  \alpha_1 \mapsto \alpha_1+1, \alpha_2 \mapsto \alpha_2-1, \alpha_4 \mapsto \alpha_4+1 $.
Using the methods of \cite{FW_2002b} we have the following result.
\begin{lemma}
The sequence of auxiliary variables $ \{g_n,f_n\}_{n=0,1,\ldots} $ defined by
\begin{align}
  g_n := & {q_n-t \over q_n - 1},
  \label{VI_L14gDefn} \\
  f_n := & {1 \over 1-t}\bigg[ (q_n-t)(q_n-1)p_n 
  \label{VI_L14fDefn} \\
  & \phantom{{1 \over 1-t}\bigg[} 
       + (1-\alpha_0-\alpha_2)(q_n-1)-\alpha_3(q_n-t)
                    - \alpha_4{(q_n-t)(q_n-1) \over q_n} \bigg]
  \nonumber
\end{align}
generated by the shift operator $ L^{-1}_{14} $ 
satisfies the discrete Painlev\'e equations associated
with the degeneration of the rational surface $ D^{(1)}_4 \to D^{(1)}_5 $
\begin{align}
  g_{n+1}g_n 
  & = t{(f_n+1-\alpha_2)(f_n+1-\alpha_0-\alpha_2) \over f_n(f_n+\alpha_3)} 
  \label{VI_L14gRecur} \\
  f_n+f_{n-1} & = -\alpha_3 + {\alpha_1 \over g_n-1} +
                  {\alpha_4 t \over g_n-t}
  \label{VI_L14fRecur}
\end{align}
\end{lemma}
\begin{proof}
Using the action of the fundamental reflections and Dynkin diagram automorphisms
given in Table 1 of \cite{FW_2002b} we compute the action of $ L^{-1}_{14} $ on
$ q $ and write it in the following way,
\begin{multline*}
{(q-t)(\hat{q}-t) \over (q-1)(\hat{q}-1)} = 
 t[ q(q-1)(q-t)p+(\alpha_1+\alpha_2)q^2
                -((\alpha_0+\alpha_1+\alpha_2)t-\alpha_0-\alpha_4)q-\alpha_4t ] \\
 \times
  [ q(q-1)(q-t)p+(\alpha_1+\alpha_2)q^2
                -((\alpha_1+\alpha_2)t-\alpha_4)q-\alpha_4t ] \\
 \div
  [ q(q-1)(q-t)p+(\alpha_1+\alpha_2)q^2
                -(-\alpha_4t+\alpha_1+\alpha_2)q-\alpha_4t ] \\
 \div
  [ q(q-1)(q-t)p+(\alpha_1+\alpha_2)q^2
                -(-(\alpha_3+\alpha_4)t+\alpha_1+\alpha_2+\alpha_3)q-\alpha_4t ] ,
\end{multline*}
where $ q:=q_n, \hat{q}:=q_{n+1} $. From the definitions 
(\ref{VI_L14gDefn},\ref{VI_L14fDefn}) this result can be readily recast as
(\ref{VI_L14gRecur}). The second (\ref{VI_L14fRecur}) follows from a computation
for $ f_n+f_{n-1} $ using the shift operator $ L_{14} $.
\end{proof}

\begin{remark}
The two systems of recurrences (\ref{VI_L01gRecur},\ref{VI_L01fRecur}) and 
(\ref{VI_L14gRecur},\ref{VI_L14fRecur}) are related by an element of the $ S_4 $ subgroup
of the $ W_a(F_4) $ transformations, namely the generator $ x^3 $ \cite{Ok_1987a}.
This has the action 
\begin{equation}
   x^3: \alpha_0 \leftrightarrow \alpha_4, t \mapsto {t \over t-1},
        q \mapsto {t-q \over t-1}, p \mapsto -(t-1)p ,
\end{equation}
and when applying these transformations to (\ref{VI_L01gDefn}), (\ref{VI_L01fDefn}),
(\ref{VI_L01gRecur}), (\ref{VI_L01fRecur}) we recover (\ref{VI_L14gDefn}), 
(\ref{VI_L14fDefn}), (\ref{VI_L14gRecur}), (\ref{VI_L14fRecur}) respectively.
\end{remark}

\begin{proposition}\label{VI_L14dPV}
Let $\{g_N\}_{N=0,1,\dots}$, $\{f_N\}_{N=0,1,\dots}$ satisfy the discrete
Painlev\'e coupled difference equations associated
with the degeneration of the rational surface $ D^{(1)}_4 \to D^{(1)}_5 $
\begin{align}
  g_{N+1}g_N 
  & = t{(f_N+N+1)(f_N+N+\mu+\omega) \over f_N(f_N-\mu-\bar{\omega})}
  \label{VI_L14Recur:a} \\
  f_N+f_{N-1} 
  & = \mu+\bar{\omega}+{N+2 \mu \over g_N-1}
      +{(N+2\omega_1)t \over g_N-t},
  \label{VI_L14Recur:b}
\end{align}
where $ t = e^{i\phi} $ subject to the initial conditions
\begin{equation*}
  g_0 = {q_0-t \over q_0-1}, \quad
  f_0 = {1 \over 1-t}\left[ (\mu+\omega)(q_0-1)+(\mu+\bar{\omega})(q_0-t)
                    - 2\omega_1{(q_0-t)(q_0-1) \over q_0} \right]
\end{equation*}
with
\begin{equation}
  q_0 = {\omega_1 \over \mu}
  { -i\dfrac{d}{d\phi}\log e^{i\mu\phi}T_1(e^{i\phi}) \over 
   \mu+\omega+i\dfrac{d}{d\phi}\log e^{i\mu\phi}T_1(e^{i\phi}) } .
\label{VI_L14Tinitial}
\end{equation}
Define $\{ q_N,p_N \}_{N=0,1,\dots}$ in terms of $\{f_N,g_N\}_{N=0,1,\dots}$ by
\begin{gather}
  q_N = {g_N-t \over g_N-1}, \\
  p_N = {g_N-1 \over (1-t)g_N} \Big[
       (g_N-1)f_N-(\mu+\bar{\omega})g_N+(N+2\omega_1){(1-t)g_N \over g_N-t}
                 -N-\mu-\omega \Big].
\end{gather}
Then with $T_0(e^{i\phi}) = 1$ and $T_1(e^{i\phi})=w_0(e^{i\phi}) $ as given by
(\ref{VI_toepM},\ref{VI_toepM2}), $\{T_N\}_{N=2,3,\dots}$ is specified by 
the recurrence
\begin{multline}
  -(N+\mu+\omega)(N+\mu+\bar{\omega})
     {T_{N+1}T_{N-1} \over T_N^2 } \\
  = q_N(q_N-1)^2p^2_N+[(2\mu-N)q_N+N+2\omega_1](q_N-1)p_N-2\mu Nq_N-N(N+2\omega_1) .
\label{VI_L14Trecur}
\end{multline}
\end{proposition}
\begin{proof}
Let $ Y_n := L^{-1}_{14}K_n - K_n = K_{n+1}-K_n $ 
and from Table 1 of \cite{FW_2002b} we have
\begin{equation*}
   Y_n = -{(t-1)q_n \over q_n-1} \left\{ (q_n-1)p_n+\alpha_0+\alpha_2-1
   + { (1-\alpha_0-\alpha_2)(\alpha_1+\alpha_2+\alpha_3) \over
        q_n(q_n-1)p_n+(\alpha_1+\alpha_2)q_n+\alpha_4 } \right\} .
\label{PVI_Y14}
\end{equation*}
Now consider
\begin{align*}
   t(t-1){d \over dt}\log{\tau_{n+1}\tau_{n-1} \over \tau_n^2}
  & = K_{n+1}+K_{n-1}-2K_n \\
  & = Y_n-L_{14}Y_n ,
\end{align*}
and this latter difference, upon again consulting Table 1 of \cite{FW_2002b}, 
turns out to be
\begin{multline*}
  Y_n-L_{14}Y_n = \\
 t(t-1){d \over dt}\log\left[
 q_n(q_n-1)^2p^2_n + [(\alpha_1+2\alpha_2)q_n+\alpha_4](q_n-1)p_n
    +\alpha_2[(\alpha_1+\alpha_2)q_n+\alpha_4] \right] .
\end{multline*}
After integrating both expressions and introducing an integration constant
Eq. (\ref{VI_L14Trecur}) follows.
\end{proof}

We now seek to relate the results of the $\tau$-function approach to the theory 
developed for the
orthogonal polynomials on the unit circle with semi-classical weights 
as given in the previous section. However we will only discuss the scheme
given in Proposition \ref{VI_L01dPV} as this is the simplest.

\begin{proposition}
The transformations linking the Hamiltonian variables $ q_N, p_N $ in Proposition
\ref{VI_L01dPV} to the reflection coefficients $ r_N, \bar{r}_N $ for the system 
of orthogonal polynomials with the weight (\ref{VI_wgt}) are given implicitly by
\begin{gather}
   q_Np_N+\mu+\bar{\omega}
   = {(N+\mu+\bar{\omega})r_N\bar{r}_N \over (N+\mu+\bar{\omega})r_N\bar{r}_N-\mu+\omega}
     {1\over q_N-1} \nonumber \\
  \times
     \left[ (N+2\omega_1)(q_N-1)
            - t{l_N \over \kappa_N}+Nt
   +(N+1+\mu+\bar{\omega})(1-r_N\bar{r}_N)t{r_{N+1} \over r_N} \right]
   \label{VIH_Ops:a} \\
   = (N+\mu+\bar{\omega})[ (N+\mu+\omega)r_N\bar{r}_N-\mu+\bar{\omega} ] \nonumber \\
  \times
     { q_N \over (N+2\omega_1)q_N
     +t\dfrac{l_N}{\kappa_N}-Nt
   -(N-1+\mu+\bar{\omega})(1-r_N\bar{r}_N)t
    \dfrac{\bar{r}_{N-1}}{\bar{r}_N}} ,
  \label{VIH_Ops:b} \\
  (q_N-1)p_N+\mu+\omega
   = (N+\mu+\omega)[ (N+\mu+\bar{\omega})r_N\bar{r}_N-\mu+\omega ] \nonumber \\
  \times
     { q_N-1 \over (N+2\omega_1)(q_N-1)
           - t\dfrac{l_N}{\kappa_N}+Nt
   +(N+1+\mu+\bar{\omega})(1-r_N\bar{r}_N)t\dfrac{r_{N+1}}{r_N} }
  \label{VIH_Ops:c} \\
   ={(N+\mu+\omega)r_N\bar{r}_N \over (N+\mu+\omega)r_N\bar{r}_N-\mu+\bar{\omega}}
     {1\over q_N} \nonumber \\
  \times
     \left[ (N+2\omega_1)q_N
    +t{l_N \over \kappa_N}-Nt
    -(N-1+\mu+\bar{\omega})(1-r_N\bar{r}_N){\bar{r}_{N-1} \over \bar{r}_N} \right] .
  \label{VIH_Ops:d}
\end{gather}
\end{proposition}
\begin{proof}
We require in addition to the primary shift operator
$ L^{-1}_{01} $ generating the $ N \mapsto N+1 $ sequence another operator which 
has the action $ i\omega_2 \mapsto i\omega_2-1 $. This is
the secondary shift operator $ T^{-1}_{34} = r_1s_4s_2s_0s_1s_2s_4 $ and has the
action $ T^{-1}_{34}: \alpha_3 \to \alpha_3+1, \alpha_4 \to \alpha_4-1 $.
From Table 1 of \cite{FW_2002b} we compute the actions of $ T^{-1}_{34}, T_{34} $ on
the Hamiltonian to be
\begin{align*}
  T^{-1}_{34}\cdot K_n - K_n 
  = & -q_n(q_n-1)p_n \\
    & +(\alpha_0+\alpha_4-1)(q_n-1)
      -(\alpha_2+\alpha_3)(\alpha_1+\alpha_2+\alpha_3){q_n-1 \over (q_n-1)p_n-\alpha_3}
  \\
  T_{34}\cdot K_n - K_n 
  = & -q_n(q_n-1)p_n \\
    & +(\alpha_0+\alpha_3-1)q_n
      -(\alpha_2+\alpha_4)(\alpha_1+\alpha_2+\alpha_4){q_n \over q_np_n-\alpha_4}
\end{align*}
However
\begin{equation*}
  T^{-1}_{34}\cdot K_n - K_n = t(t-1){d \over dt}\log{I^{1}_n \over I^{0}_n}
  = t(t-1){d \over dt}\log r_n 
\end{equation*}
and we employ the results of Corollary \ref{cor_rdot} and the evaluation of the
coefficient functions in (\ref{VI_omega:a},\ref{VI_omega:b}) to arrive at
\begin{align*}
   (t-1){\dot{r}_N \over r_N} & = {l_N \over \kappa_N}-N
   -(N+1+\mu+\bar{\omega})(1-r_N\bar{r}_N){r_{N+1} \over r_N} ,
   \\
   (t-1){\dot{\bar{r}}_N \over \bar{r}_N} & = -{l_N \over \kappa_N}+N
   +(N-1+\mu+\bar{\omega})(1-r_N\bar{r}_N){\bar{r}_{N-1} \over \bar{r}_N} .
\end{align*}
In addition we note that after recalling (\ref{ops_I0}), (\ref{VI_L01Trecur})
factorises into
\begin{equation*}  
   (N+\mu+\omega)(N+\mu+\bar{\omega})r_N\bar{r}_N 
  = [q_Np_N+\mu+\bar{\omega}][(q_N-1)p_N+\mu+\omega] .
\end{equation*}  
The stated results, (\ref{VIH_Ops:a}-\ref{VIH_Ops:d}), then follow.
\end{proof}

\section{Applications to Physical Models}
\label{Modelsection}
\setcounter{equation}{0}

\subsection{Random Matrix Averages}

A specialisation of the above results with great interest in the application of 
random matrices \cite{KS_2000b} is the quantity
\begin{equation}
   F^{\rm CUE}_N(u;\mu) := 
   \Big\langle \prod^{N}_{l=1}|u+z_l|^{2\mu} \Big\rangle_{{\rm CUE}_N} .
\label{PVI_CUE}
\end{equation}
This has the interpretation as the average of the $2\mu$-th power of the absolute 
value of the characteristic polynomial for the CUE. In the case $ |u|=1 $
(\ref{PVI_CUE}) is independent of $ u $ and has the well-known (see e.g. 
\cite{BF_1997}) Gamma function evaluation
\begin{equation}
 \Big\langle \prod^{N}_{l=1}|u+z_l|^{2\mu} \Big\rangle_{{\rm CUE}_N}\Big|_{u=e^{i\phi}}
  = \Big\langle \prod^{N}_{l=1}|1+z_l|^{2\mu} \Big\rangle_{{\rm CUE}_N}
  = \prod^{N-1}_{j=0}{j!\Gamma(j+1+2\mu) \over \Gamma^2(j+1+\mu)} ,
\end{equation}
when $ \Re(\mu) > -\half $. For $ |u|<1 $ we see by an appropriate change of variables that
\begin{align}                                       
    \Big\langle \prod^{N}_{l=1}|u+z_{l}|^{2\mu} \Big\rangle_{{\rm CUE}_N}
  & = \Big\langle \prod^{N}_{l=1}(1+|u|^2z_{l})^{\mu}(1+1/z_{l})^{\mu} \Big\rangle_{{\rm CUE}_N}
  \\
  & = {}^{\vphantom{(1)}}_{2}F^{(1)}_{1}(-\mu,-\mu;N;t_1,\ldots,t_N)|_{t_1=\ldots =t_N=|u|^2},
\label{CUEp_genH}
\end{align}
where the second equality follows from (\ref{VI_2F1}). 
For $ |u|>1 $ we can use the simple functional equation
\begin{equation}                                       
    \Big\langle \prod^{N}_{l=1}|u+z_{l}|^{2\mu} \Big\rangle_{{\rm CUE}_N}
  = |u|^{2\mu N}\Big\langle \prod^{N}_{l=1}|{1\over u}+z_{l}|^{2\mu} \Big\rangle_{{\rm CUE}_N}
\end{equation}
to relate this case back to the case $ |u|<1 $.

The weight in the first equality of (\ref{CUEp_genH}) is a special case of 
(\ref{VI_wgt}). In terms of the parameters of the form (\ref{VI_wgt}) we observe 
that $ \xi = 0 $, $ 2\mu \mapsto \mu $, $ \omega = \bar{\omega} = \mu/2 $, 
i.e. $ \omega_2 = 0 $ and $ t = |u|^2 $. The trigonometric moments are 
\begin{align}
   w_{-n} & =
   {\Gamma(\mu+1) \over n!\Gamma(\mu+1-n)}{}_2F_1(-\mu,-\mu+n;n+1;|u|^2)
   \qquad n \in \mathbb{Z}_{\geq 0} \\
   w_{n}  & = |u|^{2n}w_{-n} \qquad n \in \mathbb{Z}_{\geq 0} .
\end{align}
The results of Section \ref{PVIsection} then allow (\ref{PVI_CUE})
to be computed by a recurrence involving the corresponding reflection coefficients.

\begin{corollary}
The general moments of the characteristic polynomial $ |\det(u+U)| $ for arbitrary 
exponent $ 2\mu $ with respect to the finite CUE ensemble $ U \in U(N) $ of rank $ N $ 
is given by the system of recurrences
\begin{equation}
  {F^{\rm CUE}_{N+1}F^{\rm CUE}_{N-1} \over (F^{\rm CUE}_{N})^2} = 1-|u|^{2N}r^2_N ,
\end{equation}
with initial values
\begin{equation}
  F^{\rm CUE}_{0} = 1, \qquad F^{\rm CUE}_{1} = {}_2F_1(-\mu,-\mu;1;|u|^2) , 
\end{equation}
and the recurrence relation for the reflection coefficient $ r_N $
\begin{multline}
   2|u|^{2N}r_{N}r_{N-1}-|u|^2-1
   = {1-|u|^{2N}r^2_{N} \over r_{N}}
       \left[ (N+1+\mu)|u|^2r_{N+1} + (N-1+\mu)r_{N-1} \right]
   \\
   - {1-|u|^{2(N-1)}r^2_{N-1} \over r_{N-1}}
       \left[ (N+\mu)|u|^2r_{N} + (N-2+\mu)r_{N-2} \right],
   \label{CUEp_rRecur}
\end{multline}
subject to the initial values
\begin{equation}
   r_{0} = 1, \quad
   r_{1} = -\mu{{}_2F_1(-\mu,-\mu+1;2;|u|^2) \over {}_2F_1(-\mu,-\mu;1;|u|^2)} .
\end{equation}
\end{corollary}

\begin{proof}
From either (\ref{VI_2ndRR}), (\ref{VI_lRecur}) or (\ref{VI_2+1rRecur:a}) and 
the fact that $ \bar{r}_1 = |u|^2r_1 $ we can repeat the arguments of Corollary
\ref{realRC} to deduce that $ \bar{r}_N = |u|^{2N}r_N $ for $ N \geq 0 $.
The recurrence relation follows simply from the specialisation of 
(\ref{VI_rRecur:a}) and the initial conditions from the $ N=1 $ case.
\end{proof}
  
Another spectral statistic of fundamental importance in random matrix theory
is the gap probability for the circular unitary ensembles, and this is the
specialisation whereby $ \mu = \omega = \bar{\omega} = 0 $, 
$ |t| = 1 $ so the angle $ \phi \in [0,2\pi) $, whilst $ \xi \in \mathbb{C} $ is 
general (one is mainly interested in an open neighbourhood of $ \xi = 1 $). 
The generating function for the probability of finding exactly $ k $ eigenvalues 
$ z = e^{i\theta} $ within the sector of the unit circle 
$ \theta \in (\pi-\phi,\pi] $ is denoted by $ E^{\rm CUE}_N((0,\phi);\xi) $ and
has the definition
\begin{equation}
 E^{\rm CUE}_N((0,\phi);\xi) := {1 \over C_N}
 \left( \int^{\pi}_{-\pi} - \xi\int^{\pi}_{\pi-\phi} \right) d\theta_1 \ldots
 \left( \int^{\pi}_{-\pi} - \xi\int^{\pi}_{\pi-\phi} \right) d\theta_N
 \prod_{1 \leq j < k \leq N} |e^{i\theta_j}-e^{i\theta_k}|^2 ,
\label{PVI_CUE:b}
\end{equation}
where the normalisation $ C_N = (2\pi)^N N! $.
It is well known that Toeplitz elements with such a symbol have the form
\begin{equation}
   w_n = \delta_{n,0} + {\xi \over 2\pi i}(-1)^{n+1}{t^n-1 \over n} ,
\end{equation}
which is easily recovered from the general expression (\ref{VI_toepM}). A
recurrence scheme for the generating function (\ref{PVI_CUE:b}) involving
particular examples of the coupled discrete Painlev\'e equations 
(\ref{dPV:a}), (\ref{dPV:b}) has been presented in \cite{FW_2003a}. Here we
use recurrences found herein for $ r_N $, $ \bar{r}_N $, together with the 
fact that for (\ref{PVI_CUE:b}) one has $ r_N = t^{-N}\bar{r}_N $, to replace
the role of the coupled recurrences from \cite{FW_2003a} by a single recurrence.

\begin{corollary}\label{CUE_recur}
The generating function for the probability of finding exactly $ k $ eigenvalues 
$ z = e^{i\theta} $ from the ensemble of random $ N \times N $ unitary matrices
within the sector of the unit circle $ \theta \in (\pi-\phi,\pi] $ 
is given by the following system of recurrences in the rank of the ensemble $ N $,
\begin{equation}
   {E^{\rm CUE}_{N+1}E^{\rm CUE}_{N-1} \over (E^{\rm CUE}_{N})^2}
  = 1 - x^2_N ,
\end{equation}
where the initial values are 
\begin{equation}
  E^{\rm CUE}_{0} = 1, \quad E^{\rm CUE}_{1} = 1 - {\xi \over 2\pi}\phi ,
\end{equation}
and the auxiliary variables $ x_N $ are determined by the quasi-linear third order
recurrence relation
\begin{multline}
   2x_{N}x_{N-1} - 2\cos{\phi \over 2}
  = {1-x^2_N \over x_N}\left[ (N+1)x_{N+1}+(N-1)x_{N-1} \right] \\
    - {1-x^2_{N-1} \over x_{N-1}}\left[ Nx_{N}+(N-2)x_{N-2} \right] ,
\end{multline}
or the quadratic second order recurrence relation
\begin{multline}
 (1-x^2_N)^2\left[ (N+1)^2x^2_{N+1}+(N-1)^2x^2_{N-1} \right]
   + 2(N^2-1)(1-x^4_N)x_{N+1}x_{N-1} \\
 + 4N\cos{\phi \over 2}x_N(1-x^2_N)\left[ (N+1)x_{N+1}+(N-1)x_{N-1} \right]
    + 4N^2x^2_N\left[ \cos^2{\phi \over 2}-x^2_N \right] = 0,
\end{multline}
along with the initial values
\begin{equation}
   x_{-1} = 0, \quad x_{0} = 1, \quad
   x_{1} = -{\xi \over \pi}{\sin\dfrac{\phi}{2} \over 
                            1 - \dfrac{\xi}{2\pi}\phi } .
\end{equation} 
\begin{proof}
The first recurrence relation follows directly from the general recurrence 
(\ref{VI_rRecur:a}) and Corollary \ref{realRC} whilst the second follows from 
(\ref{VI_1+1rRecur:a}).
\end{proof}
\end{corollary}

\subsection{2-D Ising Model}
It has been known for some time that the diagonal spin correlations in the 
square lattice Ising model could be evaluated in terms of the Painlev\'e sixth 
transcendent
\cite{JM_1980} and in this work a coupled system of difference equations 
involving eleven variables were given. 
Here we give what we consider to be the simplest set of recurrence relations 
for these correlations as a special case of the general theory above.
The diagonal correlation functions are given by \cite{McCW_1973}
\begin{equation}
   \langle \sigma_{0,0}\sigma_{N,N} \rangle = 
   \begin{cases}
      \det (a_{i-j}(k))_{1 \leq i,j \leq N} 
      & \text{if $ k > 1 $ or $ T < T_c $}, \\
      \det (\tilde{a}_{i-j}(k))_{1 \leq i,j \leq N} 
      & \text{if $ k < 1 $ or $ T > T_c $},
   \end{cases}
\end{equation}
where
\begin{equation}
   a_{n}(k) := 
   {1\over 2\pi i} \int_{\TT} d\zeta \zeta^n \sqrt{k\zeta^{-1}-1 \over k\zeta-1},
   \quad
   \tilde{a}_{n}(k) := 
   {1\over 2\pi i} \int_{\TT} d\zeta \zeta^{n-1} \sqrt{1-k\zeta \over 1-k\zeta^{-1}},
\end{equation}
and the argument is defined $ k = \sinh^2(2J/k_{B}T) $, $ J $ being the coupling
strength and $ T $ the temperature.
The weight appearing in the Toeplitz determinant form of the low temperature
correlation is 
\begin{equation}
   w(z) = C z^{1/4}|1+z|^{-1/2}(1+k^{-2}z)^{1/2}
        = C z^{1/2}(1+z)^{-1/2}(1+k^{-2}z)^{1/2},
\end{equation}
which is a special case of (\ref{VI_wgt}) with $ \xi = 0 $,
$ \mu = 1/4, \omega_1 = -1/4, \omega_2 = i/2 $ and $ t = 1/k^2 $. Here 
$ a_{n} = (-1)^{n}k^{-n}w_{-n}(1/k^2) $. In the high temperature regime the
exponents $ \mu, \omega_1 $ are sign reversed and $ t = k^2 $, so
$ \tilde{a}_{n} = (-1)^{n}k^{n}w_{-n}(k^2) $.  So these cases
form an interesting example where $ \omega_2 \neq 0 $, so both $ r_{N}, \bar{r}_{N} $
are distinct and independent in contrast to the random matrix and quantum 
many-body cases. As is also well known the Toeplitz matrix elements in the low 
temperature regime are given by
\begin{align}
  w_{-n} & = {(-1)^n \over \pi}{\Gamma(n+\half)\Gamma(\half) \over \Gamma(n+1)}
             {}_2F_1(-\half,n+\half;n+1;k^{-2}), \quad n \geq 0,
         \\
  w_{n}  & = {(-1)^{n+1}k^{-2n} \over \pi}{\Gamma(n-\half)\Gamma(\thalf) \over \Gamma(n+1)}
             {}_2F_1(\half,n-\half;n+1;k^{-2}), \quad n > 0,
\end{align}
whilst those in the high temperature regime are
\begin{align}
  w_{-n} & = {(-1)^nk^{2n+1} \over \pi}{\Gamma(n+\half)\Gamma(\thalf) \over \Gamma(n+2)}
             {}_2F_1(\half,n+\half;n+2;k^{2}), \quad n \geq 0,
         \\
  w_{n}  & = {(-1)^{n-1} \over \pi k}{\Gamma(n-\half)\Gamma(\half) \over \Gamma(n)}
             {}_2F_1(-\half,n-\half;n;k^{2}), \quad n > 0.
\end{align}

\begin{corollary}
The diagonal correlation function for the Ising model valid in both the low and
high temperature phases (with $ k \mapsto 1/k $ in the latter case) is determined by
\begin{equation}
 \frac{\langle \sigma_{0,0}\sigma_{N+1,N+1} \rangle\langle \sigma_{0,0}\sigma_{N-1,N-1} \rangle}
      {\langle \sigma_{0,0}\sigma_{N,N} \rangle^2}
  = 1-r_{N}\bar{r}_{N},
\end{equation}
along with the quasi-linear $ 2/1 $ 
\begin{multline}
   (2N+3)k^{-2}(1-r_{N}\bar{r}_{N})r_{N+1}
     + 2N\left[ k^{-2}+1-(2N-1)k^{-2}r_{N}\bar{r}_{N-1} \right]r_{N} \\
     + (2N-3)\left[ (2N-1)r_{N}\bar{r}_{N}+1 \right]r_{N-1} = 0,
   \label{ising_rRecur:a}
\end{multline}
and $ 1/2 $ recurrence relations 
\begin{multline}
   (2N+1)(1-r_{N}\bar{r}_{N})\bar{r}_{N+1}
     + 2N\left[ (2N-3)\bar{r}_{N}r_{N-1}+k^{-2}+1 \right]\bar{r}_{N} \\
     + (2N-1)k^{-2}\left[ -(2N+1)r_{N}\bar{r}_{N}+1 \right]\bar{r}_{N-1} = 0,
   \label{ising_rRecur:b}
\end{multline}
subject to initial conditions for the low temperature regime
\begin{equation}                                       
   r_{0} = 1, \quad \bar{r}_{0} = 1, \quad
   r_{1} = {2-k^2 \over 3}+{k^2-1 \over 3}{{\rm K}(k^{-1}) \over {\rm E}(k^{-1})},
   \quad
   \bar{r}_{1} = -1+{k^2-1 \over k^2}{{\rm K}(k^{-1}) \over {\rm E}(k^{-1})},
\end{equation}
or to the initial conditions for the high temperature regime given by
\begin{equation}                                       
   r_{0} = 1, \quad \bar{r}_{0} = 1, \quad
   r_{1} = {1 \over 3}\left\{
            {2 \over k^2}-{{\rm E}(k) \over (k^2-1){\rm K}(k)+{\rm E}(k)} \right\},
   \quad
   \bar{r}_{1} = -{k^2{\rm E}(k) \over (k^2-1){\rm K}(k)+{\rm E}(k)},
\end{equation}
where $ {\rm K}(k), {\rm E}(k) $ are the complete elliptic integrals of the first
and second kind respectively.
\end{corollary}

\begin{proof}
(\ref{ising_rRecur:a},\ref{ising_rRecur:b}) follow from 
(\ref{VI_2+1rRecur:a}) and its "conjugate" upon the specialisation to the parameters
above. The initial conditions follow from explicit evaluation of the Toeplitz
determinants.
\end{proof}

The correlation function and reflection coefficients have particularly simple, yet
general forms when expressed in terms of generalised hypergeometric functions. 

\begin{corollary}
In the low temperature phase the diagonal correlation function is given by
\begin{equation}                                       
    \langle \sigma_{0,0}\sigma_{N,N} \rangle =
  {}^{\vphantom{(1)}}_{2}F^{(1)}_{1}(-\half,\half;N;t_1,\ldots,t_N)|_{t_1=\ldots =t_N=1/k^2},
\label{ising_genH:a}
\end{equation}
whilst the reflection coefficients are given by
\begin{align}
   r_{N} 
  & = (-1)^{N}{(-\half)_{N} \over N!}
  { {}^{\vphantom{(1)}}_{2}F^{(1)}_{1}(-\half,\thalf;N+1;t_1,\ldots,t_N)
      \over 
    {}^{\vphantom{(1)}}_{2}F^{(1)}_{1}(-\half,\half;N;t_1,\ldots,t_N) 
  }\Bigg|_{t_1=\ldots =t_N=1/k^2},
  \label{}\\
   \bar{r}_{N} 
  & = (-1)^{N}{(N-1)! \over (\half)_{N}}
  { \lim_{\epsilon \to 0}\epsilon
    {}^{\vphantom{(1)}}_{2}F^{(1)}_{1}(-\half,-\half;N-1+\epsilon;t_1,\ldots,t_N)
      \over 
    {}^{\vphantom{(1)}}_{2}F^{(1)}_{1}(-\half,\half;N;t_1,\ldots,t_N)
  }\Bigg|_{t_1=\ldots =t_N=1/k^2}.
  \label{}
\end{align}
In the high temperature phase the diagonal correlation function is
\begin{equation}                                       
    \langle \sigma_{0,0}\sigma_{N,N} \rangle =
   \frac{(2N-1)!!}{2^NN!}k^N
  {}^{\vphantom{(1)}}_{2}F^{(1)}_{1}(\half,\half;N+1;t_1,\ldots,t_N)|_{t_1=\ldots =t_N=k^2},
\label{ising_genH:b}
\end{equation}
and the reflection coefficients are given by
\begin{align}
   r_{N} 
  & = (-1)^{N}{(-\half)_{N} \over (N+1)!}
  { {}^{\vphantom{(1)}}_{2}F^{(1)}_{1}(\half,\thalf;N+2;t_1,\ldots,t_N)
      \over 
    {}^{\vphantom{(1)}}_{2}F^{(1)}_{1}(\half,\half;N+1;t_1,\ldots,t_N) 
  }\Bigg|_{t_1=\ldots =t_N=k^2},
  \label{}\\
   \bar{r}_{N} 
  & = (-1)^{N}{N! \over (\half)_{N}}
  { {}^{\vphantom{(1)}}_{2}F^{(1)}_{1}(\half,-\half;N;t_1,\ldots,t_N)
      \over 
    {}^{\vphantom{(1)}}_{2}F^{(1)}_{1}(\half,\half;N+1;t_1,\ldots,t_N)
  }\Bigg|_{t_1=\ldots =t_N=k^2}.
  \label{}
\end{align}
\end{corollary}

\begin{proof}
The evaluations in the low temperature phase follow from 
(\ref{VI_genH:a},\ref{VI_genH:b}), although some care needs to be taken with 
$  \bar{r}_{N} $ because $ -\mu+\bar{\omega} = 0 $. 
The limit that arises has a series development
\begin{multline}
\lim_{\epsilon \to 0}\epsilon
    {}^{\vphantom{(1)}}_{2}F^{(1)}_{1}(-\half,-\half;N-1+\epsilon;t_1,\ldots,t_N) \\
  = \sum^{\infty}_{\kappa:l(\kappa) = N}
    {([-\half]^{(1)}_{\kappa})^2 \over [N]^{(1)}_{\kappa}}
    {\prod^{N}_{j=1}(N-j+\kappa_{j}) \over (N-1)!}
    {s_{\kappa}(t_1,\ldots,t_N) \over h_{\kappa}}
\end{multline}
so that only those terms with lengths $ l(\kappa) = N $ contribute to the sum.
The high temperature expressions follow from the low temperature ones through 
the transformation $ \mu \leftrightarrow \omega_1 $.
\end{proof}

It is of interest to note that as $ N $ grows more of the leading order terms in 
the expansion of (\ref{ising_genH:a}) become independent of $ N $, and the
following limit becomes explicit
\begin{equation}                                       
  \lim_{N \to \infty} \langle \sigma_{0,0}\sigma_{N,N} \rangle
   = (1-k^{-2})^{1/4}.
\label{}
\end{equation}

At zero temperature, $ k = \infty $, the solutions simplify to
\begin{equation}                                       
   r_{N} = (-1)^{N}{(-\half)_{N} \over N!}, \quad
   \bar{r}_{N} = 0 \; (N \geq 1), \quad 
    \langle \sigma_{0,0}\sigma_{N,N} \rangle = 1,
\end{equation}
whilst at the critical point, $ k = 1 $, we have the simple solutions
\begin{gather}                                       
   r_{N} = {(-1)^{N-1} \over (2N+1)(2N-1)}, \quad
   \bar{r}_{N} = (-1)^{N}, \quad l_{N} = {N \over 2N+1}, \\
    \langle \sigma_{0,0}\sigma_{N,N} \rangle = 
    \prod^{N}_{j=1}{\Gamma^2(j) \over \Gamma(j+\half)\Gamma(j-\half)},
\end{gather}
and at infinite temperature they become
\begin{equation}                                       
   r_{N} = (-1)^{N}{(-\half)_{N} \over (N+1)!}, \quad
   \bar{r}_{N} = (-1)^{N}{ N! \over (\half)_{N}}, \quad 
    \langle \sigma_{0,0}\sigma_{N,N} \rangle = 0 \; (N \geq 1),
\end{equation}
in agreement with the known results \cite{McCW_1973}.

\subsection*{Acknowledgments}
This research has been supported by the Australian Research Council. NSW appreciates
the generosity of Will Orrick in supplying expansions of the Toeplitz determinants for
the diagonal correlations of the Ising model and the assistance of Paul Leopardi in 
calculating gap probabilities for the CUE. Our manuscript has benefited from the critical
reading by Alphonse Magnus and we thank him, Mourad Ismail and Percy Deift for their
advice and suggestions.

\bibliographystyle{amsplain}
\bibliography{moment,nonlinear,random_matrices}

\end{document}